\documentclass[aps,prd,11pt,twoside,tightenlines,nofootinbib,showpacs,preprint,superscriptaddress]{revtex4-1}

\usepackage[colorlinks=true]{hyperref}
\usepackage{xcolor}
\hypersetup{colorlinks=true, citecolor=blue, urlcolor=blue, linkcolor=blue}
\usepackage{amsmath,amssymb}
\usepackage[final]{graphicx}
\usepackage{subcaption}

\begin{document}

\title{Study the molecular nature of $\sigma$, $f_{0}(980)$, and $a_{0}(980)$ states}

\author{Hiwa A. Ahmed}
\affiliation{School of Physics and Electronics, Central South University, Changsha 410083, China}
\affiliation{Physics Department, College of Science, University of Sulaimani, Kurdistan Region 46001, Iraq}

\author{C. W. Xiao}
\email{xiaochw@csu.edu.cn}
\affiliation{School of Physics and Electronics, Central South University, Changsha 410083, China}

\date{\today}

\begin{abstract}

We investigate the characteristics of $\sigma$, $f_{0}(980)$, and $a_{0}(980)$ with the formalism of chiral unitary approach. With the dynamical  generation of them, we make a further study of their properties by evaluating the couplings, the compositeness, the wave functions and the radii. We also research their properties in the single channel interactions, where the $a_{0}(980)$ can not be reproduced in the $K\bar{K}$ interactions with isospin $I=1$ since the potential is too weak. In our results, the states of $\sigma$ and $f_{0}(980)$ can be dynamically reproduced stably with varying cutoffs both in the coupled channel and the single channel cases. We find that the $\pi\eta$ components is much important in the coupled channel interactions to dynamically reproduce the $a_{0}(980)$ state, which means that $a_{0}(980)$ state can not be a pure $K\bar{K}$ molecular state. We obtain their radii as: $|\langle r^2 \rangle|_{f_0(980)} = 1.80 \pm 0.35$ fm, $|\langle r^2 \rangle|_{\sigma} = 0.68 \pm 0.05$ fm and $|\langle r^2 \rangle|_{a_0(980)} = 0.94 \pm 0.09$ fm. Based on our investigation results, we conclude that the $f_{0}(980)$ state is mainly a $K\bar{K}$ bound state, the $\sigma$ state a resonance of $\pi\pi$ and the $a_{0}(980)$ state a loose $K\bar{K}$ bound state. From the results of the compositeness, they are not pure molecular states and have something non-molecular components, especially for the $\sigma$ state.

\end{abstract}

\maketitle

\section{Introduction}

Even though Quantum Chromodynamics (QCD) is the fundamental theory of strong interaction and governs the high energy region, the nature and the structure of the lowest scalar mesons still problematic and under debate. One of the main topics of the high energy physics is to comprehend the properties of the hadronic resonances. The conventional picture of the hadrons based on the quark model is the baryon made of $qqq$ and the meson $q\bar{q}$. However, that is not the whole picture of the observed hadrons, with the development of the experiments, many resonances have been found, which may have complex structures since their nature cannot be interpreted by the conventional ways, such as tetraquarks \cite{tetra}, hybrids \cite{hybrid}, and glueballs \cite{Frere:2015xxa} for mesons, and pentaquarks and heptaquarks for baryons, or molecular states. These exotic states have drawn much attention both in theories and experiments to understand their structure and decay properties, see more details in the reviews \cite{Chen:2016qju,Hosaka:2016pey,Chen:2016spr,Lebed:2016hpi,Esposito:2016noz,Guo:2017jvc,Ali:2017jda,Olsen:2017bmm,Karliner:2017qhf,Yuan:2018inv,Brambilla:2019esw}. In the low energy region the perturbative QCD failed because of the confinement, so we need to explore a non-perturbative QCD, such as Lattice QCD \cite{Kogut,Luscher,Mohler}, QCD sum rules \cite{Shifman,Reinders,Dias,Hidalgo,Agaev:2017cfz,Agaev:2018sco,Agaev:2018fvz}, Effective Field Theory \cite{Politzer,Georgi,Epelbaum}, Chiral Unitary Approach (ChUA) \cite{Kaiser,Oller1,Oller2,Hyodo1,Oset1}, and so on. In case of meson-meson and meson-baryon interaction, chiral dynamics is crucial in understanding the structure and nature of the resonances, and it has shown that many known resonances are dynamically generated as an outcome of the hadron-hadron interaction \cite{Francesco}.  

Following the work of Ref. \cite{Xiao:2019lrj}, we continue to study the properties of the $\sigma$ [or $f_{0}(500)$], $f_{0}(980)$ \cite{f0}, and $ a_{0}(980)$ \cite{a0} states. Although the states of $f_{0}(980) $ and $ a_{0}(980)$ are nearly degenerated, they have different isospin and other properties. Several proposals were made about the nature of these scalar particles, such as $q \bar{q}$ state \cite{Morgan1,Morgan2,Morgan3,Tornqvist}, multiquark states \cite{tetra,Agaev:2017cfz,Agaev:2018sco,Agaev:2018fvz,tetra2,Achasov:2003cn}, or $K\bar{K}$ molecules \cite{Weinstein1,Weinstein2,Weinstein3,Janssen}. The Evidence of four-quark nature for the $f_{0}(980)$ and $a_{0}(980)$ states are found in the $\phi$ meson radiative decay \cite{Achasov:2003cn} where more experimental informations and discussions can be referred to Refs. \cite{Achasov:2009ee,Achasov:2017ozk,Achasov:2019vcs}. The nature of the $ \sigma$ resonance is different from the other two. The masses of the $f_{0}(980) $ and $ a_{0}(980)$ are close to the $K\bar{K}$ threshold, conversely $\sigma$ is far above $\pi\pi$ threshold. Moreover the decay width of the sigma is very large, which does not behave like an ordinary Breit-Wigner resonance \cite{Pelaezreport}. Furthermore, from the large $ N_{c}$ limit calculations \cite{Pelaez1,Pelaez2} and Regge theory \cite{Nakamura} are confirmed that $ \sigma$ is not an ordinary $q\bar{q}$ structure. In the work of Ref. \cite{npa620}, using the ChUA, the potential of the pseudoscalars calculated from the chiral Lagrangians \cite{Gasser,Meissner2,Pich,Ecker,Bernard}, and then by applying the unitarity in coupled channel scattering amplitudes, the $\sigma$, $f_{0}(980)$ and $ a_{0}(980)$ are dynamically generated. Along the line of Ref. \cite{npa620}, we make a further investigation of the properties of the $\sigma$, $f_{0}(980)$, and $ a_{0}(980)$ states by evaluating their compositeness, the wave functions and the radii both in the coupled channel and the single channel interactions.
 
In the present work, we will firstly introduce the formalism of the interactions of $K\bar{K}$ and its coupled channels. Then, we discuss the definition of the couplings and how to calculate the compositeness, the wave functions and the radii for a resonance in ChUA. Following, we show our results in details for the cases of the coupled channel and the single channel, respectively. Finally, we close with our conclusions.

\section{Formalism}

In this section, we firstly revisit the formalism of Ref. \cite{npa620}, where the interaction potentials for the coupled channels are derived from the lowest order chiral Lagrangian,  and then performing the S-wave projection, the scattering amplitudes are evaluate with a set of on-shell Bethe-Salpeter equations. Next, we introduce the definitions of the  couplings in the coupled channel, the wave functions, compositeness and radii of the generated resonances.

\subsection{S-wave scattering amplitude in the coupled channels and single channel}

The most general chiral Lagrangian can be written in a perturbative manner according to the powers of the momenta of the pseudoscalar mesons \cite{Gasser,Meissner2,Pich,Ecker},
\begin{equation} 
\mathcal{L}_{\mathrm{ChPT}}(U)=\sum_{n} \mathcal{L}_{2 n}=\mathcal{L}_{2}+\mathcal{L}_{4}+\mathcal{O}\left(p^{6}\right),
 \end{equation}
where the lowest order chiral Lagrangian $\mathcal{L}_{2}$ contains the most general low energy interactions of the pseudoscalar meson octet, which is given by,
\begin{equation}
\mathcal{L}_{2}=\frac{1}{12 f^{2}}\left\langle\left(\partial_{\mu} \Phi \Phi-\Phi \partial_{\mu} \Phi\right)^{2}+M \Phi^{4}\right\rangle,
\label{eq:l2}
\end{equation}
where $f$ is the pion decay constant, the value of which is taken as $92.4$ MeV \cite{pdg2018}, $\left\langle \texttt{  } \right\rangle$ stands for the trace of matrices, and $\Phi$ is the pseudo Goldstone boson fields, defined as
\begin{equation} 
\Phi(x)=\frac{1}{\sqrt{2}} \phi^{a} \lambda^{a}=\left( \begin{array}{ccc}{\frac{1}{\sqrt{2}} \pi^{0}+\frac{1}{\sqrt{6}} \eta_{8}} & {\pi^{+}} & {K^{+}} \\ {\pi^{-}} & {\pi^{0}+\frac{1}{\sqrt{6}} \eta_{8}} & {K^{0}} \\ {K^{-}} & {\overline{K}^{0}} & {-\frac{2}{\sqrt{6}} \eta_{8}}\end{array}\right).
\end{equation}
Besides, the pseudoscalar meson mass matrix $M$ is given by
\begin{equation}
M=\left(\begin{array}{ccc}{m_{\pi}^{2}} & {0} & {0} \\ {0} & {m_{\pi}^{2}} & {0} \\ {0} & {0} & {2 m_{K}^{2}-m_{\pi}^{2}}\end{array}\right),
\end{equation}
where we have taken the isospin limit ($m_{u}=m_{d}$).

From this Lagrangian, Eq. \eqref{eq:l2}, we can derive the tree level amplitudes for $K\bar{K}$ , $\pi\pi$ and $\pi\eta$ channels,  which will be used in the coupled channel Bethe-Salpeter equations. After performing the S-wave projection, the interaction potentials in the isospin basises are given by \cite{npa620}, 
\begin{equation}
\begin{array}{l} 
 {V_{11}^{I=0} =-\frac{1}{4 f^{2}}\left(3 s+4 m_{K}^{2}-\Sigma_{i} p_{i}^{2}\right)}, \\  
 {V_{21}^{I=0} =-\frac{1}{3 \sqrt{12} f^{2}}\left(\frac{9}{2} s+3 m_{K}^{2}+3 m_{\pi}^{2}-\frac{3}{2} \sum_{i} p_{i}^{2}\right)}, \\ 
 {V_{22}^{I=0} =-\frac{1}{9 f^{2}}\left(9 s+\frac{15 m_{\pi}^{2}}{2}-3 \sum_{i} p_{i}^{2}\right)},
 \end{array}
\end{equation}

\begin{equation}
\begin{array}{l} 
{V_{11}^{I=1} =-\frac{1}{12 f^{2}}\left(3 s-\sum_{i} p_{i}^{2}+4 m_{K}^{2}\right)}, \\ 
{V_{21}^{I=1}=\frac{\sqrt{3 / 2}}{12 f^{2}}\left(6 s-2 \sum_{i} p_{i}^{2}+\frac{4}{3} m_{\pi}^{2}-\frac{4}{3} m_{K}^{2}\right)}, \\ 
{V_{22}^{I=1}=-\frac{1}{3 f^{2}} m_{\pi}^{2}},
\end{array}
\end{equation}
where we specify the $K\bar{K}$ and $\pi\pi$ channels with the labels 1 and 2, respectively, for the case of isospin $I=0$, and the $K\bar{K}$ and $\pi\eta$ channels for the case of $I=1$.  For the on shell amplitudes, one can take $ p_{i}^{2} = m_{i}^{2}$.

\begin{figure}
\begin{subfigure}{.45\textwidth}
  \centering
  \includegraphics[width=1\linewidth]{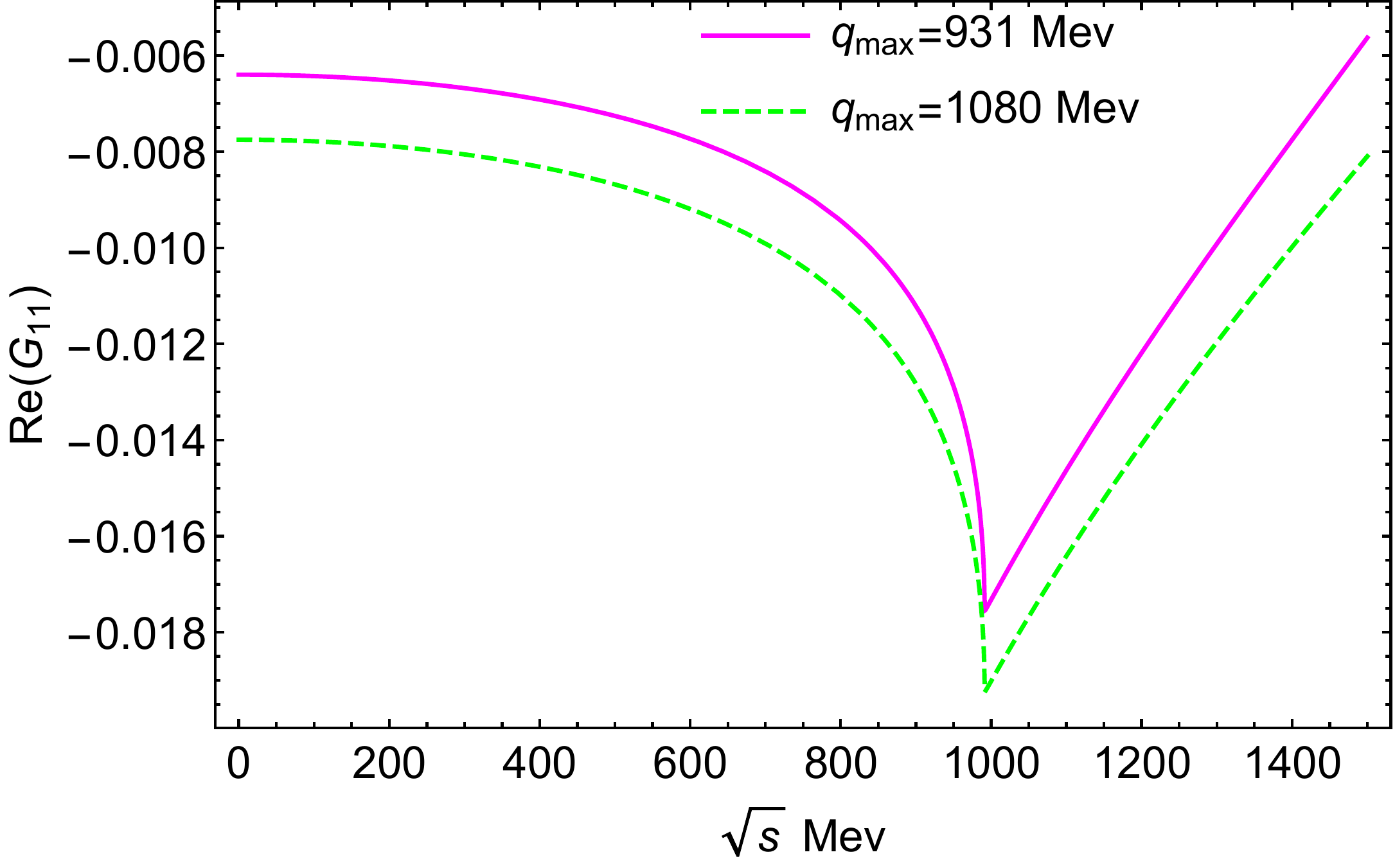}
  \caption{\footnotesize Real part of $ G_{11}$ }
  \label{fig:fig1a}
\end{subfigure}
\begin{subfigure}{.45\textwidth}
  \centering
  \includegraphics[width=1\linewidth]{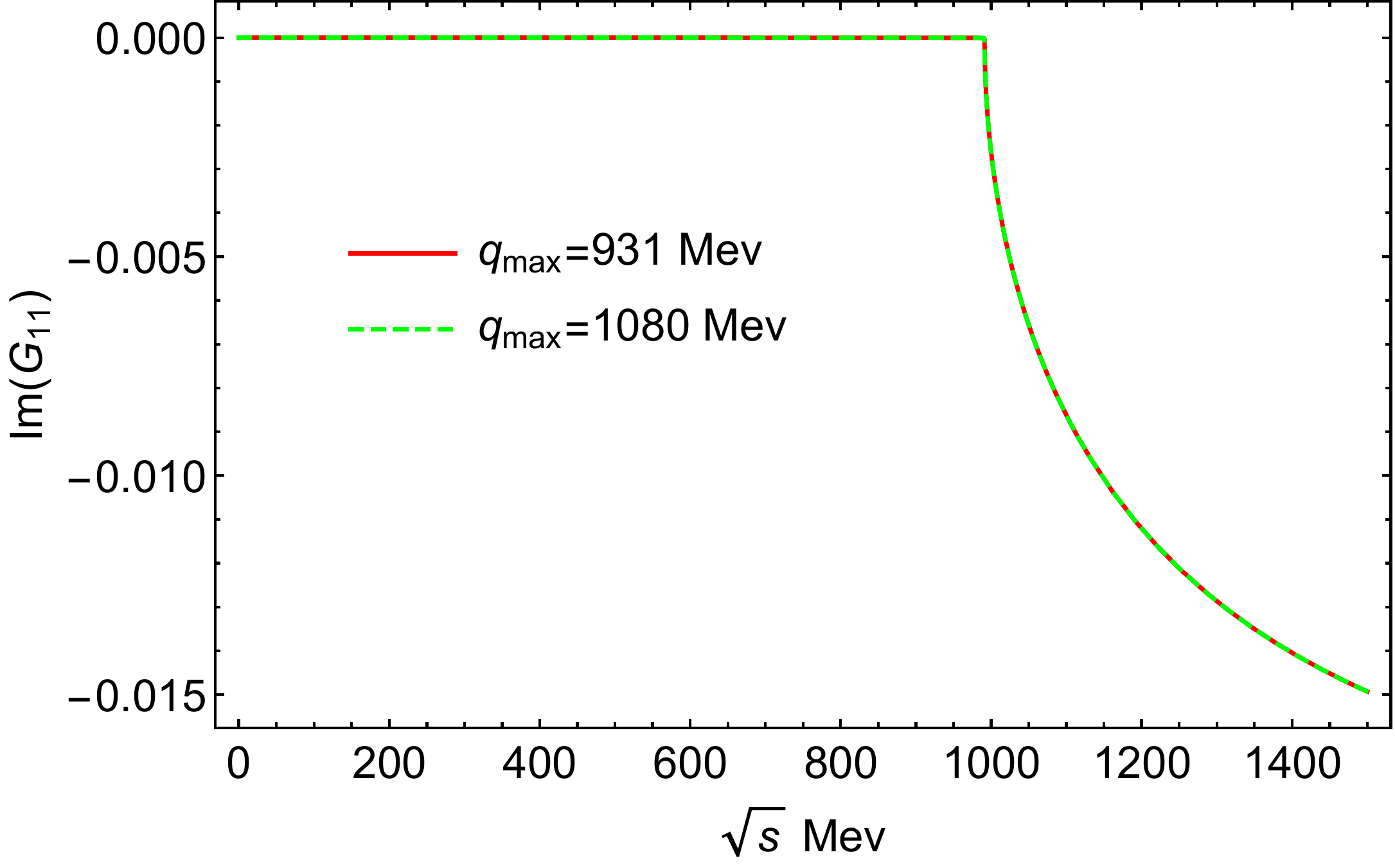}
  \caption{\footnotesize Imaginary part of $ G_{11}$}
  \label{fig:fig1b}
\end{subfigure} 

\begin{subfigure}{.45\textwidth}
  \centering
  \includegraphics[width=1\linewidth]{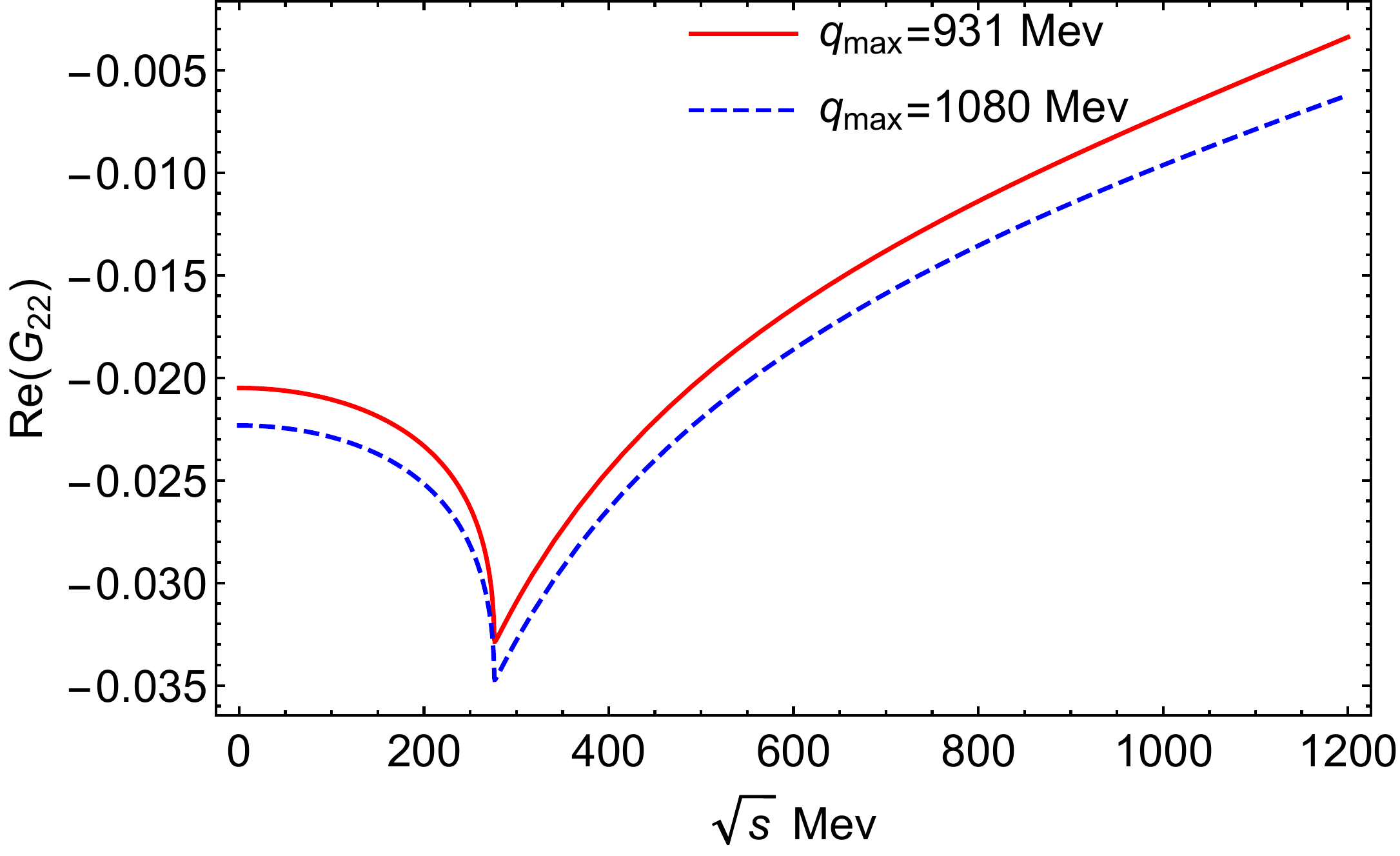}
  \caption{\footnotesize Real part of $ G_{22}$}
  \label{fig:fig1c}
\end{subfigure}
\begin{subfigure}{.45\textwidth}
  \centering
  \includegraphics[width=1\linewidth]{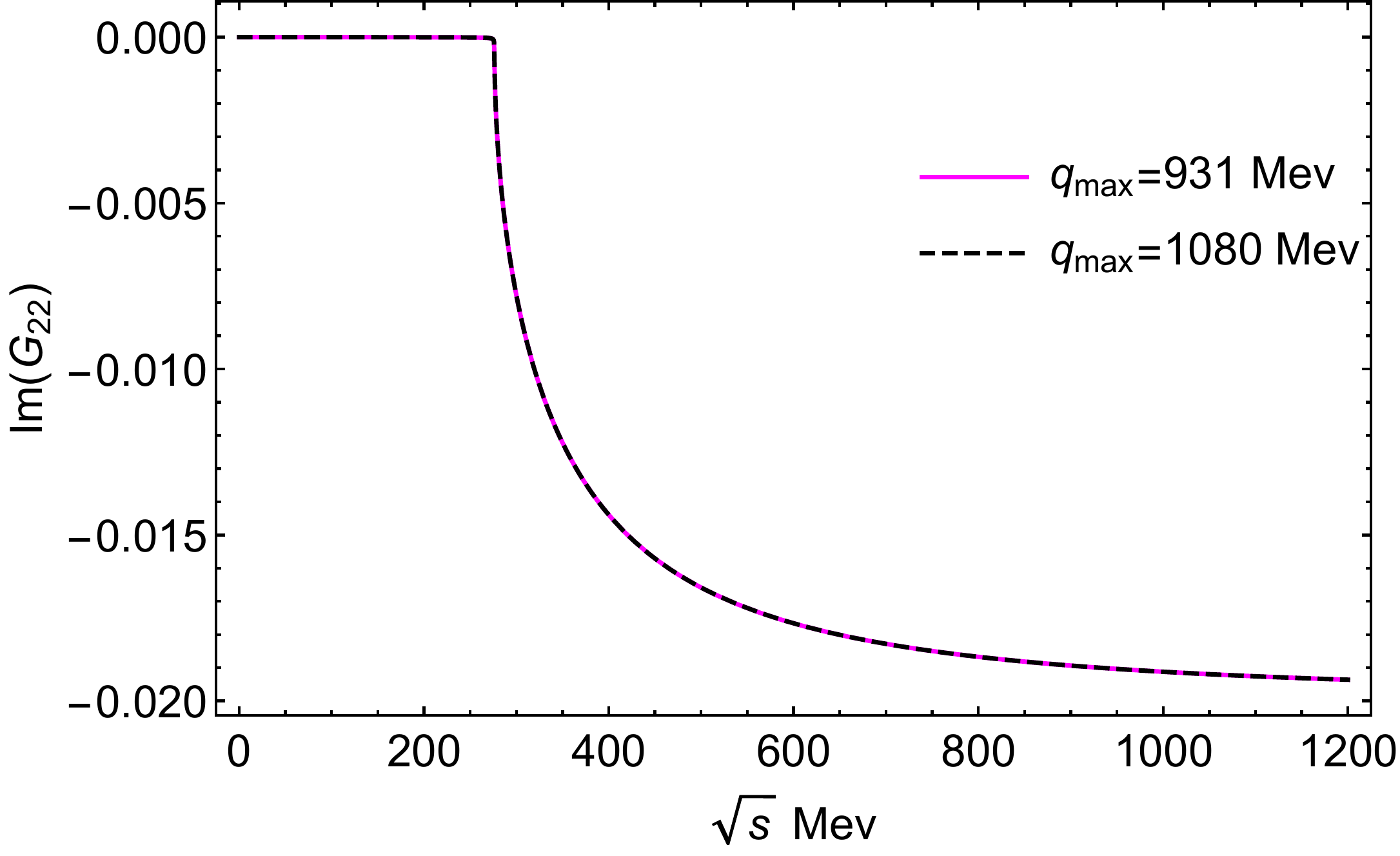}
  \caption{\footnotesize Imaginary part of $ G_{22}$}
  \label{fig:fig1d}
\end{subfigure}%
\caption{Real part and imaginary part of the propagator for the case of isospin $I=0$ with two different cutoffs $q_{max}$.}
\label{fig:fig1}
\end{figure}

For the scattering amplitudes of the coupled channels, one can solve the Bethe-Salpeter equations factorized on shell \cite{npa620},
\begin{equation}
 T= [1- VG]^{-1} V \text{   .}
 \label{eq8}
\end{equation}
It is worth to note that in the present case $T$, $V$, and $G$ are $2 \times 2$ matrices. The element of  the diagonal $G$ matrix is the loop function of two intermediate mesons in the $i$-th channel, given by
\begin{equation}
 G _ { i i } ( s ) = i \int \frac { d ^ { 4 } q } { ( 2 \pi ) ^ { 4 } } \frac { 1 } { q ^ { 2 } - m _ { 1 } ^ { 2 } + i \varepsilon } \frac { 1 } { \left( p _ { 1 } + p _ { 2 } - q \right) ^ { 2 } - m _ { 2 } ^ { 2 } + i \varepsilon }  \text{ ,}
 \label{eq9}
\end{equation}
where $p_{1}$ and $p_{2}$ are the four-momenta of the two initial particles, respectively, and $m_{1}$, $m_{2}$ are the masses of the two intermediate particles appearing in the loop. Note that the G function is logarithmically divergent. There are two methods to solve this singular integral, either using the three-momentum cut-off method \cite{npa620}, where the analytic expression is given by Ref \cite{analyticG}, or the dimensional regularization method \cite{dimentionalG}. Using the cut-off method  we can rewrite Eq. (\ref{eq9}) as 
\begin{equation}
G _ { ii } ( s ) = \int _ { 0 } ^ { q _ { \max } } \frac { q ^ { 2 } d q } { ( 2 \pi ) ^ { 2 } } \frac { \omega _ { 1 } + \omega _ { 2 } } { \omega _ { 1 } \omega _ { 2 } \left[ s - \left( \omega _ { 1 } + \omega _ { 2 } \right) ^ { 2 } + i \epsilon \right] }   \text{  ,}
 \label{eq10}
\end{equation}
where $q=| \vec { q } |$, $ \omega _ { i } = ( \vec {\; q } ^ { 2 } + m _ { i } ^ { 2 } ) ^ { 1 / 2 } \text { and } s = (p_1 + p_2)^2 $, and the cutoff, $q_{max}$, is the only one free parameter. We show our results of the real part and the imaginary part of the $G$ functions in the isospin $I=0$ case in Fig. \ref{fig:fig1} with two different cutoffs (about their values see the discussions at the beginning of next section), where one can see that the imaginary part of the loop function is independent with the the cutoff, which leads to extrapolate to the Second Riemann sheet easily, see the discussions below.

Using ChUA, one also can easily determine the masses and the decay widths of the resonances produced in the coupled channel interactions just by looking for the poles in the second Riemann sheets. Thus, one need to extrapolate the analytical structure of the scattering amplitudes in the complex $s$ plane. To fulfil these, one can extrapolate the $G(s)$ function into the second Riemann sheet by 
\begin{equation}
  G_{i i}^{(II)}\left(s\right) =G_{i i}^{(I)}\left(s\right)-2 i \operatorname{Im} G_{i i}^{(I)}\left(s\right)
  =G_{i i}^{(I)}\left(s\right)+\frac{i}{4 \pi} \frac{p_{cmi} (s)}{\sqrt{s}}  \textbf{ ,}
\end{equation}
where the three momentum in center-of-mass (CM) frame is given by 
\begin{equation}
p_{cmi} (s)=\frac{\lambda^{1/2}(s, m_1^2, m_2^2)}{2\sqrt{s}}\, ,
\end{equation}
with the usual K\"allen triangle function $\lambda(a,b,c)=a^2+b^2+c^2-2(ab+ac+bc)$, see more details in Ref. \cite{npa620}.

In the present case, only two coupled channels, the elements of the scattering amplitudes $T$ matrix can be written as \cite{npa620},
\begin{equation}
\begin{array}{l} 
{T_ { 11 } = \frac { 1 } { \Delta _ { c } } \left( \Delta _ { \pi } V _ { 11 } + V _ { 12 } ^ { 2 } G _ { 22 } \right)}, \\ 
{T_ { 21 } = \frac { 1 } { \Delta _ { c } } \left( V _ { 21 } G _ { 11 } V _ { 11 } + \Delta _ { k } V _ { 21 } \right)}, \\ 
{T_ { 22 } = \frac { V _ { 22 } } { \Delta \pi } + \frac { V _ { 12 } ^ { 2 } G _ { 11 } } { \Delta _ { \pi } \Delta _ { c } }},
\end{array}  
 \label{eq11}
\end{equation}
where one defines
\begin{equation}
\begin{array}{c} 
{\Delta _ { \pi } = 1 - V _ { 22 } G _ { 22 }}, \\
{\Delta _ { K } = 1 - V _ { 11 } G _ { 11 }}, \\
{\Delta _ { c } = \Delta _ { K } \Delta _ { \pi } - V _ { 12 } ^ { 2 } G _ { 11 } G _ { 22 }}.
\end{array}
\end{equation}

For the single channel cases, we have the $K\bar{K}$ and $\pi\pi$ ($\pi\eta$) interaction channel separately for $I=0$ ($I=1$). Just by taking $V_{21}=0$, one can easily reduce Eq. (\ref{eq11}) to,
\begin{equation}
{T _ { 11 } = \frac {  V _ { 11 } } { \Delta _ { k } }} \textbf{ , } 
{T _ { 22 } = \frac { V _ { 22 } } { \Delta \pi }} \text{ .} 
 \label{eq13}
\end{equation} 

\subsection{The  couplings and the wave functions}

By applying the Laurent expansion of the amplitude close to the pole, the scattering amplitudes can be written as  \cite{Oller,Feng-Kun}
\begin{equation} 
T_{i j}=\frac{g_{i} g_{j}}{s-s_{p o l e}}+\gamma_{0}+\gamma_{1}\left(s-s_{p o l e}\right)+\cdots  \text{  ,}
\label{eq21}
\end{equation}
where $g_{i}$ and  $g_{j}$ are the coupling constants of the $ i $-th and $ j $-th channels, which can be calculated from the residue of the pole \cite{hidden,n/d}
\begin{equation} 
g_{i} g_{j}=\lim _{s \rightarrow s_{p o l e}}\left(s-s_{p o l e}\right) T_{i j}   \text{  .}
\label{eq22}
\end{equation}
Using Cauchy Integral formula, we can evaluate the residue as a loop integral in the complex $s$ plane,
\begin{equation} 
g_{i}^{2}=\frac{1}{2 \pi i} \oint T_{i i} d s   \text{  .}
 \end{equation}
where the integral is over a closed path in the complex $s$ plane around the pole $ s=s_{pole}$. Furthermore, with the couplings of the corresponding poles, one can generalize Weinberg's rule \cite{Weinrule} for bound state or resonance to the ChUA  \cite{Aceti}
\begin{equation} 
- \sum_{i} g_{i}^{2}\left[\frac{d G_{i}}{d s}\right]_{s=s_{pole}}=1   \text{  ,}
 \end{equation}
where an alternative derivation of this relationship can be found in Ref. \cite{Hyodo2}. This equation refer as the sum rule for the bound states or the resonances dynamically generated by the coupled channel interactions. More discussions and applications of this rule can be found in Refs. \cite{Sekihara,Xiao,Hyodo3,Aceti2,guooller}. This equation holds for the resonance or the  bound state which is a pure molecular state. However, in some cases, if a physical state couples not only to hadron-hadron pairs, but also to a different component of non-molecular type, this relation becomes for the composite states
\begin{equation} 
- \sum_{i} g_{i}^{2}\left[\frac{d G_{i}}{d s}\right]_{s=s_{pole}} =1-Z   \text{  ,}
\label{eq25}
 \end{equation}
where $Z$ represents the probability that the system is not a molecule components but something else. As discussed in Ref. \cite{Aceti2}, the interpretation of $Z$ as a probability non-molecular (meson-meson or meson-baryon state in ChUA) component is strict for bound states, which is related to the genuine component in the wave function of the state omitted from the coupled channels. Note that for a specified channel the $G_i$ function should be extrapolated to the right Riemann sheet for a corresponding pole of the state.

To understand more about the sources of the resonances, we study the wave function of the resonance at small distances. Once we have the wave function of a resonance, one can also investigate its form factor, which response the state to external sources. Following the formalism of Ref. \cite{Yamagata}, the wave function of a resonance in coordinate space is given by 
\begin{equation}
\phi(\vec{r})=\int_{q_{\max }} \frac{d^{3} \vec{p}}{(2 \pi)^{3 / 2}} e^{i \vec{p} \cdot \vec{r}}\langle\vec{p} | \Psi\rangle \text{  .}
\end{equation}
After performing the angle integration of the momentum, we obtained \cite{hidden}
\begin{equation}
\begin{aligned}  \phi(\vec{r})=& \frac{1}{(2 \pi)^{3 / 2}} \frac{4 \pi}{r} \frac{1}{C} \int_{q_{\max }} p d p \sin (p r)\times \frac{\Theta\left(q_{\max }-|\vec{p}|\right)}{E-\omega_{1}(\vec{p})-\omega_{2}(\vec{p})} \frac{m_{V}^{2}}{\vec{p}^{2}+m_{V}^{2}} \end{aligned},
 \label{eq18}
\end{equation}
where $C$ is the normalization constant, and $E \equiv \sqrt{s_{pole}}$, thus, which is real for a pure bound state with zero width and otherwise complex for the general cases in ChUA. Note that here we put an extra form factor $f(\vec{q}) = \frac{m_{V}^{2}}{\vec{p}^{2}+m_{V}^{2}}$ to regulate the scale of the wave function, and our conclusions do not change if we remove it. Using the wave functions that we have, one can evaluate the form factor of the states with its definition \cite{Yamagata},
\begin{equation}
\begin{aligned}
F(\vec{q})&=\int d^3 \vec{r} \phi(\vec{r}) \phi^*(\vec{r}) e^{-i \vec{q}\, ' \cdot \vec{r}} \\
&= \int d^3 \vec{p} \frac{\theta(\Lambda - p)\, \theta(\Lambda-|\vec{p}-\vec{q}|)}{[E-\omega_1(p)-\omega_2(p)]\,[E-\omega_1(\vec{p}-\vec{q})-\omega_2(\vec{p}-\vec{q})]},
\end{aligned}
\end{equation}
with a normalization to keep $F(q=0)\equiv 1$. For a generated state in ChUA, a pole with its width, which is complex, the form factor is complex too, see the results below. Finally, the radii of the states (or mean square distance) can be evaluated from the form factor, 
\begin{equation}
\left\langle r^{2}\right\rangle=-6\left[\frac{\mathrm{d} F(q)}{\mathrm{d} q^{2}}\right]_{q^{2}=0} \text{  .}
\label{eq33}
\end{equation}
Note that a soft step function needed to make the form factor converge in this case. On the other hand, for the case of a weakly bound state, the radii of the state can also be obtained from the tail of the wave functions as done in Ref. \cite{Sekihara}
\begin{equation}
\left\langle r^{2}\right\rangle_{i}=\frac{-g_i^{2}\left[\frac{\mathrm{d} G_i(s)}{\mathrm{d} s}\right]_{s=s_{pole}}}{4 \mu_{i} B_{\mathrm{E}, i}} \text{  ,}
\label{eq32}
\end{equation}
where the binding energy $ B_{\mathrm{E}, i} = m_{i} + m_{i}^{\prime} - M_{\mathrm{B}}  $, and the reduced mass $ \mu_{i}= \frac{m_{i}m_{i}^{\prime}}{m_{i} + m_{i}^{\prime}}$. Conceptually, $ \left\langle r^{2}\right\rangle_{i}$ is the mean-squared distance of the bound state in the $i$-th channel.

\section{Results}

We first revisit the $K\bar{K}$ interactions with its coupled channels of $\pi\pi$ or $\pi\eta$, where the states of $\sigma$, $f_{0}(980)$, and $a_{0}(980)$ are dynamically generated in the coupled channel approach as done in Ref. \cite{npa620}. But, we make a further study of the couplings, the compositeness, the wave functions and the radii for these states to investigate more details on their properties, as show the results as below. To find more information about the structure of the poles corresponding these states, we examine the single channel interactions. Note that, for the only one free parameter in our approach, what we used below for the value of the cutoff is the one determined in Ref. \cite{Xiao:2019lrj} by dong a combined fit of the experimental data, $q_{max}=931$ MeV, which is a bit different with the ones used in Ref. \cite{npa620}. To see the uncertainties of our calculations, we also show the results with the one of about 15\% division to the upper limits, $q_{max}=1080$ MeV and varying the values between 15\% division in some cases.

\subsection{Coupled channel approach }

\begin{figure}
\begin{subfigure}{.45\textwidth}
  \centering
  \includegraphics[width=1\linewidth]{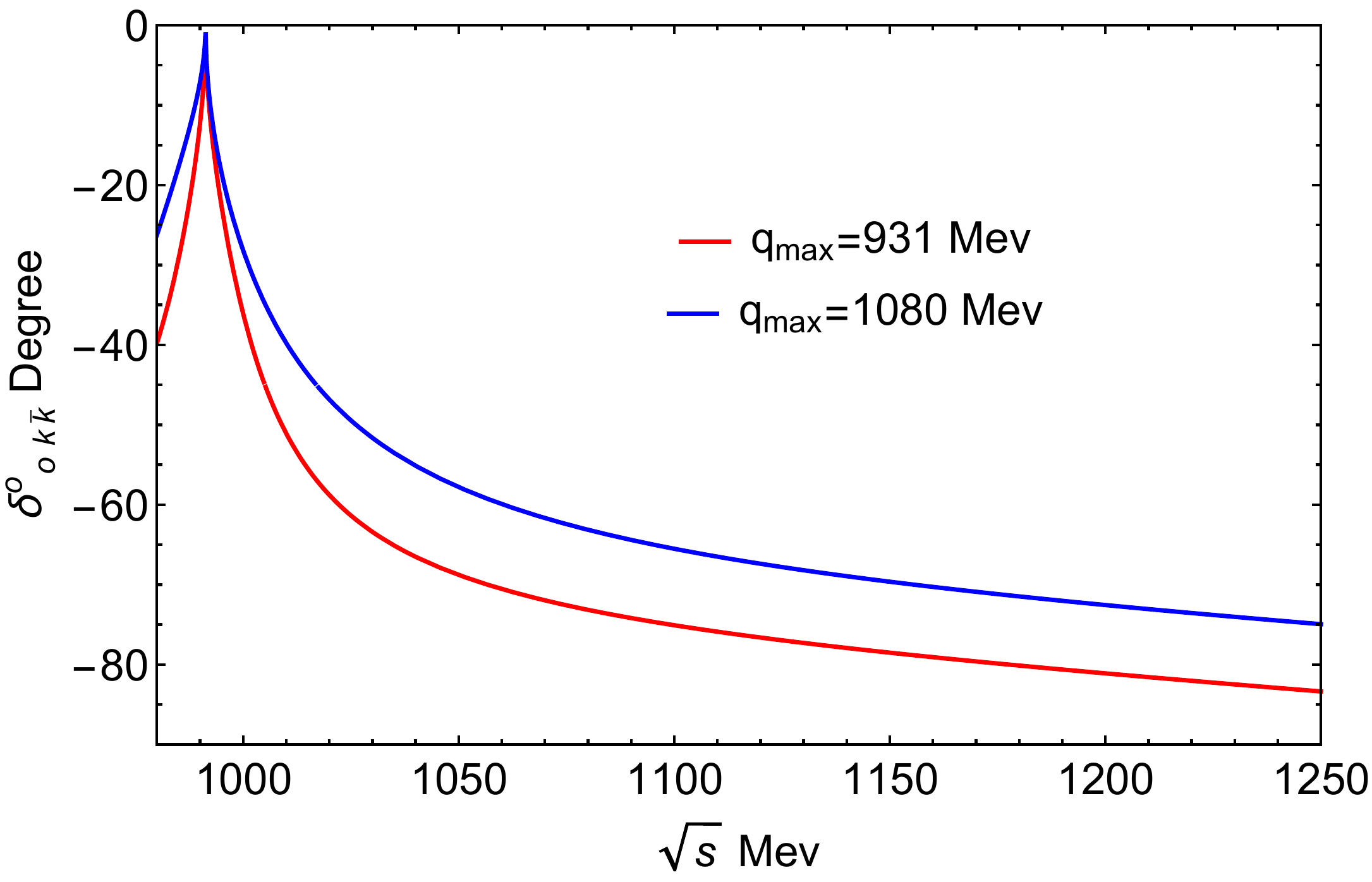}
  \caption{\footnotesize Results of $ K\bar{K}$ phase shifts.}
\end{subfigure}%
\begin{subfigure}{.45\textwidth}
  \centering
  \includegraphics[width=1\linewidth]{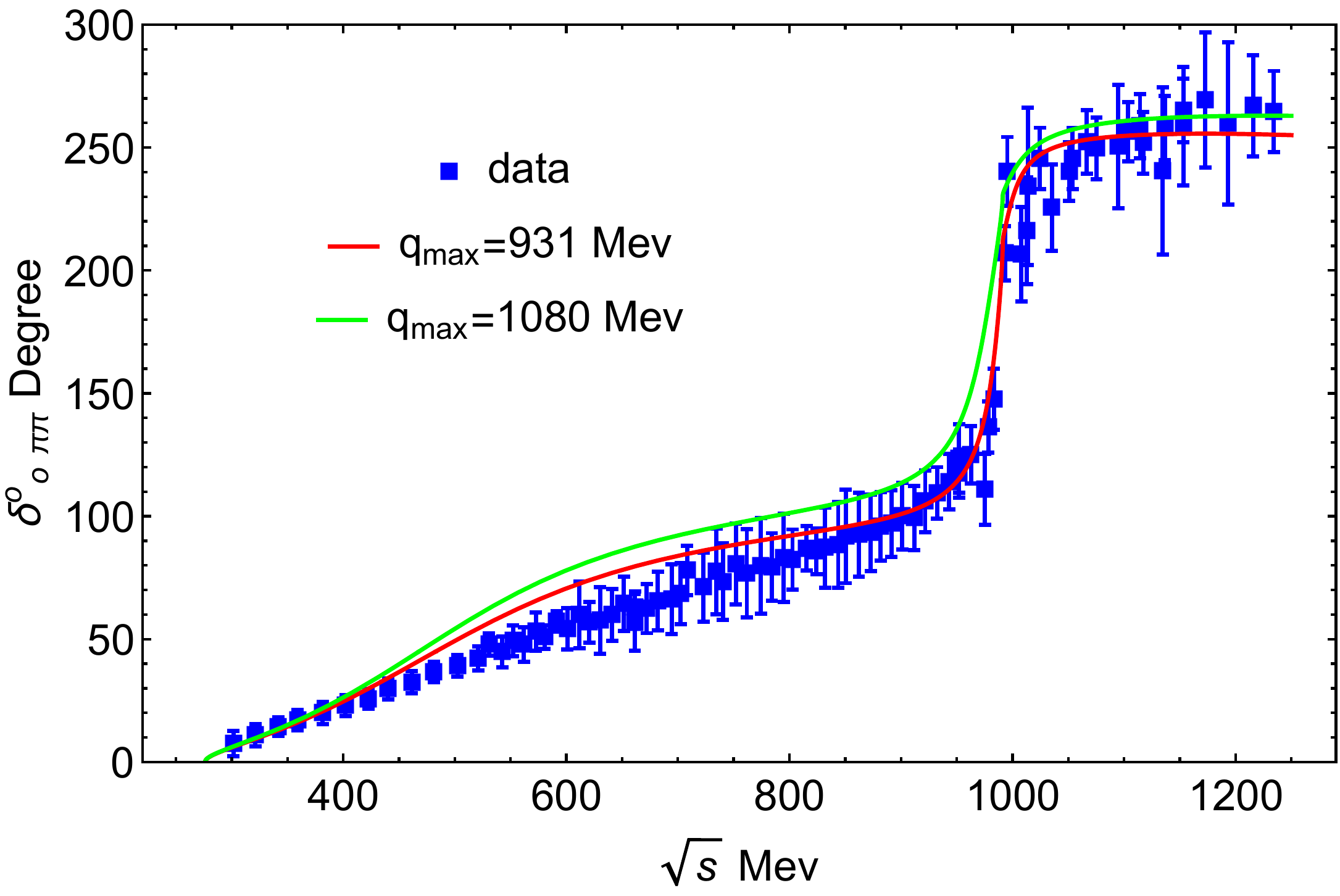}
  \caption{\footnotesize Results of $\pi\pi$ phase shifts.}
  \label{fig:fig2b}
\end{subfigure} \\
\begin{subfigure}{.45\textwidth}
  \centering
  \includegraphics[width=1\linewidth]{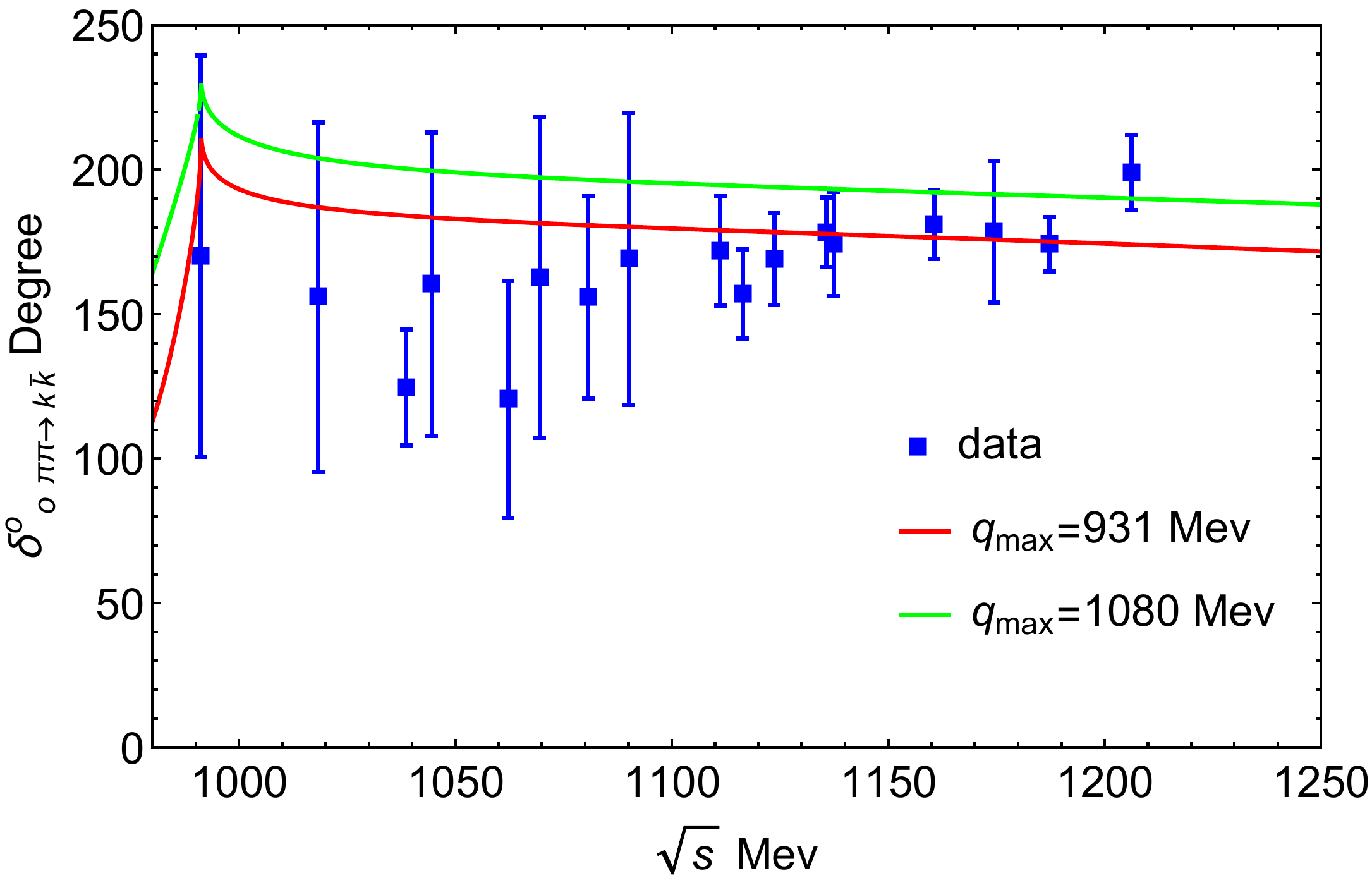}
  \caption{\footnotesize Results of $K\bar{K}\rightarrow \pi\pi$ phase shifts.}
\end{subfigure}
\begin{subfigure}{.45\textwidth}
  \centering
  \includegraphics[width=1\linewidth]{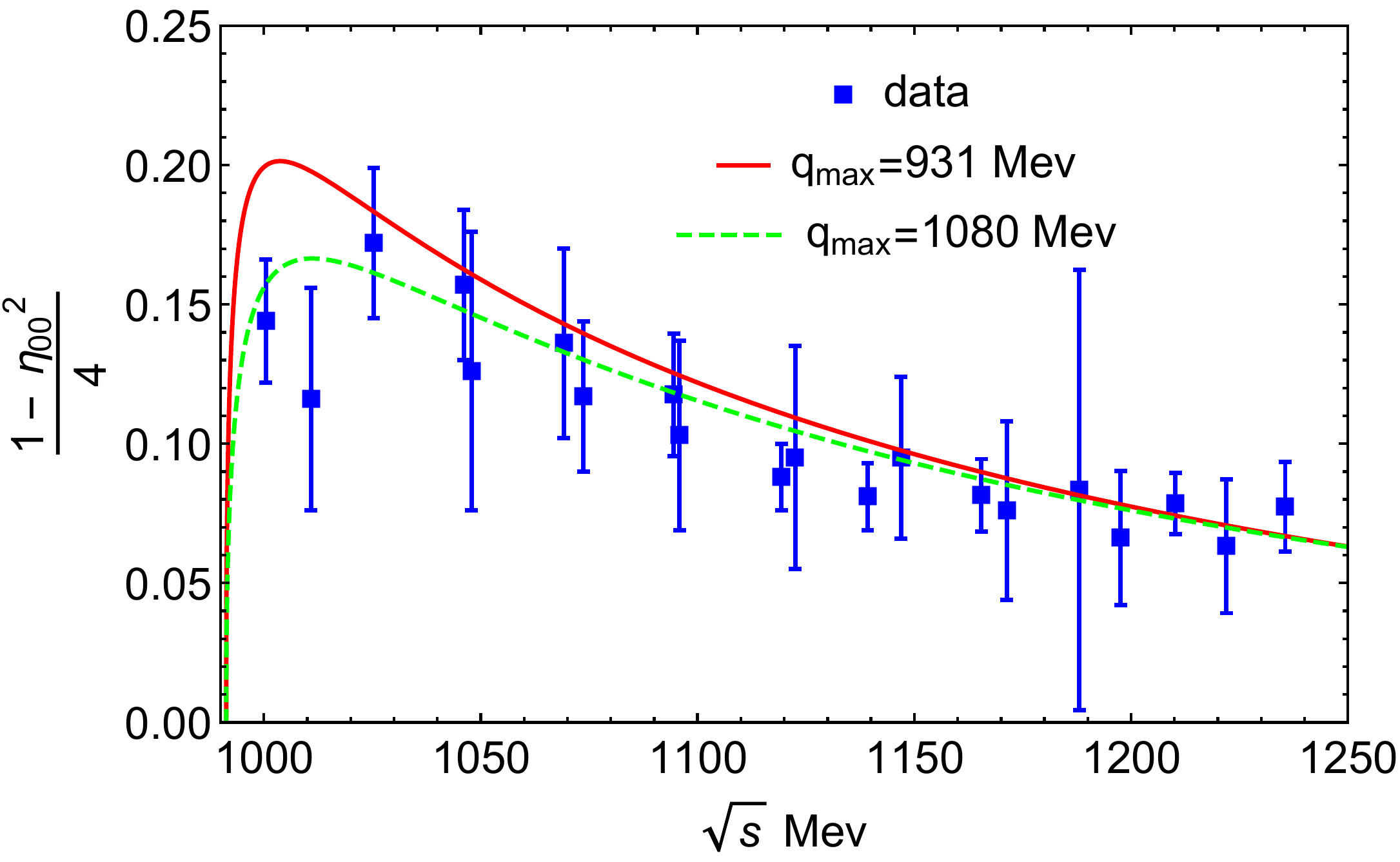}
  \caption{\footnotesize Results of inelasticities $\frac{(1-\eta_{00}^{2})}{4}$.}
\end{subfigure}%
\caption{Our resutls for the sector of isospion $I=0$ with two different values of $q_{max}$.}
\label{fig:fig2}
\end{figure}

\begin{figure}
\begin{subfigure}{.45\textwidth}
  \centering
  \includegraphics[width=1\linewidth]{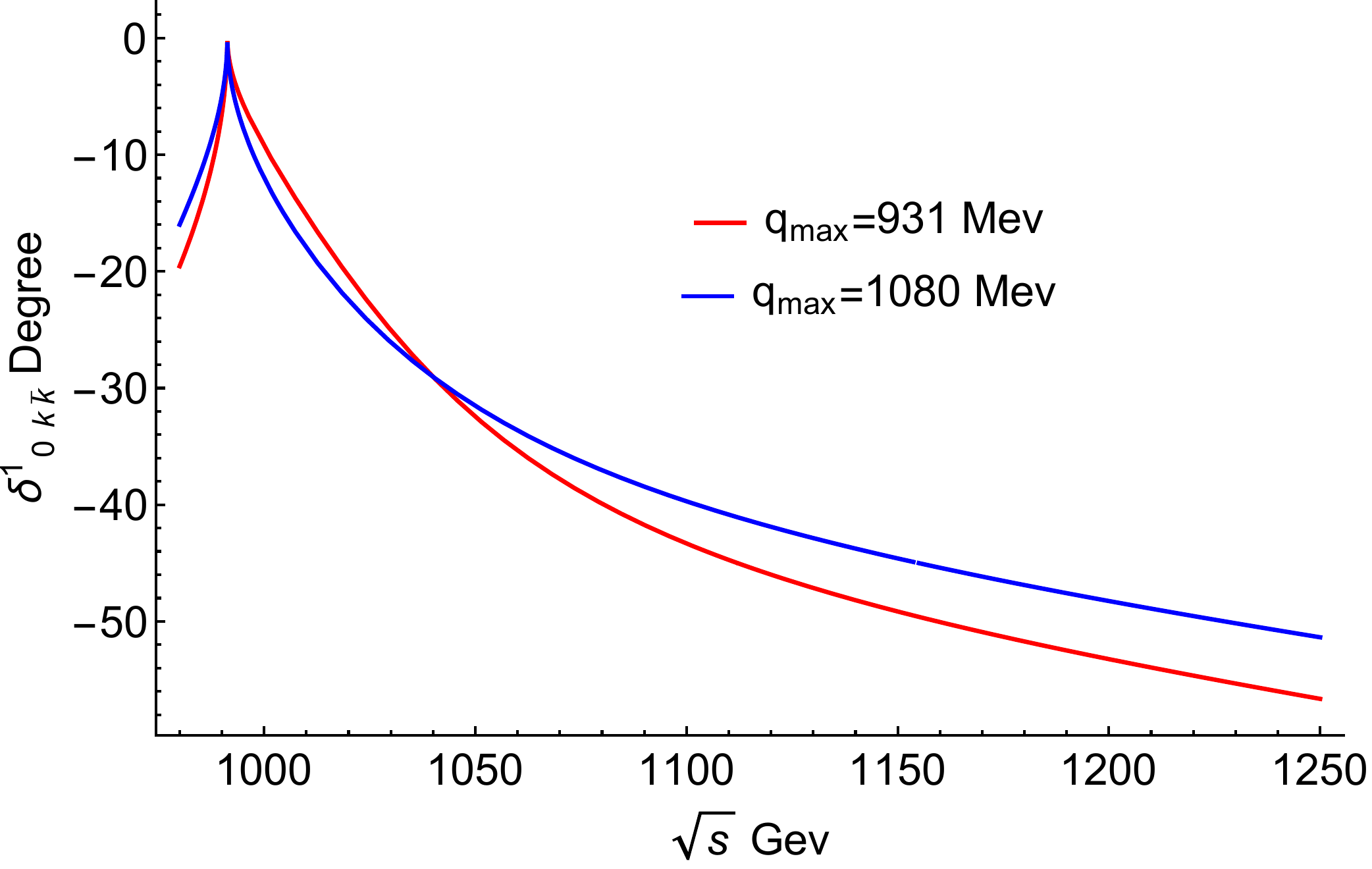}
  \caption{\footnotesize Results of $ K\bar{K}$ phase shift.}
\end{subfigure}%
\begin{subfigure}{.45\textwidth}
  \centering
  \includegraphics[width=1\linewidth]{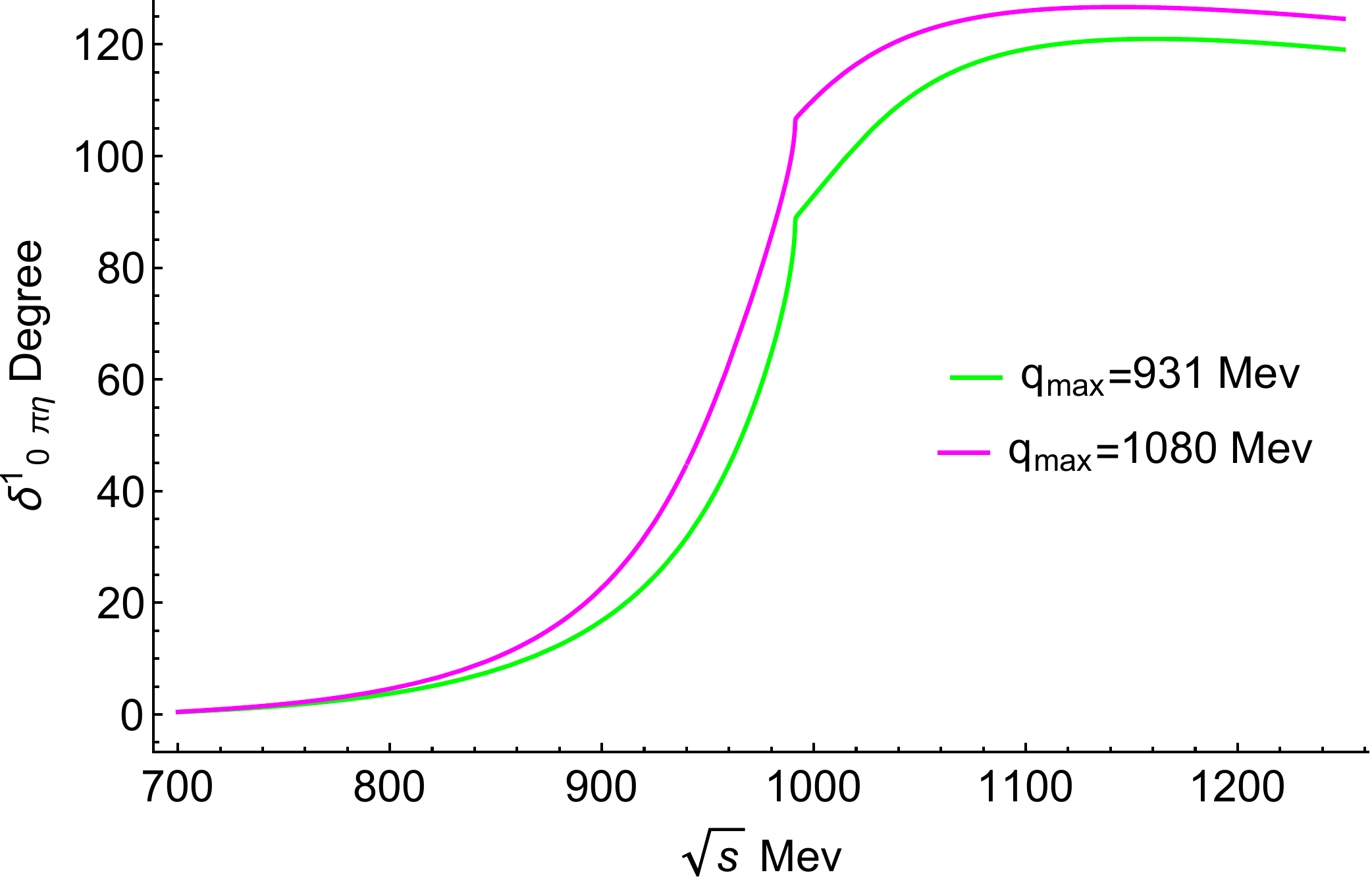}
  \caption{\footnotesize Results of $\pi\eta$ phase shift.}
\end{subfigure} \\
\begin{subfigure}{.45\textwidth}
  \centering
  \includegraphics[width=1\linewidth]{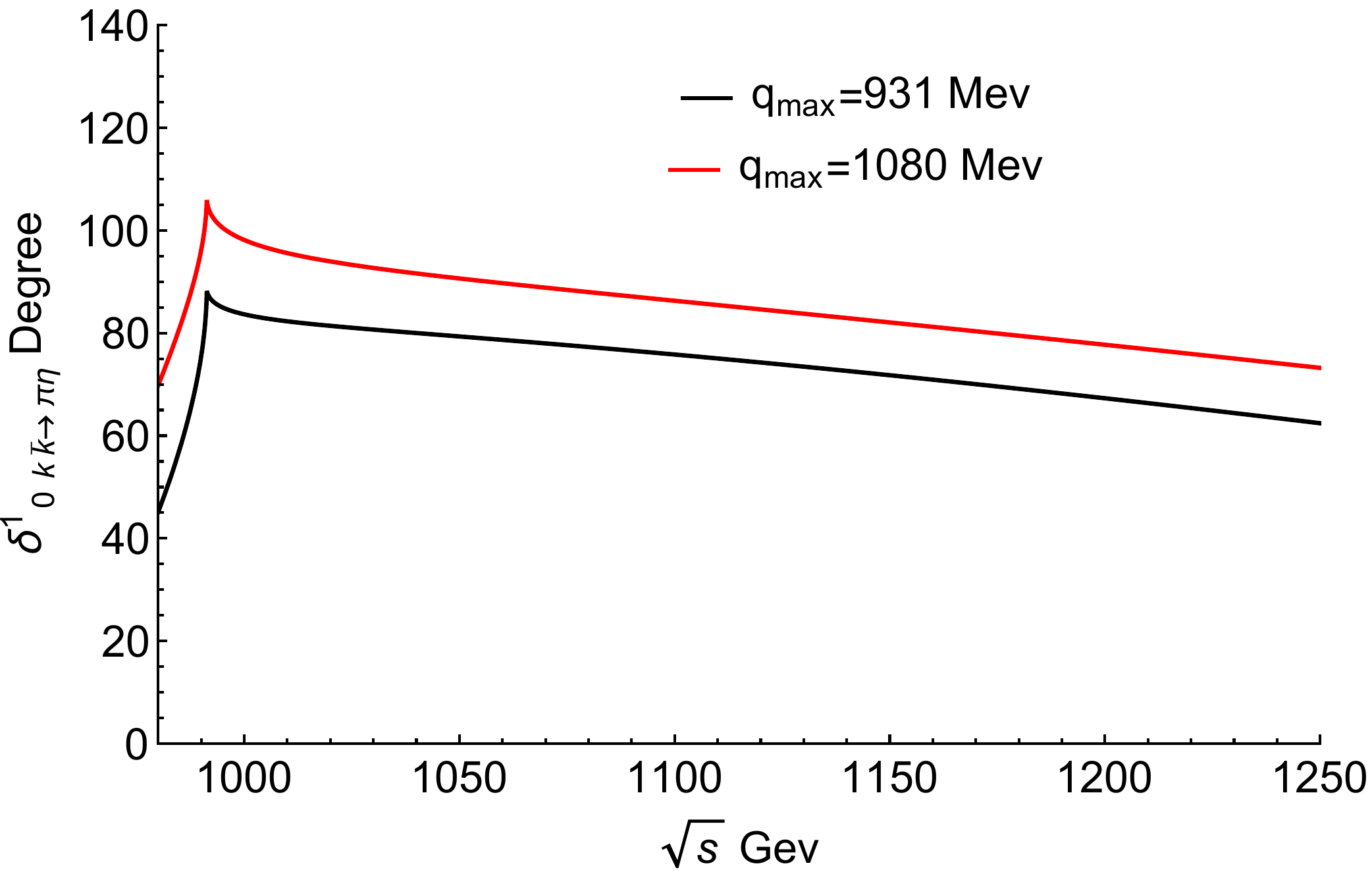}
  \caption{\footnotesize Results of $K\bar{K}\rightarrow \pi\eta$ phase shift.}
\end{subfigure}
\begin{subfigure}{.45\textwidth}
  \centering
  \includegraphics[width=1\linewidth]{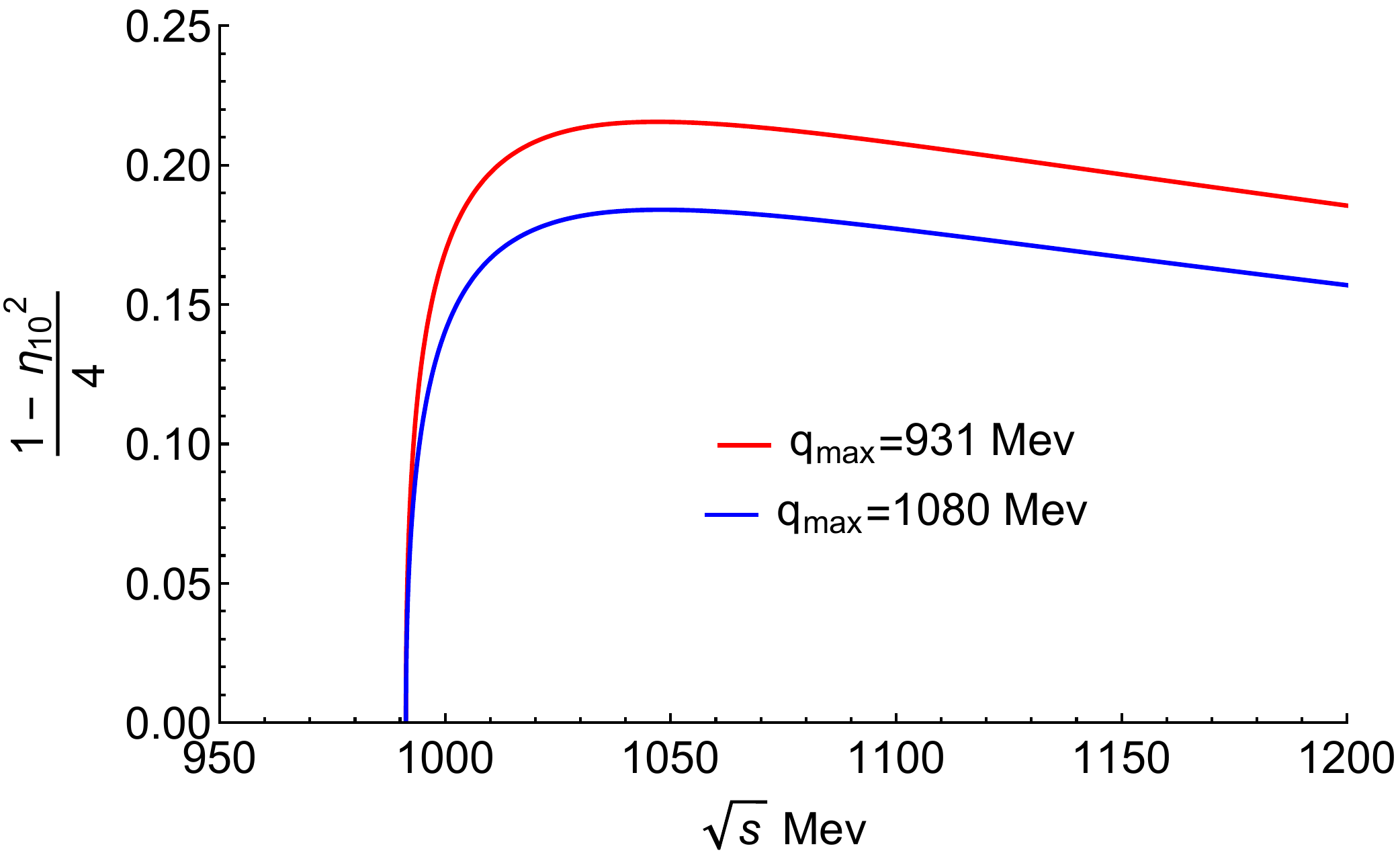}
  \caption{\footnotesize Results of inelasticities $\frac{(1-\eta_{01}^{2})}{4}$.}
\end{subfigure}%
\caption{Our resutls for the sector of $I=1$.}
\label{fig:fig3}
\end{figure}

We first calculate the phase shifts and the inelasticities. As done in Ref. \cite{npa620}, the two-channels $S$-matrix are used,
\begin{equation}
 S = \left[ \begin{array} { l l } 
 { \eta e ^ { 2 i \delta _ { 1 } } } & { i \left( 1 - \eta ^ { 2 } \right) ^ { 1 / 2 } e ^ { i \left( \delta _ { 1 } + \delta _ { 2 } \right) } } \\ 
 { i \left( 1 - \eta ^ { 2 } \right) ^ { 1 / 2 } e ^ { i \left( \delta _ { 1 } + \delta _ { 2 } \right) } } & { \eta e ^ { 2 i \delta _ { 2 } } } 
 \end{array} \right]  \texttt{,}
 \label{eq26}
\end{equation}
where the observables of $\delta _ { 1 }$, $\delta _ { 2 }$ correspond to the phase shifts of the channel 1, 2, respectively, and the one of $ \eta$ is the inelasticity. These observables can be calculated from the relationship between $S$-matrix and the scattering amplitude $T$-matrix, having
\begin{equation}
\begin{array}{l}
{T_{11}=-\frac{8 \pi \sqrt{s}}{2 i p_{cm1}}\left(S_{11}-1\right)}, \\ 
{T_{22}=-\frac{8 \pi \sqrt{s}}{2 i p_{cm2}}\left(S_{22}-1\right)}, \\ 
{T_{12}=t_{21}=-\frac{8 \pi \sqrt{s}}{2 i \sqrt{p_{cm1} p_{cm2}}}\left(S_{12}-1\right)}\end{array},
\end{equation}
where $p_{cmi}$ is the corresponding three momentum in the CM frame as discussed above. The results of the phase shifts and the inelasticities in isospin of  $I=0$ and $I=1$ sectors are shown in Figs. \ref{fig:fig2} and \ref{fig:fig3}, respectively. In Fig. \ref{fig:fig2}, we can see that the results of $I=0$ sector are in good agreement with the experimental data up to $\sqrt{s}=1.2$ GeV even with the upper limit of the cutoff. From Fig. \ref{fig:fig2b}, one can see that the $\sigma$ structure is a wide bump and the signal of $f_0(980)$ is in the sharp increasing region which crosses $90^\circ$ \cite{Pelaezreport}. However, in $I=1$ sector because of the lake of experimental data for phase shifts and inelasticities, we make some predictions for them, where the structure of $a_0(980)$ can be clearly seen in the phase shifts. 

\begin{figure}
\begin{subfigure}{.5\textwidth}
  \centering
  \includegraphics[width=1\linewidth]{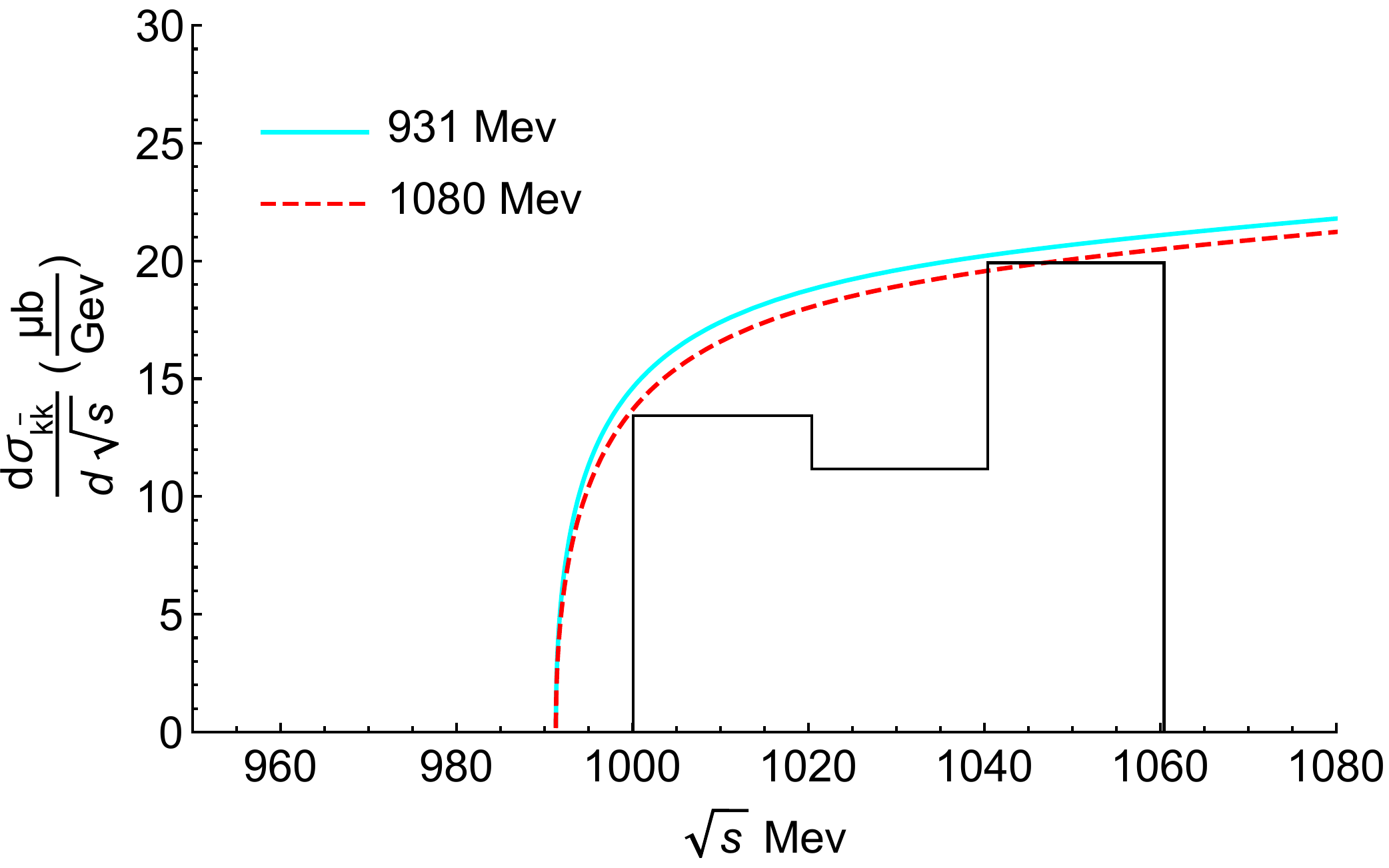}
  \caption{\footnotesize The invariant mass distribution of $K\bar{K}$ ($C=2.5\times 10^{-5}$).}
\end{subfigure}%
 \begin{subfigure}{.5\textwidth}
  \centering
\includegraphics[width=1 \linewidth]{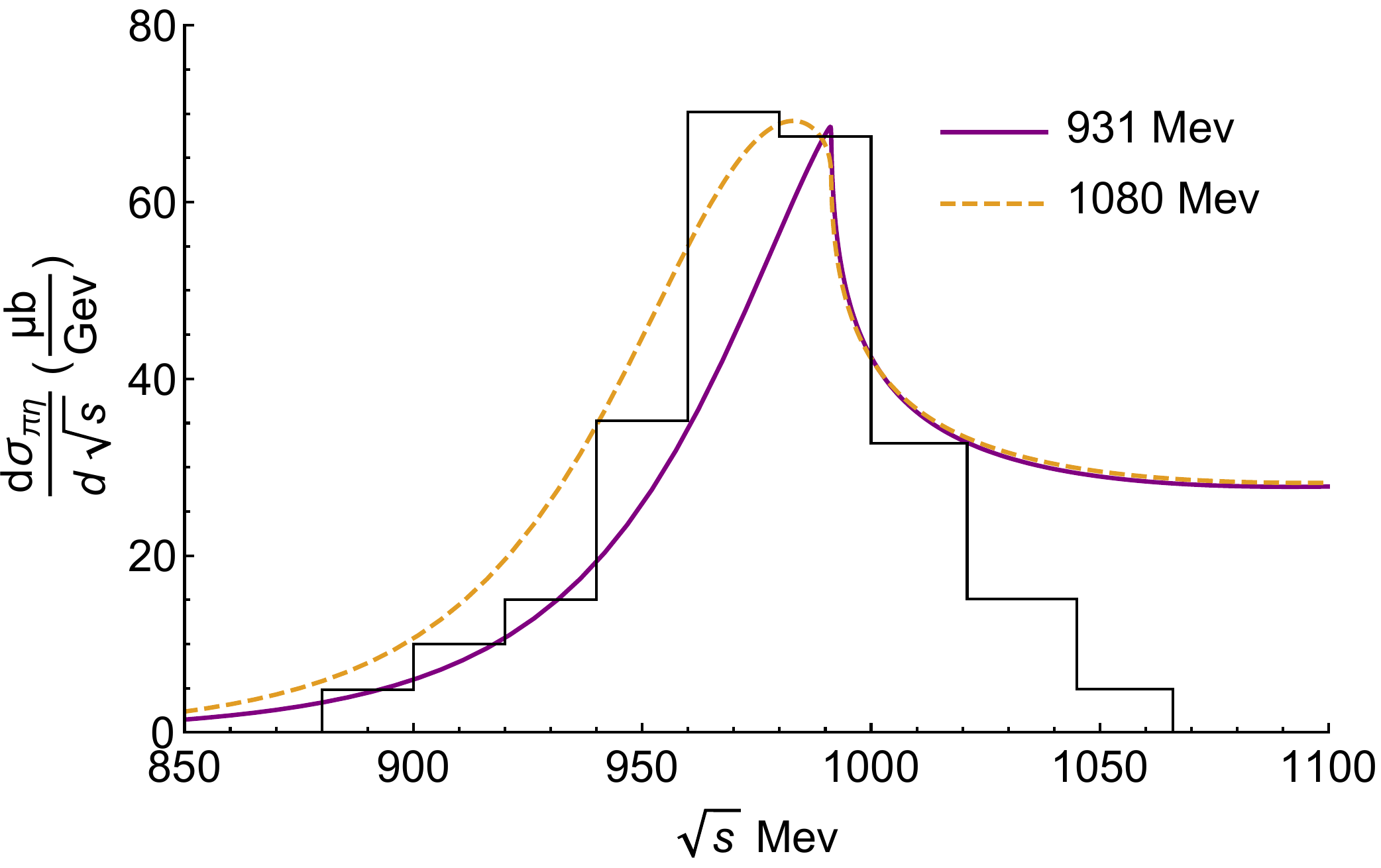}
\caption{\footnotesize The invariant mass distribution of $\pi\eta$ ($C=3.63\times 10^{-5}$).}
\end{subfigure}%
  \caption{Results for the invariant mass distribution with two different $q_{max}$.}
  \label{fig:fig4}
\end{figure}

\begin{figure}
\begin{subfigure}{.5\textwidth}
  \centering
  \includegraphics[width=1\linewidth]{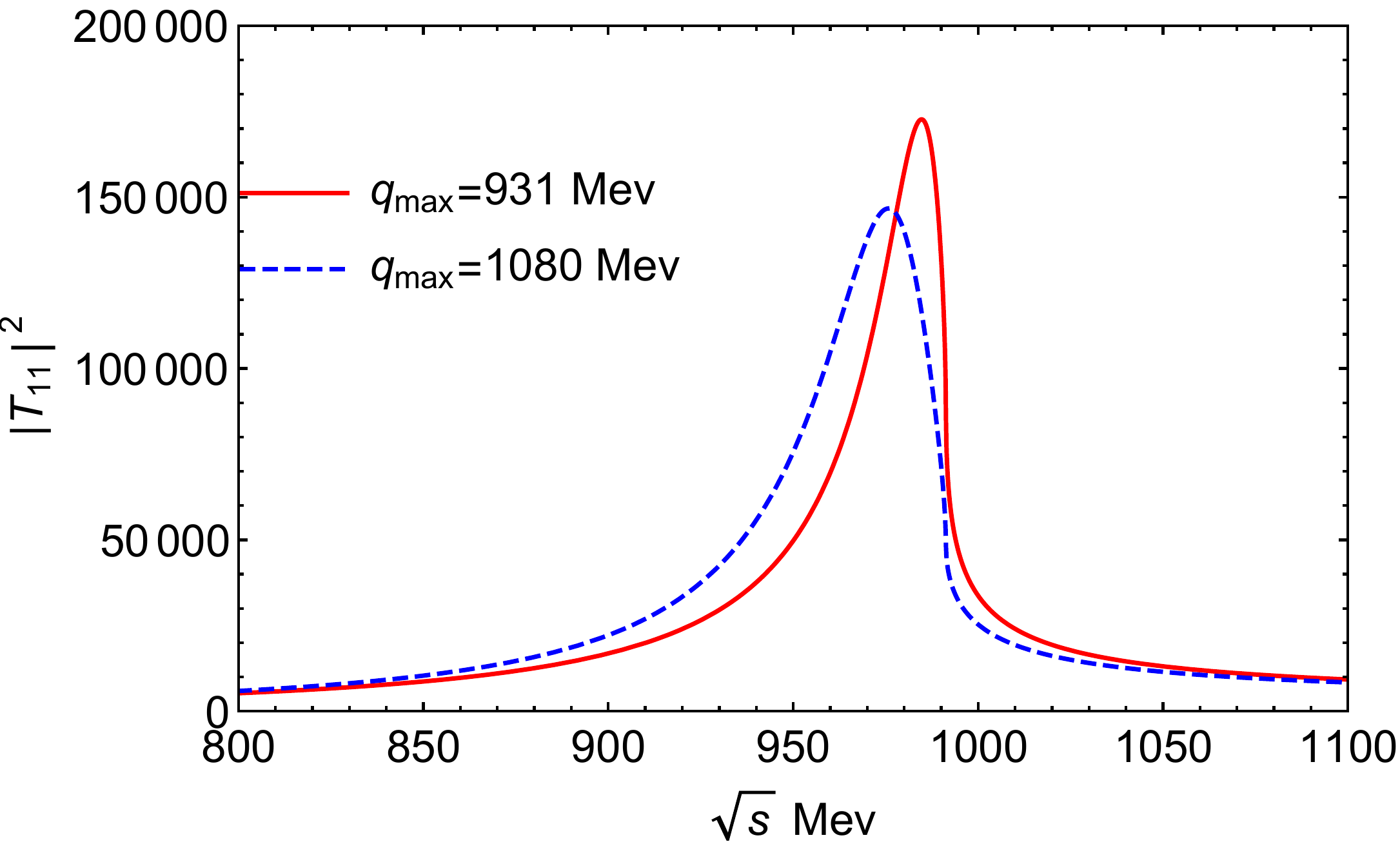}
  \caption{\footnotesize Results of $|T_{11}|^2$ for the $K\bar{K}$ channel.} \label{fig:fig5a}
\end{subfigure}%
\begin{subfigure}{.5\textwidth}
  \centering
  \includegraphics[width=1\linewidth]{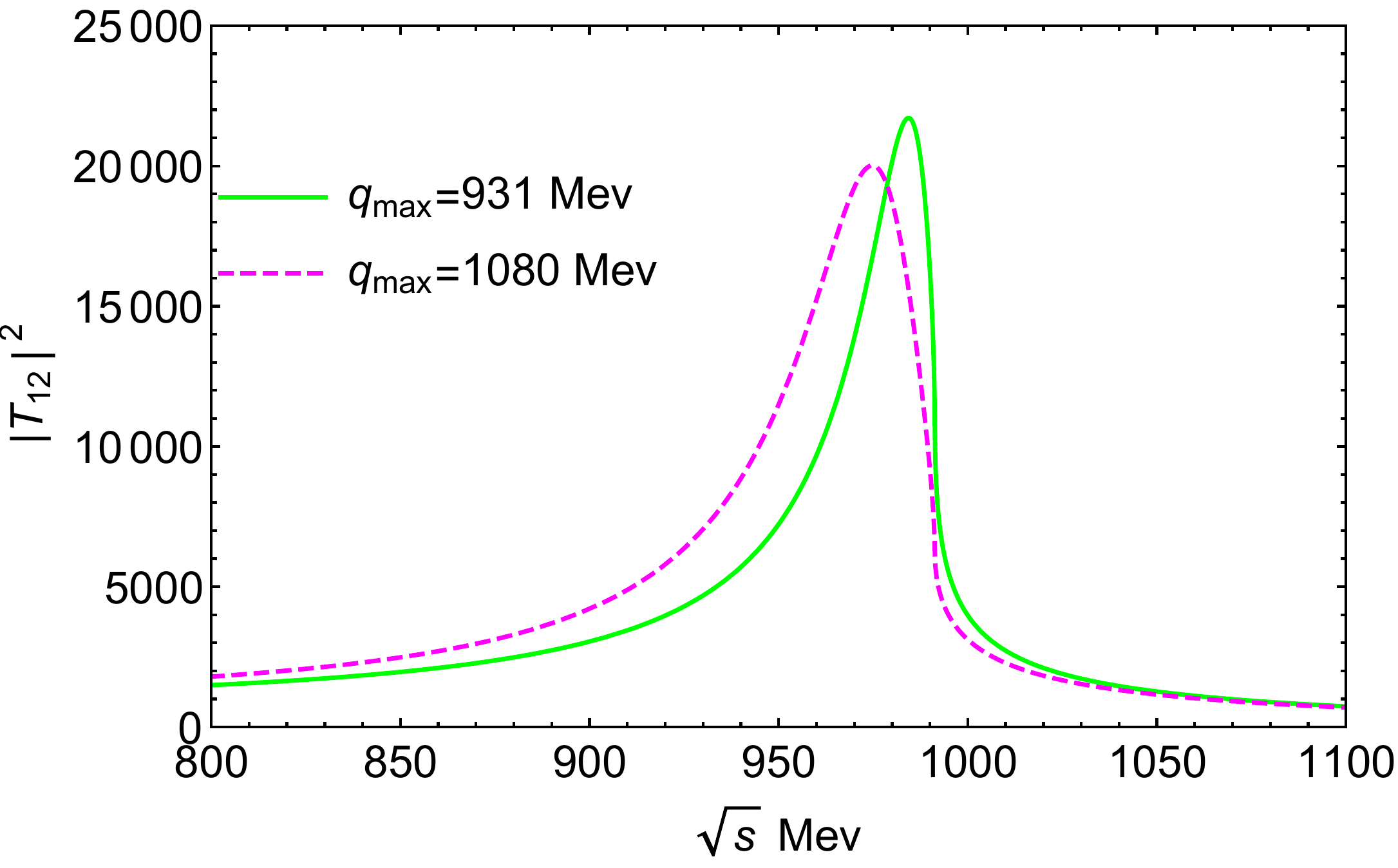}
  \caption{\footnotesize Results of $|T_{12}|^2$ for the $K\bar{K}\rightarrow \pi\pi$ channel.} \label{fig:fig5b}
\end{subfigure} 
\begin{subfigure}{.5\textwidth}
  \centering
  \includegraphics[width=1\linewidth]{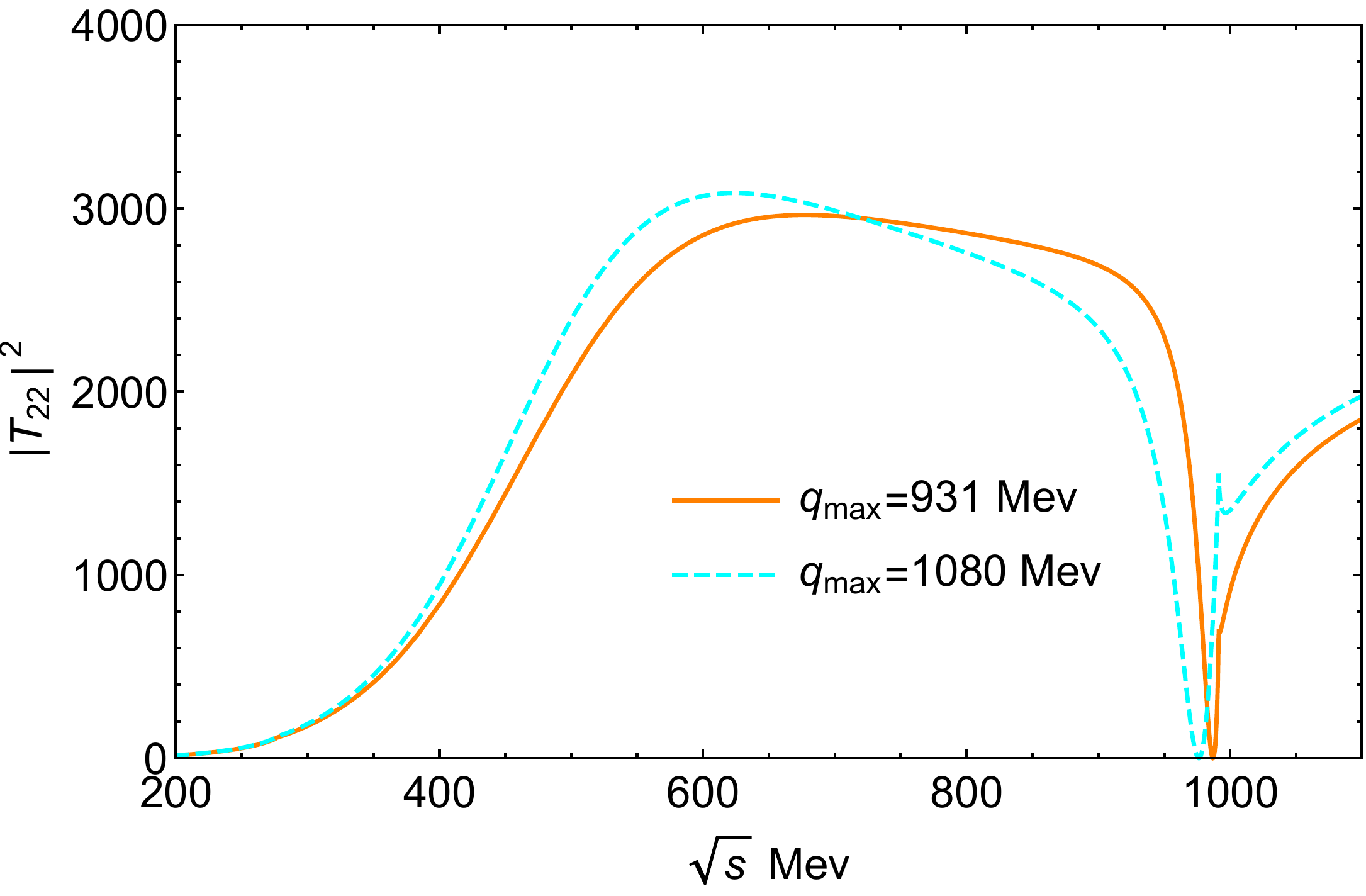}
  \caption{\footnotesize Results of $|T_{22}|^2$ for the $\pi\pi$ channel.} \label{fig:fig5c}
\end{subfigure}
\caption{Results of the modulus squared of the scattering amplitudes in $I=0$ sector.}
\label{fig:fig5}
\end{figure}

\begin{figure}
\begin{subfigure}{.5\textwidth}
  \centering
  \includegraphics[width=1\linewidth]{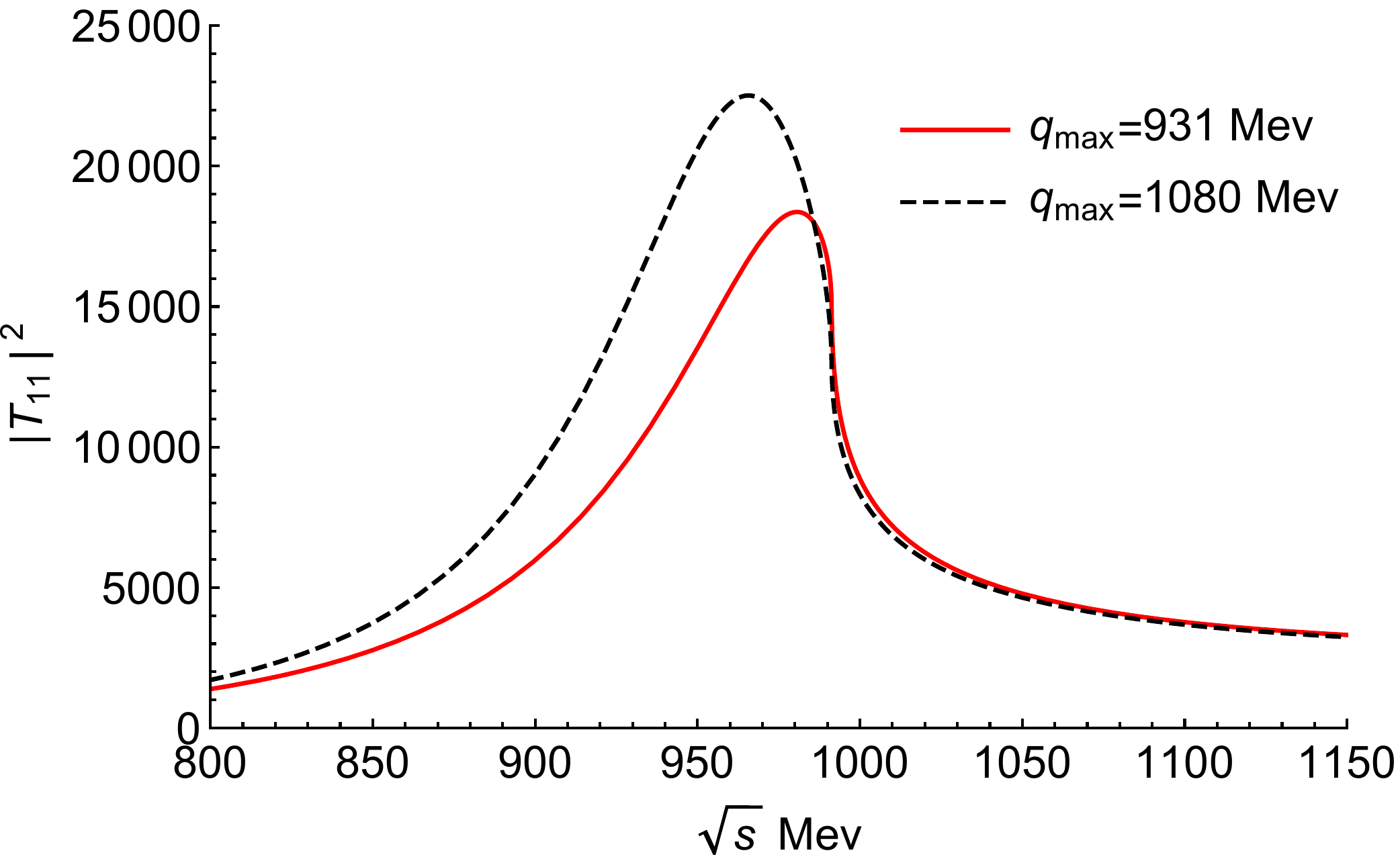}
  \caption{\footnotesize Results of $|T_{11}|^2$ for the $K\bar{K}$ channel.}
\end{subfigure}%
\begin{subfigure}{.5\textwidth}
  \centering
  \includegraphics[width=1\linewidth]{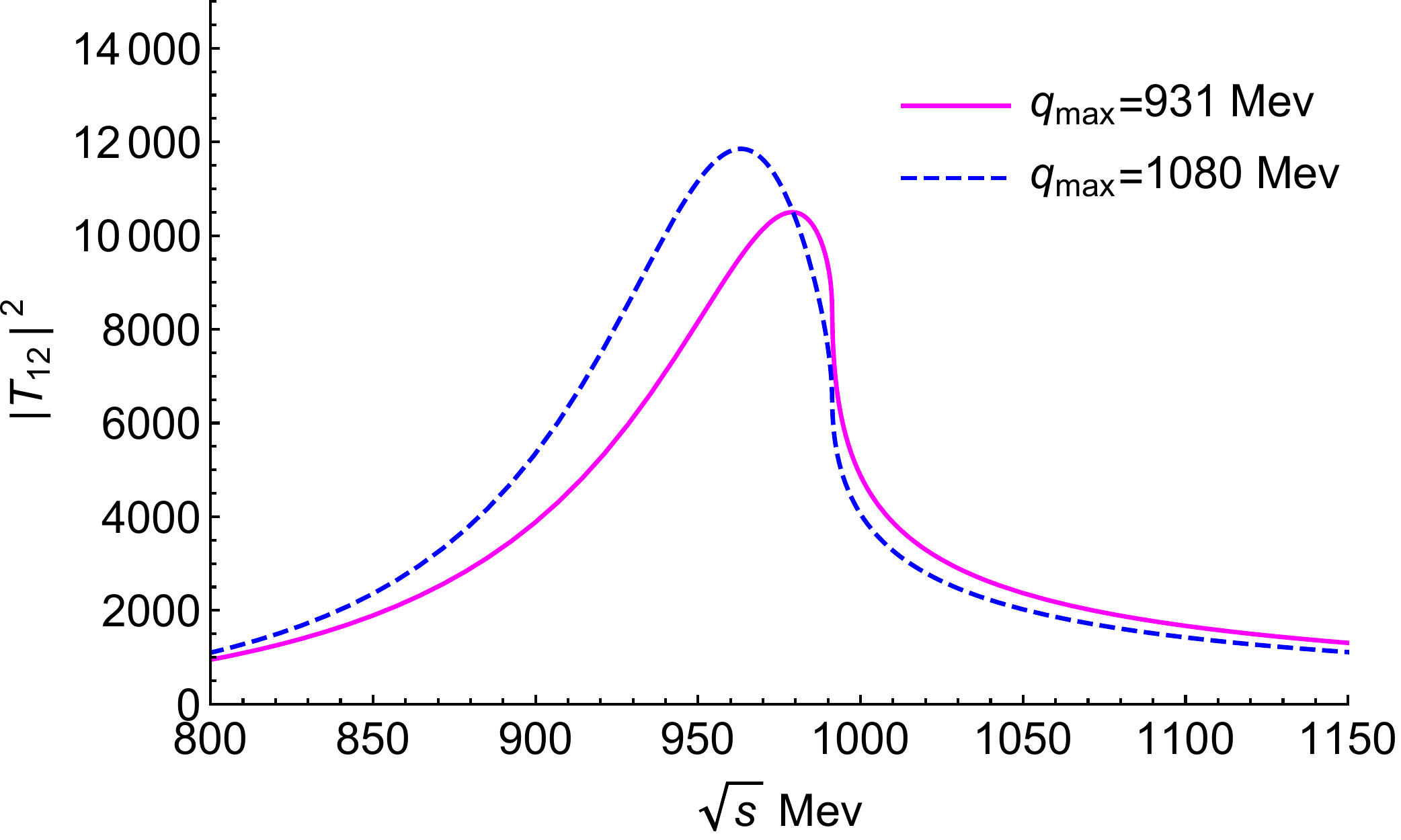}
  \caption{\footnotesize Results of $|T_{12}|^2$ for the $K\bar{K}\rightarrow \pi\eta$ channel.}
\end{subfigure} 
\begin{subfigure}{.5\textwidth}
  \centering
  \includegraphics[width=1\linewidth]{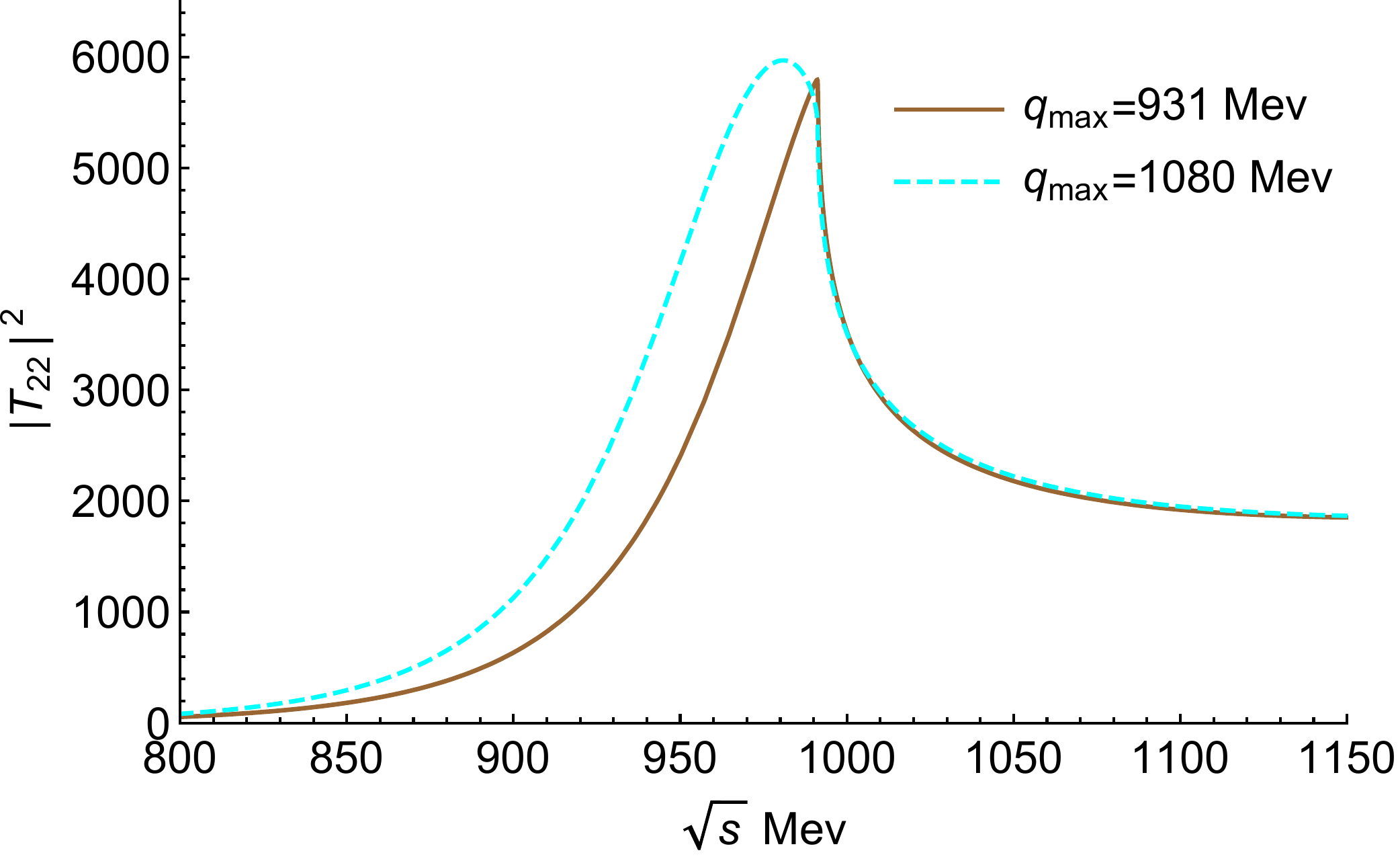}
  \caption{\footnotesize Results of $|T_{22}|^2$ for the $\pi\eta$ channel.} \label{fig:fig6c}
\end{subfigure} 
\caption{Results of the modulus squared of the scattering amplitudes in $I=1$ sector.}
\label{fig:fig6}
\end{figure}

\begin{figure}
\begin{subfigure}{.5\textwidth}
  \includegraphics[width=1\linewidth]{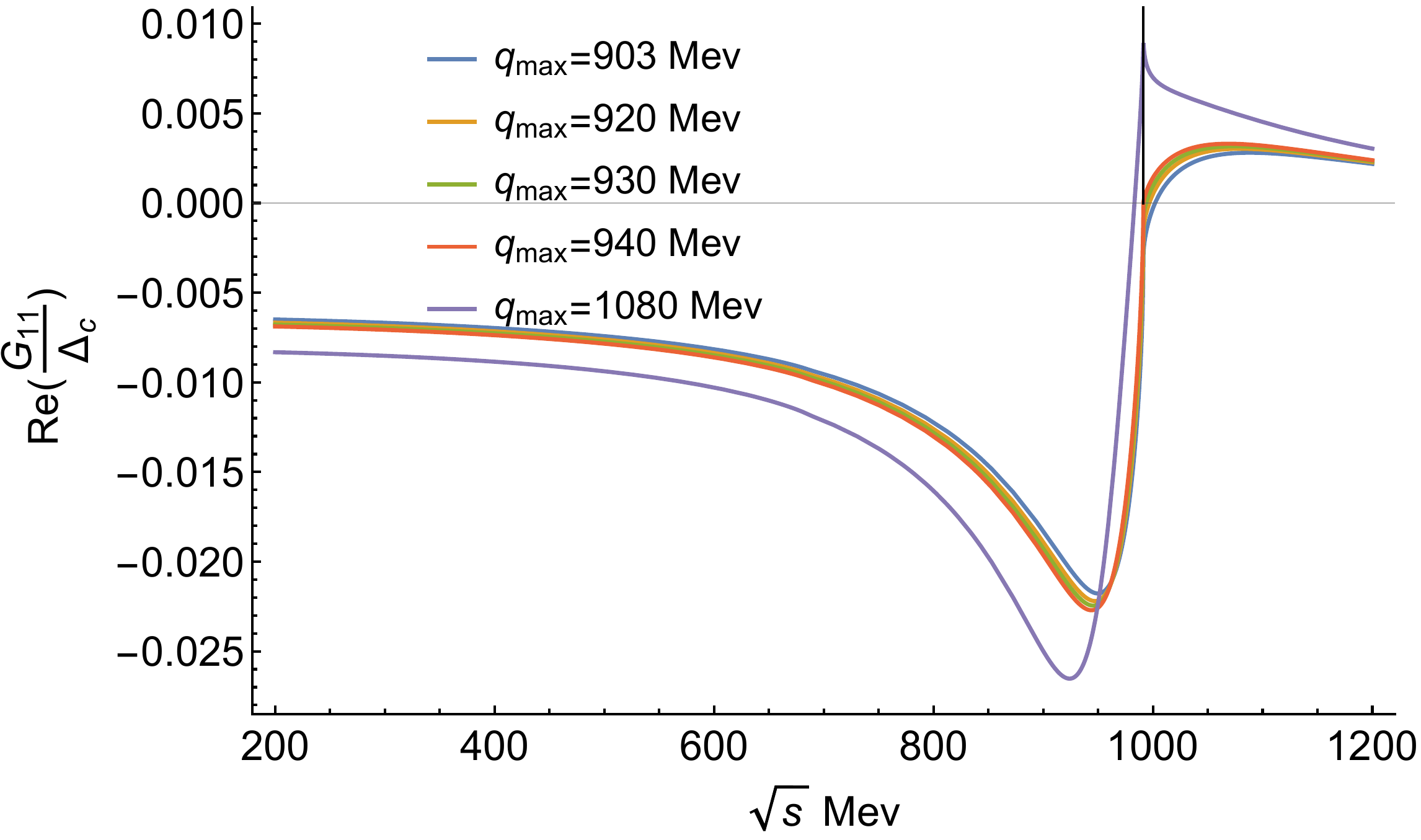}
\end{subfigure}%
\begin{subfigure}{.5\textwidth}
 \includegraphics[width=1\linewidth]{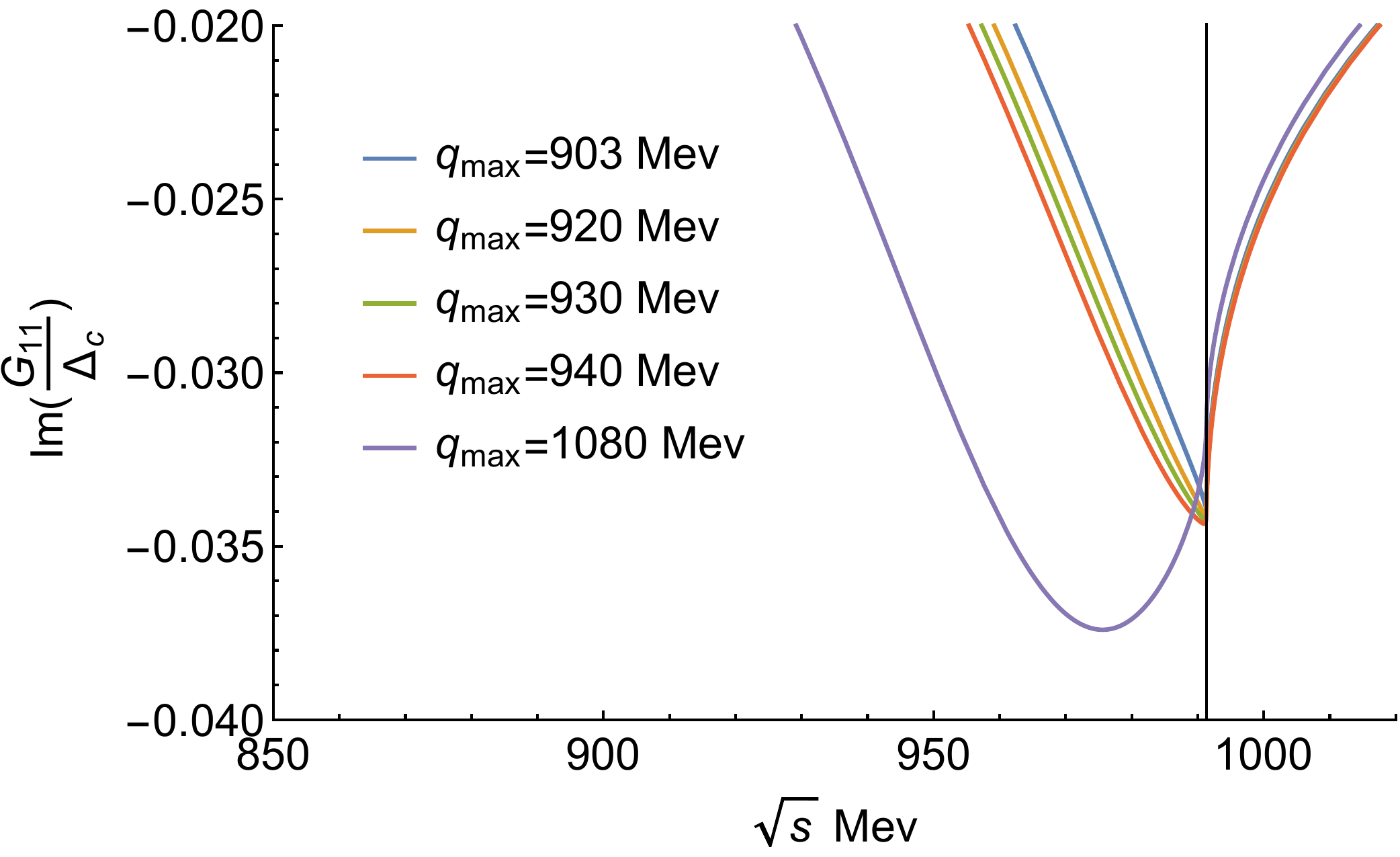} 
\end{subfigure}  
\caption{\footnotesize Results of the real part (left) and imaginary (right) part of ($\frac{G_{11}}{\Delta_c}$) with different cutoffs.}
\label{fig:fig7}
\end{figure}

Next, we show our results for the invariant mass distributions. As done in Ref\cite{npa620}, we compare our results with the data of $\pi\eta$ invariant mass distribution from the reaction $K^{-}p \rightarrow \Sigma(1385) \pi^{-} \eta$ and the ones of $K\bar{K}$ from the reaction $K^{-}p \rightarrow \Sigma^{+}K^{-}K^{0}$, see Fig. \ref{fig:fig4}, where we use
\begin{equation}
 \frac { d \sigma _ {ii} } { d \sqrt{s} } = C \left| T _ { ii } \right| ^ { 2 } q _ {cmi}  \text{  ,}
\end{equation}
where $T_{ii}$ is the scattering amplitude of the $K\bar{K}$ or $\pi\eta$ channel, $q_ {cmi}$ is three momentum in CM frame and $C$ the normalization factor. To see more clearly the resonances dynamically produced in the coupled channel interactions, we plot the modulus squared of the scattering amplitudes in $I=0$ and $I=1$  sectors as shown in Figs. \ref{fig:fig5} and \ref{fig:fig6}. From Figs. \ref{fig:fig5a} and \ref{fig:fig5b} of $|T_{11}|^{2}$ and $|T_{12}|^{2}$ for $I=0$, the peak of $f_0(980)$ state is clearly seen. In Fig. \ref{fig:fig5c}, the broad structure of $T_{22}$ are $\sigma$ resonance, where the dip is the signal of $f_0(980)$ state closed to the $K\bar{K}$ threshold and the structure of the amplitudes are consistent with the ones calculated with dispersion method \cite{Dai:2019zao}. Likewise, the $a_{0}(980)$ resonance can be clearly seen in $|T_{11}|^{2}$,  $|T_{12}|^{2}$, and  $|T_{22}|^{2}$ in $I=1$ sectors in Fig. \ref{fig:fig6}. In spite of showing the $a_{0}(980)$ resonance in the results of $|T_{22}|^{2}$, see Fig. \ref{fig:fig6c}, there is an extra feature, which is called threshold effect \cite{diffcross,npa620}, of which more details can be seen a recent review \cite{guofk}. This feature is due to the strong coupling of the resonance $a_{0}(980)$ to the $K\bar{K}$ channel which cause to dwindle the width of the scattering amplitude and change the location of the maximum. This effect is originated from the second term of $T_{22}$ and precisely comes from the imaginary part of the term ($\frac{G_{11}}{\Delta_c}$) as shown in Fig. \ref{fig:fig7}.
 
\begin{figure}
\begin{subfigure}{.45\textwidth}
  \centering
  \includegraphics[width=1\linewidth]{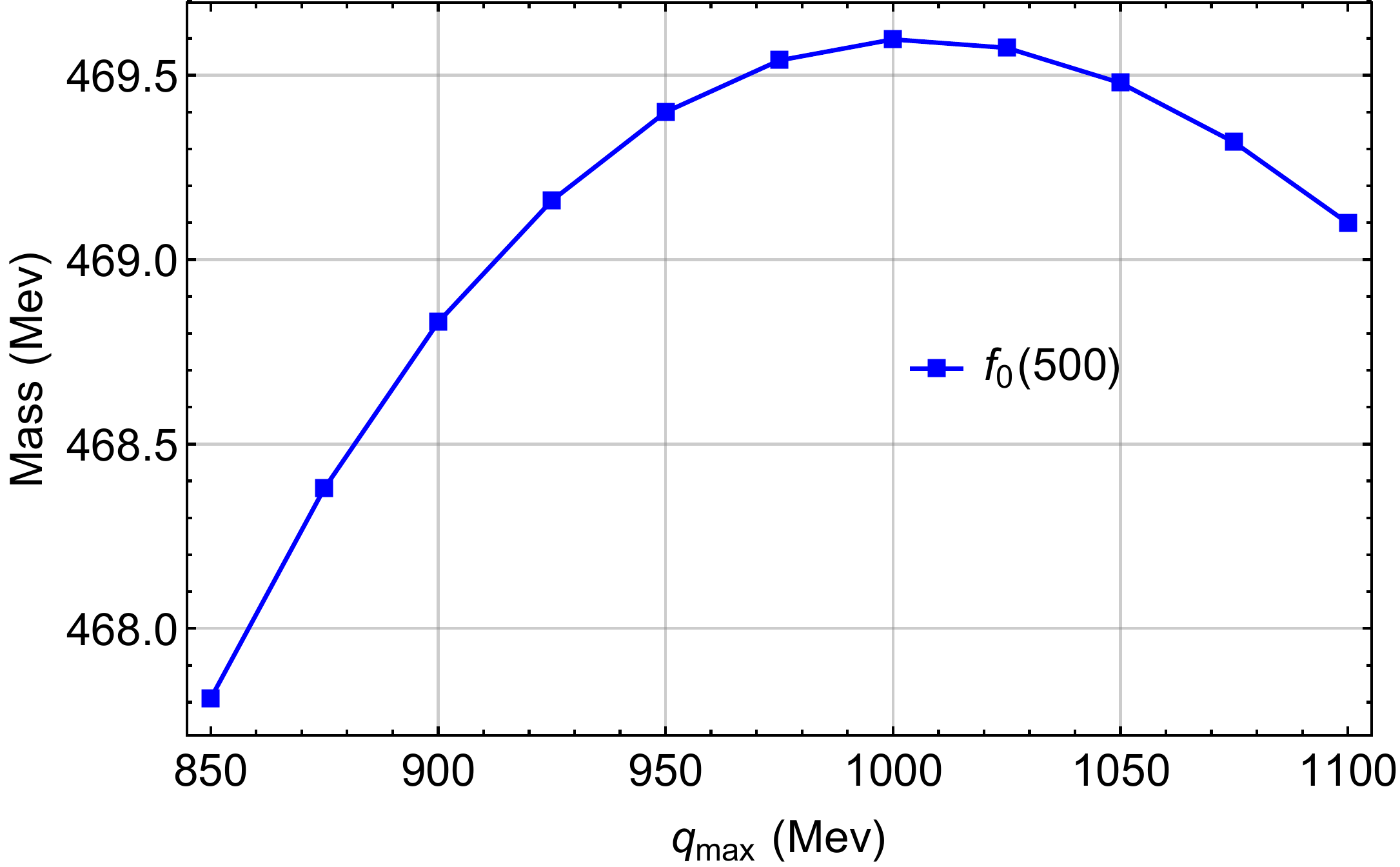} 
\end{subfigure} 
\begin{subfigure}{.45\textwidth}
  \centering
  \includegraphics[width=1\linewidth]{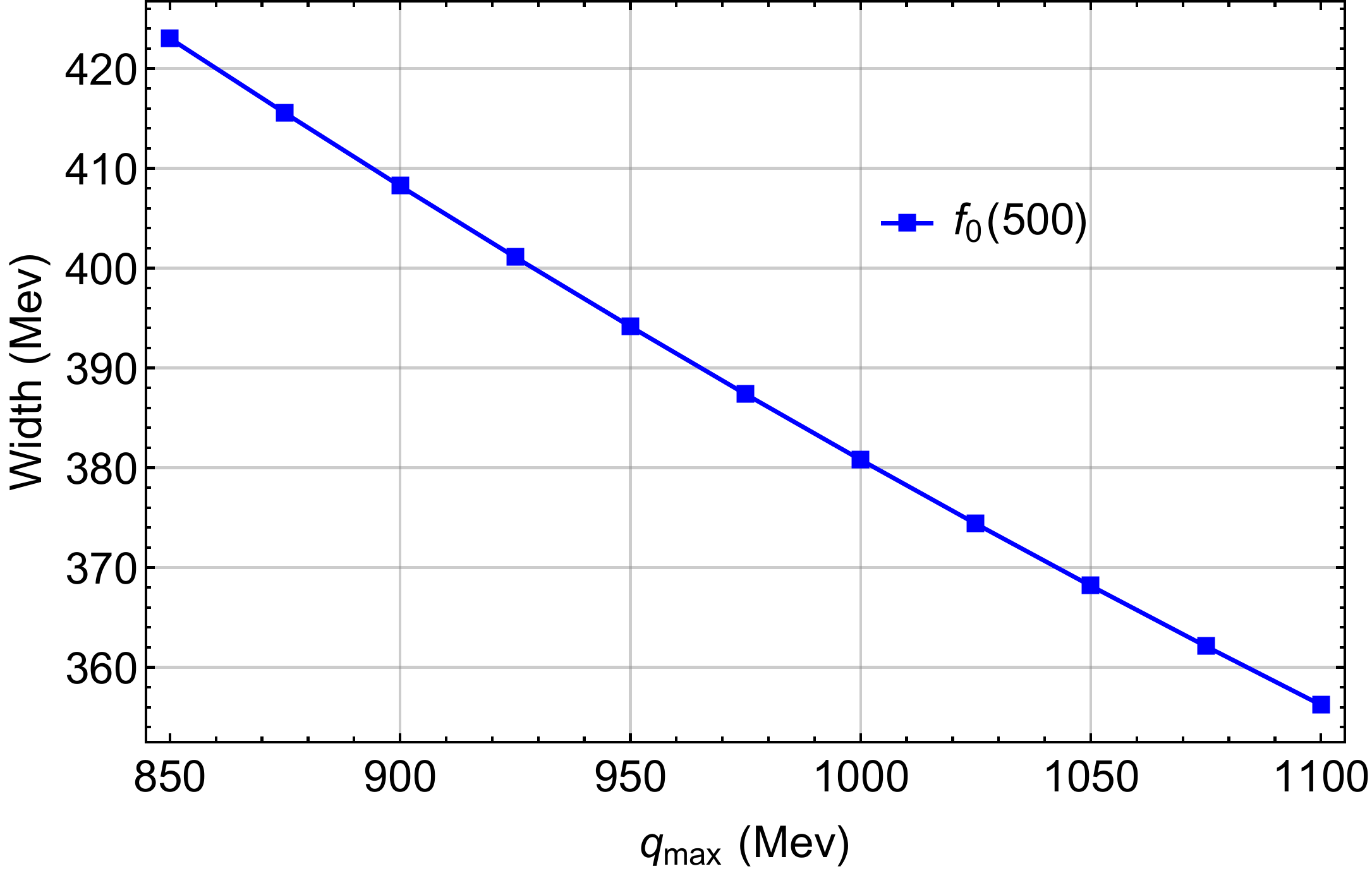} 
\end{subfigure} 
\begin{subfigure}{.45\textwidth}
  \centering
  \includegraphics[width=1\linewidth]{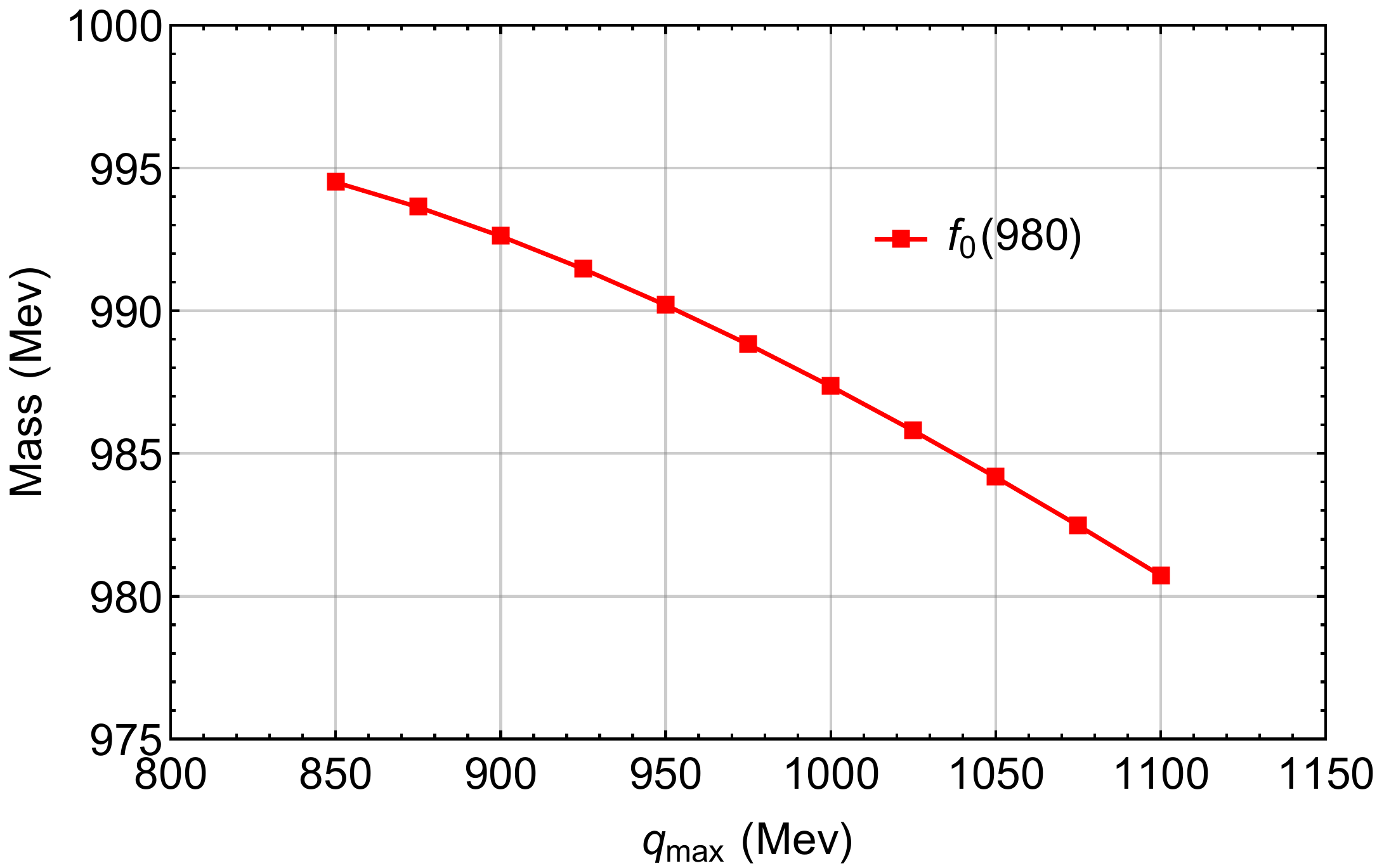} 
  \end{subfigure}
  \begin{subfigure}{.45\textwidth}
  \centering
  \includegraphics[width=1\linewidth]{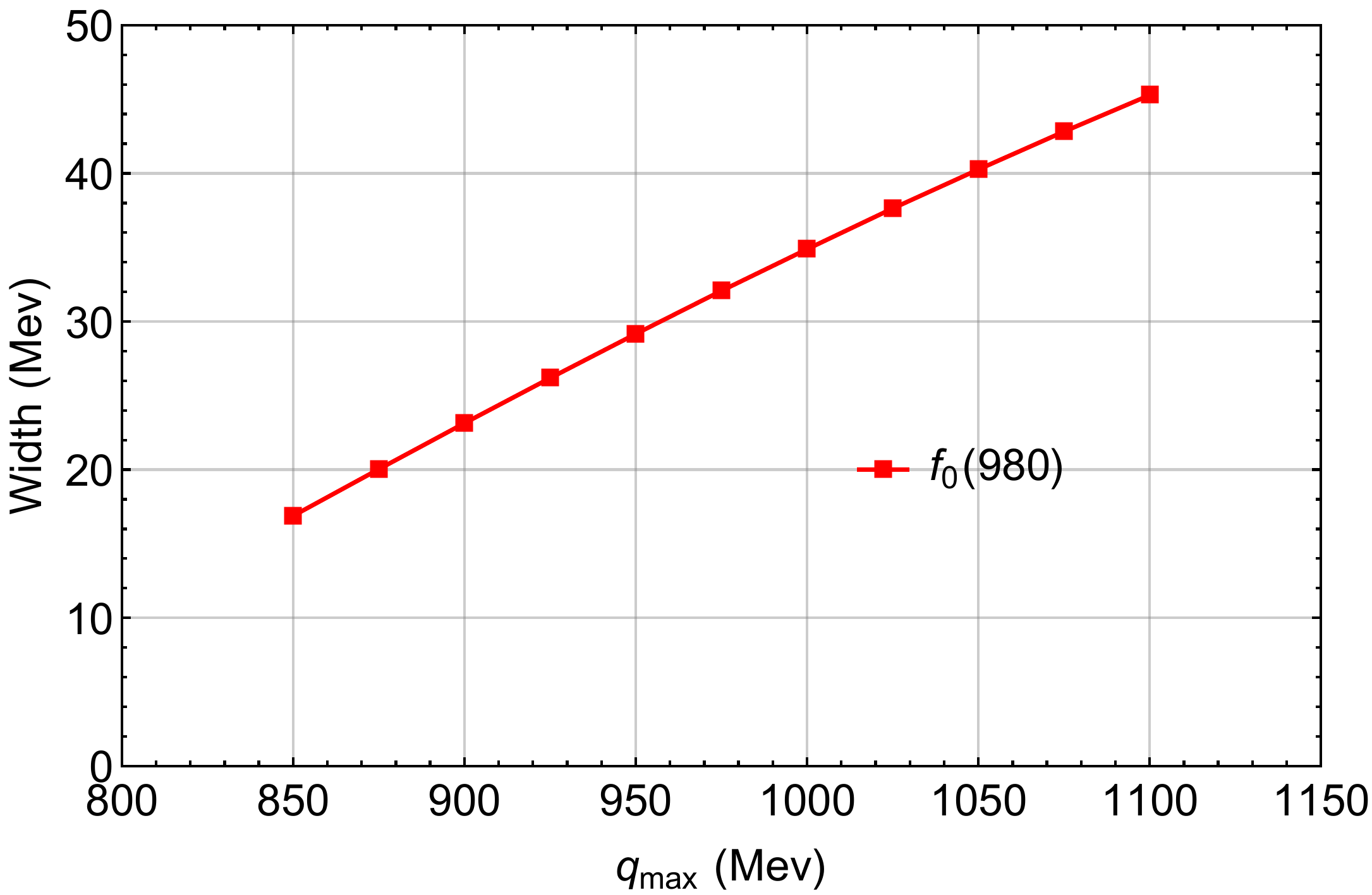} 
  \end{subfigure}
\begin{subfigure}{.45\textwidth}
  \centering
  \includegraphics[width=1\linewidth]{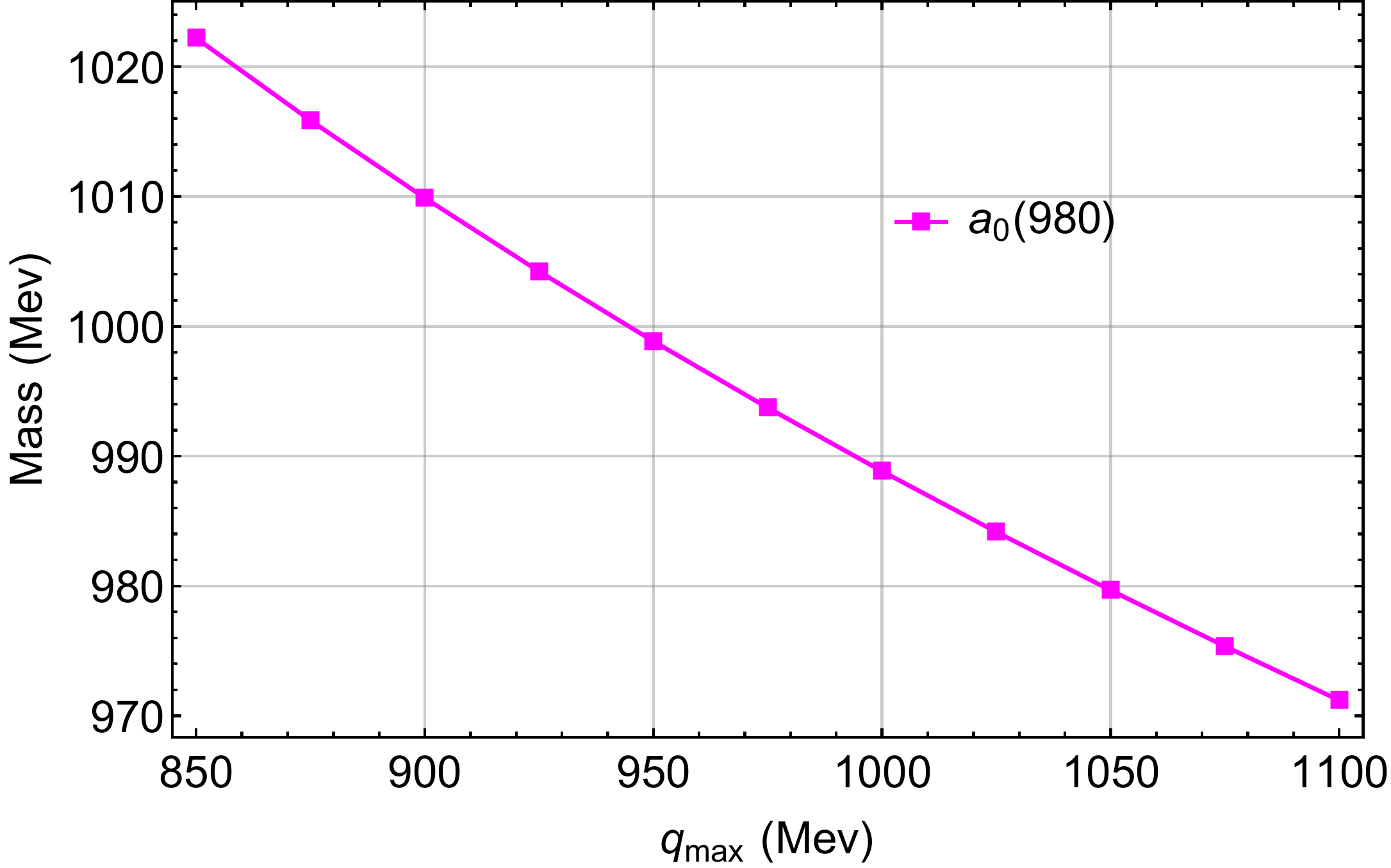} 
  \end{subfigure}
  \begin{subfigure}{.45\textwidth}
  \centering
  \includegraphics[width=1\linewidth]{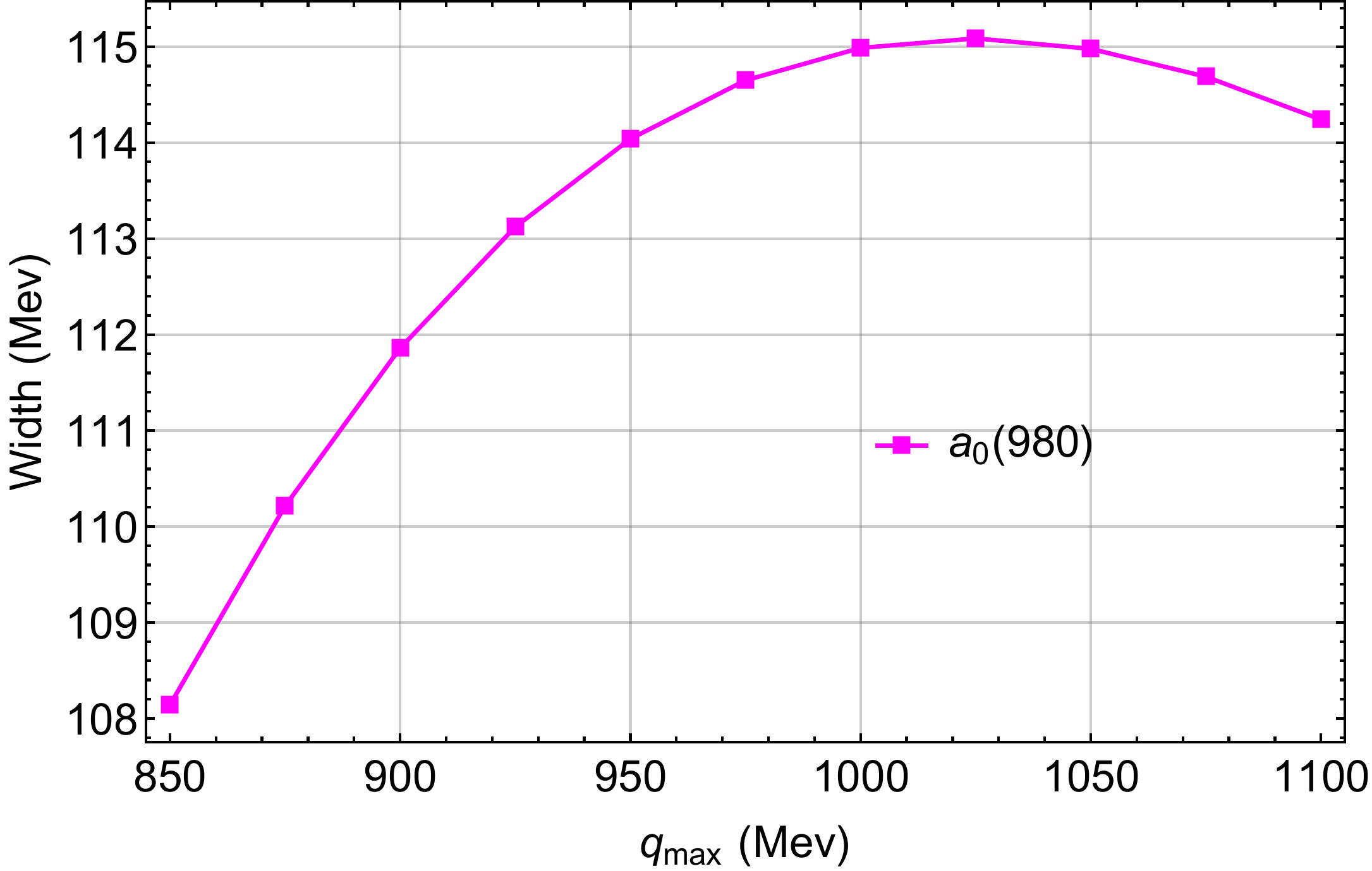} 
  \end{subfigure}  
  \caption{Trajectories for the masses and the widths of the poles in the second Riemann sheets corresponding to $\sigma$ ($f_{0}(500)$), $f_{0}(980)$, and  $a_{0}(980)$ by varying cut-off ($q_{max}$).}
\label{fig:fig8}
\end{figure}

In Figs. \ref{fig:fig5} and \ref{fig:fig6}, we have dynamically produced the states of $\sigma$, $f_0(980)$ and $a_0(980)$ in the modulus squared of the scattering amplitudes. Thus, we can search for their corresponding poles in the second Riemann sheets to determine their masses and widths. To see their poles stable or not,  we plot the trajectories for the masses and the widths of their poles by changing the value of $q_{max}$, see Fig. \ref{fig:fig8}. From Fig. \ref{fig:fig8}, we can find that even varying the free parameter, $q_{max}$, the corresponding poles for them are stably produced in the second Riemann sheets. But, one can see that the behaviour of pole for $\sigma$ is different from the other two, $f_0(980)$ and $a_0(980)$. When the cutoff increases, the mass of $\sigma$ slightly increases to a maximum and then declines, and its width decreases, whereas the masses of $f_{0}(980)$ and $a_{0}(980)$ always decline from above $K\bar{K}$ threshold to below threshold, and their widths increase (the width of $a_{0}(980)$ increase to an upper limit). These differences mean that the properties of $\sigma$ look like different from the ones of $f_0(980)$ and $a_0(980)$. On the other hand, the mass of $f_0(980)$ is more stable than the one of $a_0(980)$, whereas, the width of $a_0(980)$ does not change much when the cutoff varies. Thus, we continue to make further investigations about their different properties. 

For the sake of the complete investigations about the characteristics of these resonances, we continue to study the couplings, the compositeness, the wave functions and the radii as well. The couplings to various channels for isospin $I=0$ and $I=1$ sectors have been calculated using Eq. (\ref{eq22}), as presented in Tables \ref{tab:tab1} and \ref{tab:tab2}, respectively. From these results  in $I=0$ sector, it is observed that the $\sigma$ state couples to the $\pi\pi$ channel strongly, while $f_{0}$ strongly couples to the $ K\bar{K}$ channel. Thus, the pole of the $\sigma$ state dominates by the $\pi\pi$ channel whereas the one of the $f_{0}$ state mainly by the $K\bar{K}$ channel. In $I=1$ sector, the $a_{0}$ state is tightly coupled to both the $K\bar{K}$ and $\pi\eta$ channels, whereas it has more tendency to the $K\bar{K}$ channel, which means that the pole of $a_{0}$ is dominated by the $K\bar{K}$ channel but the contributions of $\pi\eta$ is significant too.

\begin{table}
\center
\caption{Couplings of $\sigma$ and $f_{0}$ to every channel for $I=0$ sector.}
\begin{tabular}{|c|c|c|c|c|}
\hline
  $ q_{max} =  931$ MeV & $g_{K\bar{K}}g_{K\bar{K}}(\text{GeV}^{2}) $ & $|g_{K\bar{K}}|(\text{GeV}) $  & $g_{\pi\pi}g_{\pi\pi}(\text{GeV}^{2}) $ & $|g_{\pi\pi}|(\text{GeV}) $ \\  \hline
    $\sigma$ : $469.23 +199.70 i$  & $-1.05 + 1.72 i $           &   1.42        &    $ - 3.49 + 8.20 i$           &   2.98                \\   \hline 
    $f_{0}$ : $991.17 + 13.45 i$      &   $10.92 - 10.91 i$          &   3.92          &      $ -1.76 + 0.70 i $             &   1.37     \\   \hline  
  $ q_{max} =1080$  MeV                                                                        \\   \hline
 $\sigma$ : $469.28 +180.46 i$  &  $-0.80 +1.86 i$            &   1.42                    &     $-2.0+8.28 i $          &   2.92            \\  \hline
  $f_{0}$ : $982.13 +21.67 i$      &   $16.15 - 10.55 i $             &   4.39                 &   $-2.34 + 1.11 i $             &   1.60      \\   \hline
\end{tabular}
\label{tab:tab1}
\end{table}

\begin{table}
\center
\caption{Couplings of $a_{0}$ for $I=1$ sector.}
\begin{tabular}{|c|c|c|c|c|}
\hline
  $ q_{max} = 931$ MeV & $g_{K\bar{K}}g_{K\bar{K}}(\text{GeV}^{2}) $ & $|g_{K\bar{K}}|(\text{GeV}) $  & $g_{\pi\eta}g_{\pi\eta}(\text{GeV}^{2}) $ & $|g_{\pi\eta}|(\text{GeV}) $ \\   \hline  
  $a_{0}$ : $1002.90 + 56.68 i$  &   $24.17 - 9.22 i$         &   5.08           &     $10.30 + 5.71 i $             &   3.43                 \\    \hline   
  $ q_{max} = 1080$ MeV                                                                       \\   \hline  
  $a_{0}$ : $974.50 +57.31 i$  &   $21.83 - 3.28 i $          &   4.78        &     $8.16 + 5.20 i $                    &   3.11                 \\    \hline  
\end{tabular}
\label{tab:tab2}
\end{table}

Using the sum rule of Eq. (\ref{eq25}), the compositeness can be calculated from the couplings of the dynamically generated resonances, where one can check whether $f_{0}$ and $a_{0}$ are a pure molecular state or have something else. Our results are given in Tables \ref{tab:tab3} and \ref{tab:tab4}. From the results of Table \ref{tab:tab3}, once again we can conclude that the structure of $f_{0}$ is highly dominated by the $K\bar{K}$ molecular components, which is up to 80\% with the central value of $q_{max}=931$ MeV, and has very small parts of the $\pi\pi$ components even though the coupling to the $\pi\pi$ channel is not so small, which is more than 1/3 of the one to the $K\bar{K}$ channel, see Table. \ref{tab:tab1}. By contrast, the $\sigma$ state has large part components of $\pi\pi$ about 40\% and quite tiny parts of $K\bar{K}$, where one can find that this state still has much large parts of non-molecular components. Our resuts of the compositeness in Table. \ref{tab:tab3} for the states of $\sigma$ and $f_{0}$ are consistert with the ones obtained in Ref. \cite{guooller} with the inverse amplitude method. The $a_{0}$ state has a main components of $K\bar{K}$ and some contributions from the $\pi\eta$ component, see Table \ref{tab:tab4}, but it stil has something else about 30\%. These results are comparable with the work of Ref. \cite{Sekihara} where the properties of these resonances are investigated with the formalism of finite-volume. Therefore, these states are not pure molecular states and have something else, where Ref. \cite{Baru:2003qq} also conclude that the $f_0(980)$ and $a_0(980)$ states are not elementary states based with a Flatt\'e parameterization analysis.

\begin{table}
\center
\caption{Results of the compositeness of the poles in $I=0$ sector.}
\begin{tabular}{|c|c|c|c|c|}
\hline
  $ q_{max}=931$ MeV    &   $( 1- Z)_{K\bar{K}}$  &   $|( 1- Z)_{K\bar{K}}|$  &     $ (1- Z)_{\pi\pi} $  &   $ |(1- Z)_{\pi\pi}| $    \\ \hline
  $\sigma$ : $469.23 +199.70 i$  &   $-0.01 + 0.01 i $        &   0.01                &     $-0.13 - 0.37 i $     &      0.40                      \\ \hline
  $f_{0}$ : $991.17 + 13.45 i$      &  $0.79 + 0.12 i $        &   0.80                 &     $0.02- 0.01 i$     &      0.02                    \\ \hline
    $ q_{max}=1080$ MeV                                                                         \\ \hline
  $\sigma$ : $469.28 +180.46 i$  &   $- 0.00 + 0.01 i$         &   0.01      &     $-0.16- 0.36 i $     &      0.39                       \\ \hline 
  $f_{0}$ : $982.13 +21.67 i$      &   $0.70 + 0.11 i $        &   0.70                 &     $0.02 - 0.01 i $    &      0.02                   \\ \hline
\end{tabular}
\label{tab:tab3}
\end{table}

\begin{table}
\center
\caption{Results of the compositeness of the poles in $I=1$ sector.}
\begin{tabular}{|c|c|c|c|c|}
\hline
  $ q_{max}=931$ MeV   &  $( 1- Z)_{K\bar{K}}$  &   $|( 1- Z)_{K\bar{K}}|$  &    $ (1- Z)_{\pi\eta} $  &   $ |(1- Z)_{\pi\eta}| $    \\ \hline
  $a_{0}$ : $1002.90 + 56.68 i$     &   $0.37 + 0.41 i $      &        0.55           &     $ -0.09 - 0.13 i$    &       0.16                 \\ \hline 
  $ q_{max}=1080$ MeV                                                               \\ \hline
  $a_{0}$ : $974.50 +57.31 i$   &   $0.34 + 0.29 i$       &        0.45              &      $-0.07 - 0.12 i $   &       0.14                 \\ \hline
\end{tabular}
\label{tab:tab4}
\end{table}
    
\begin{figure}
\begin{subfigure}{.45\textwidth}
  \centering
  \includegraphics[width=1\linewidth]{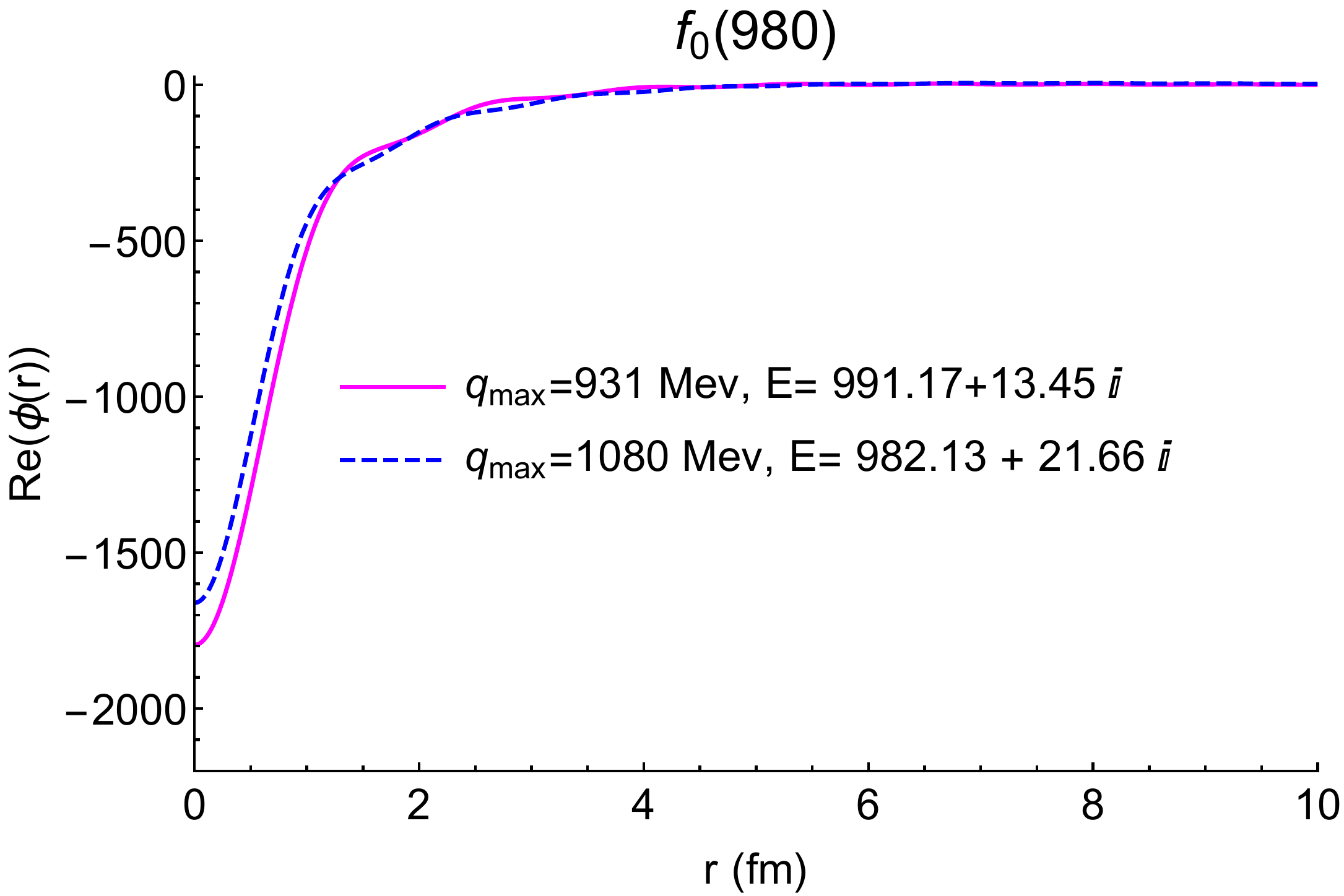} 
\end{subfigure} 
\begin{subfigure}{.45\textwidth}
  \centering
  \includegraphics[width=1\linewidth]{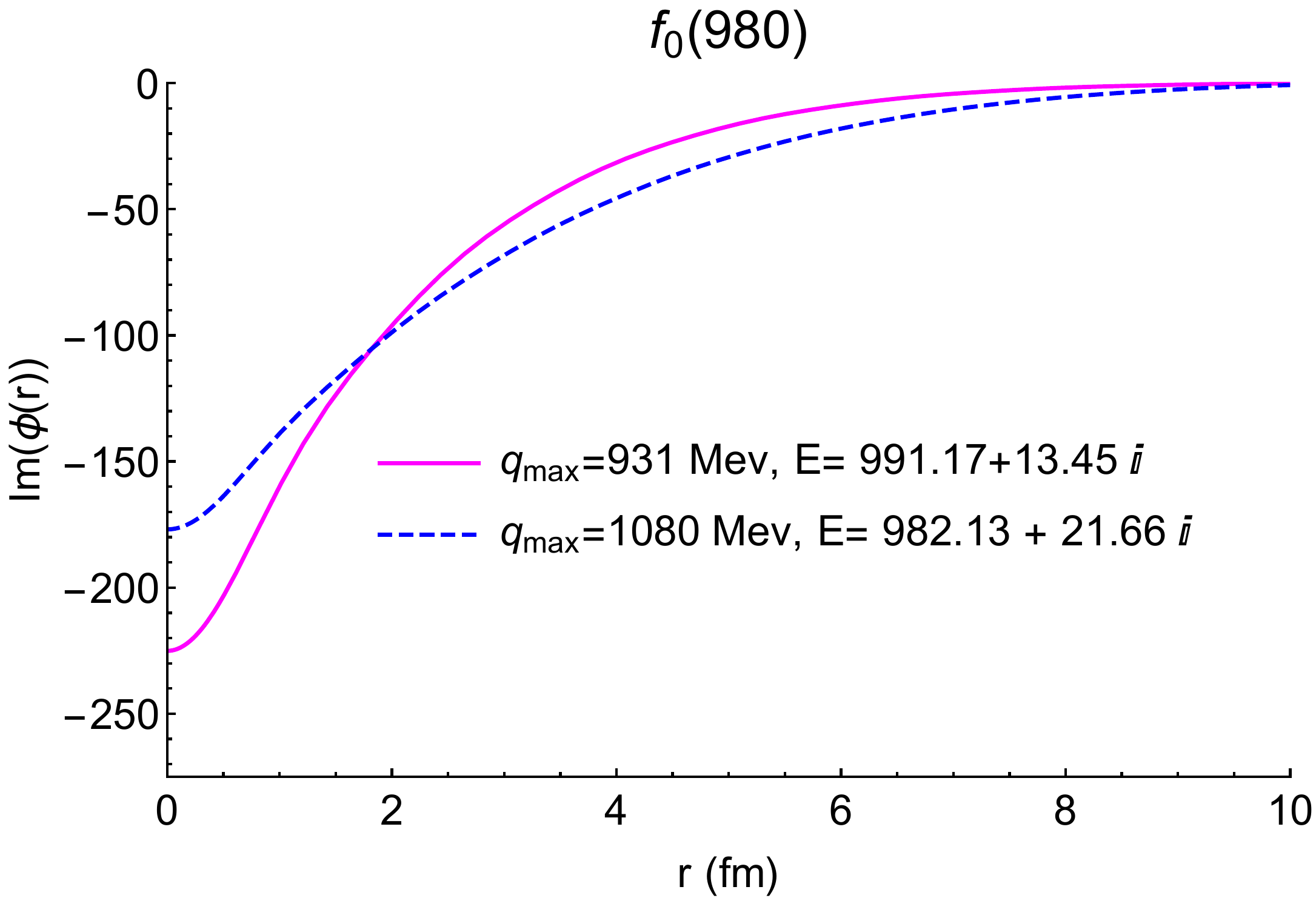} 
  \end{subfigure}
    \begin{subfigure}{.45\textwidth}
  \centering
  \includegraphics[width=1\linewidth]{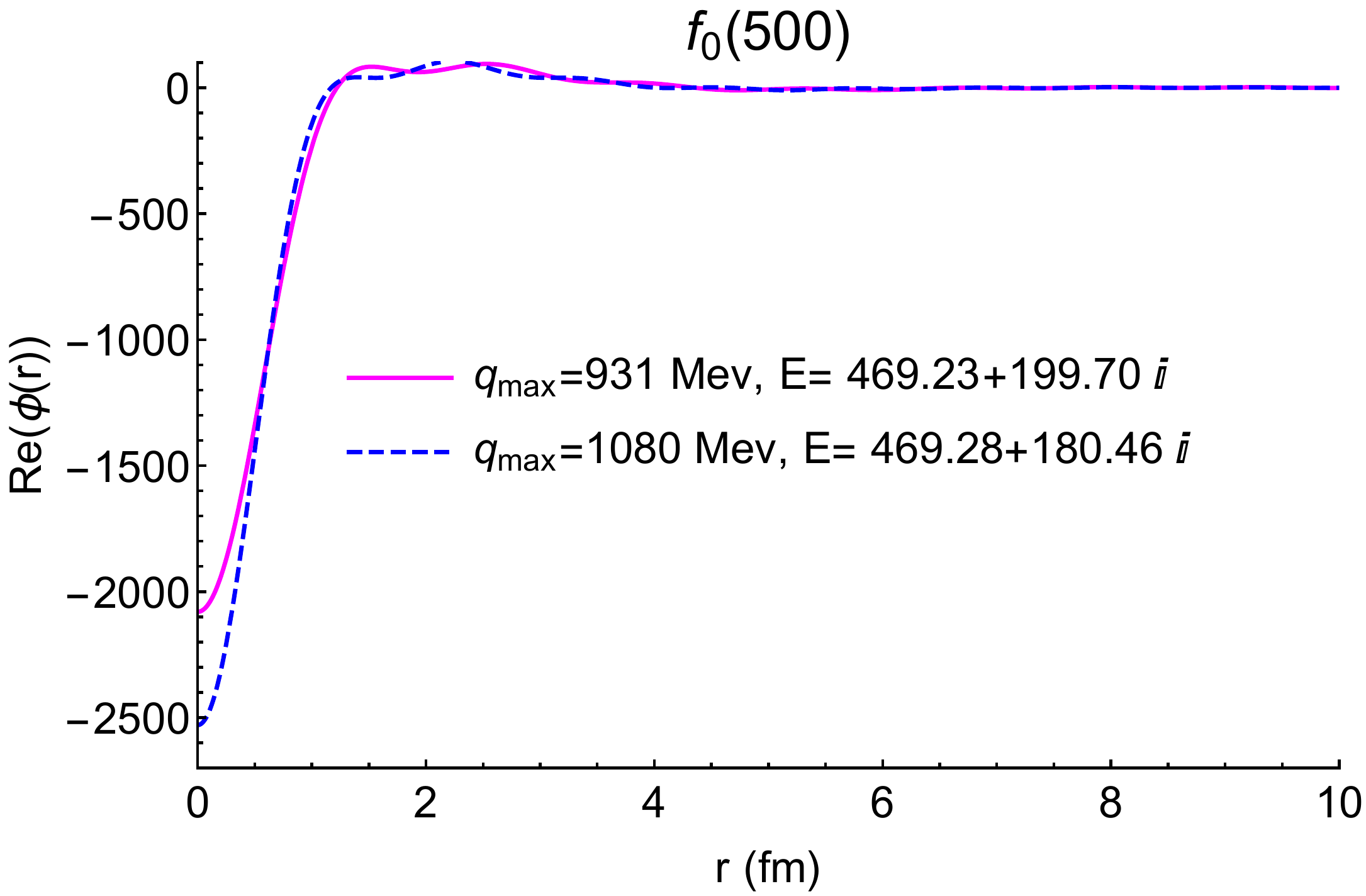} 
\end{subfigure} 
\begin{subfigure}{.45\textwidth}
  \centering
  \includegraphics[width=1\linewidth]{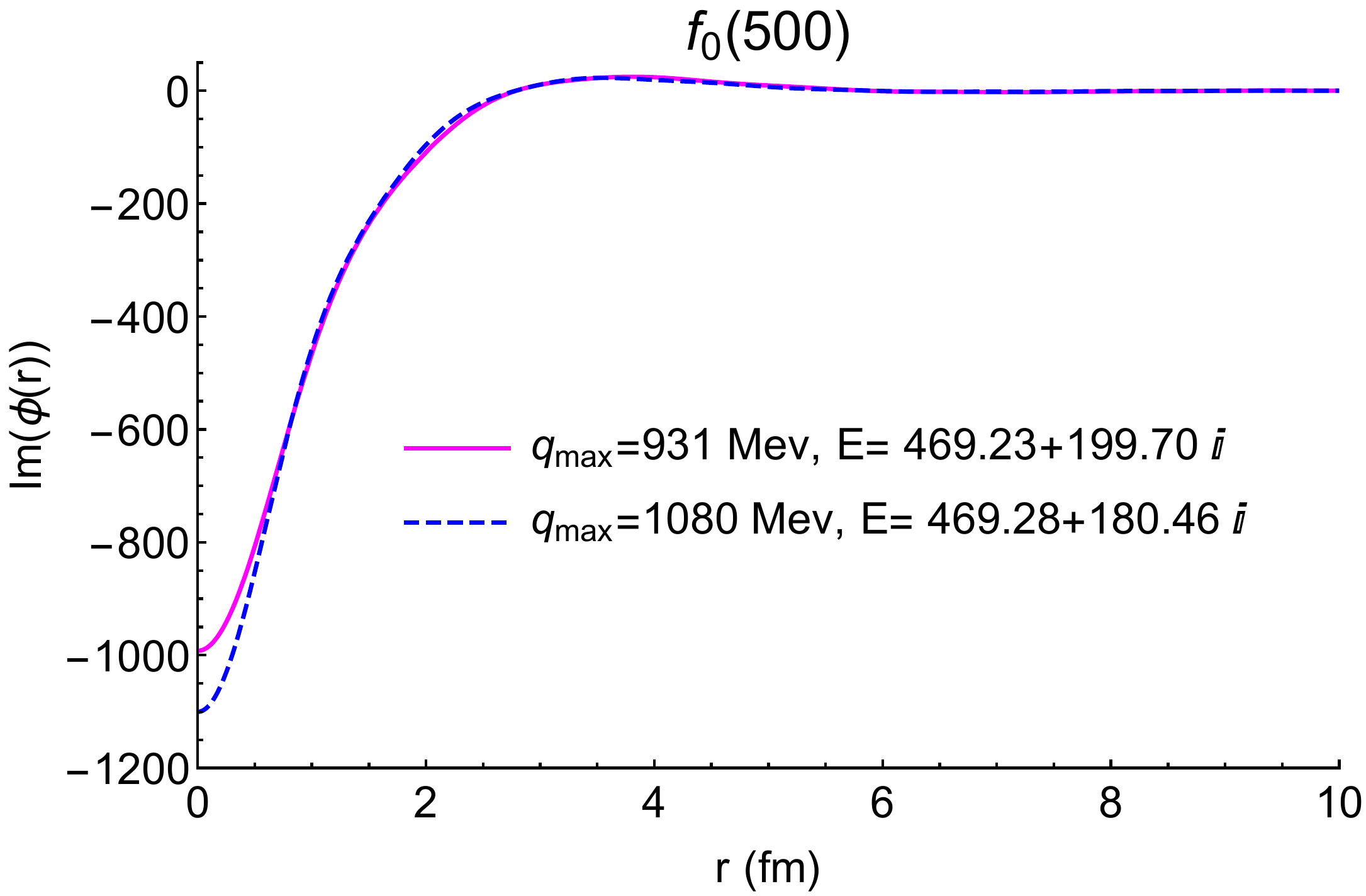} 
  \end{subfigure}
  \begin{subfigure}{.45\textwidth}
  \centering
  \includegraphics[width=1\linewidth]{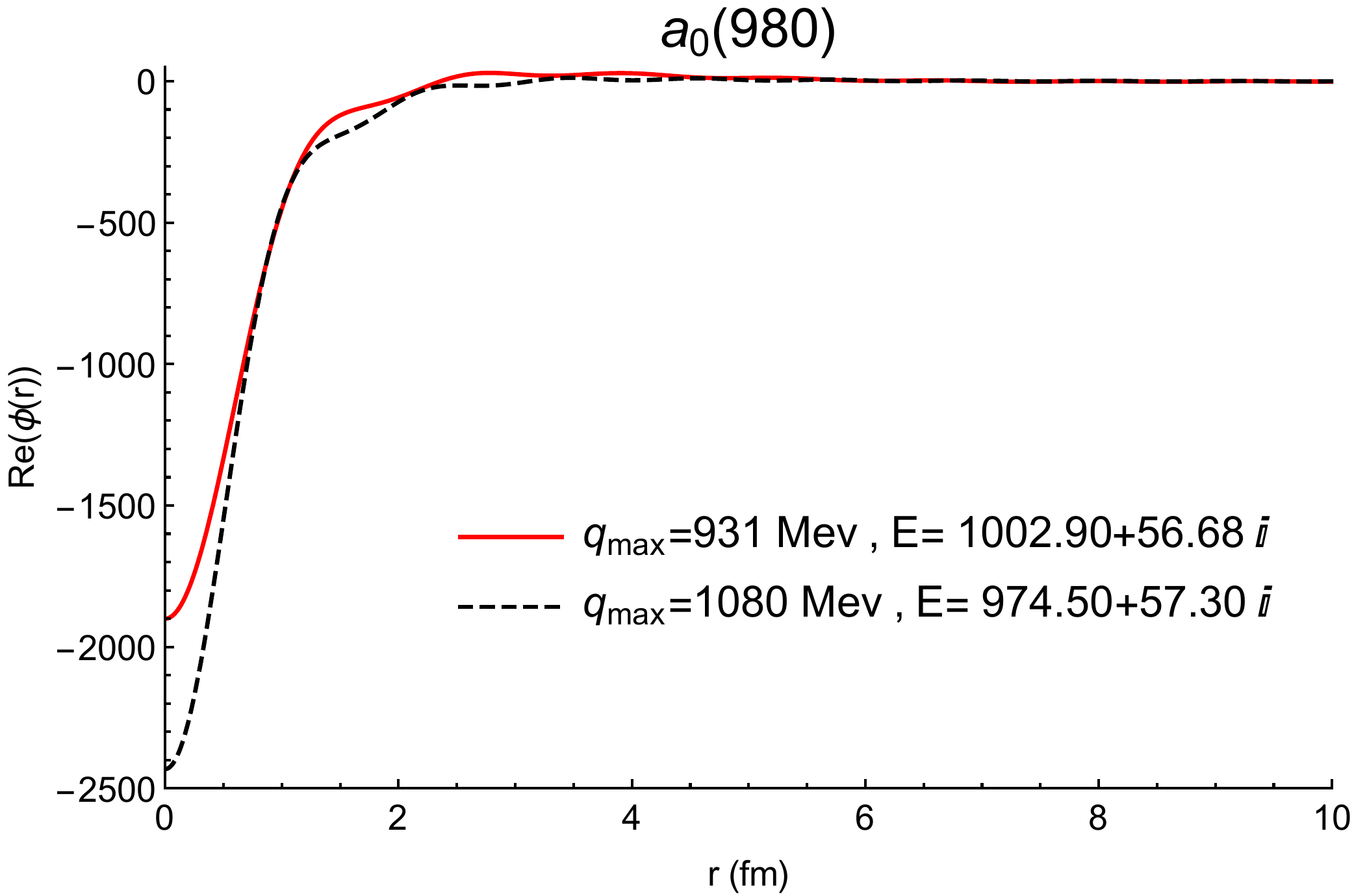} 
\end{subfigure} 
\begin{subfigure}{.45\textwidth}
  \centering
  \includegraphics[width=1\linewidth]{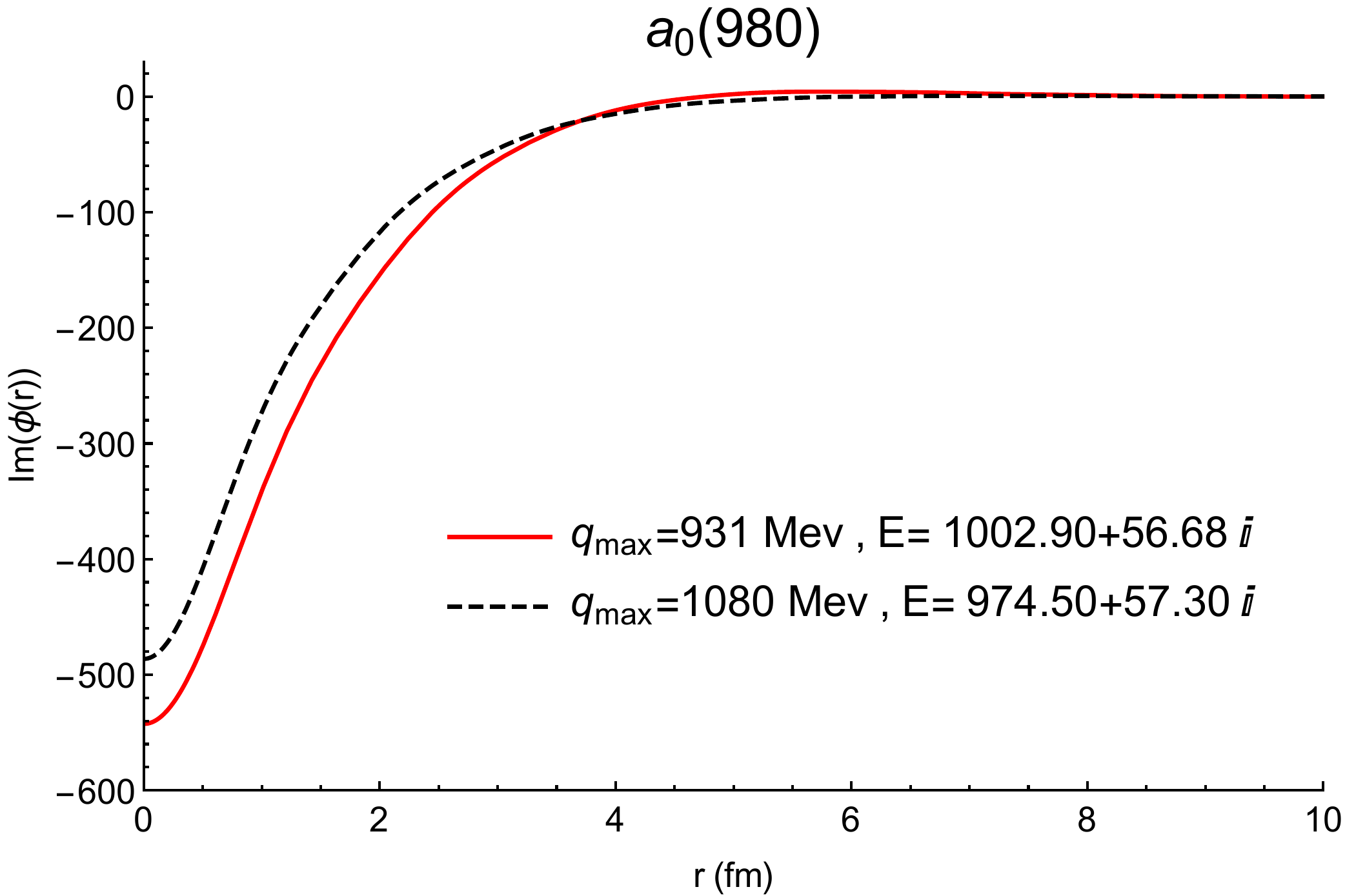} 
  \end{subfigure} 
  \caption{Results of the real part and imaginary part of the wave functions for the $f_{0}$ (upper parts),  $\sigma$ (middle parts) and $a_{0}$ (lower parts) states.}
\label{fig:fig9}
\end{figure}

\begin{figure}
  \begin{subfigure}{.45\textwidth}
  \centering
  \includegraphics[width=1\linewidth]{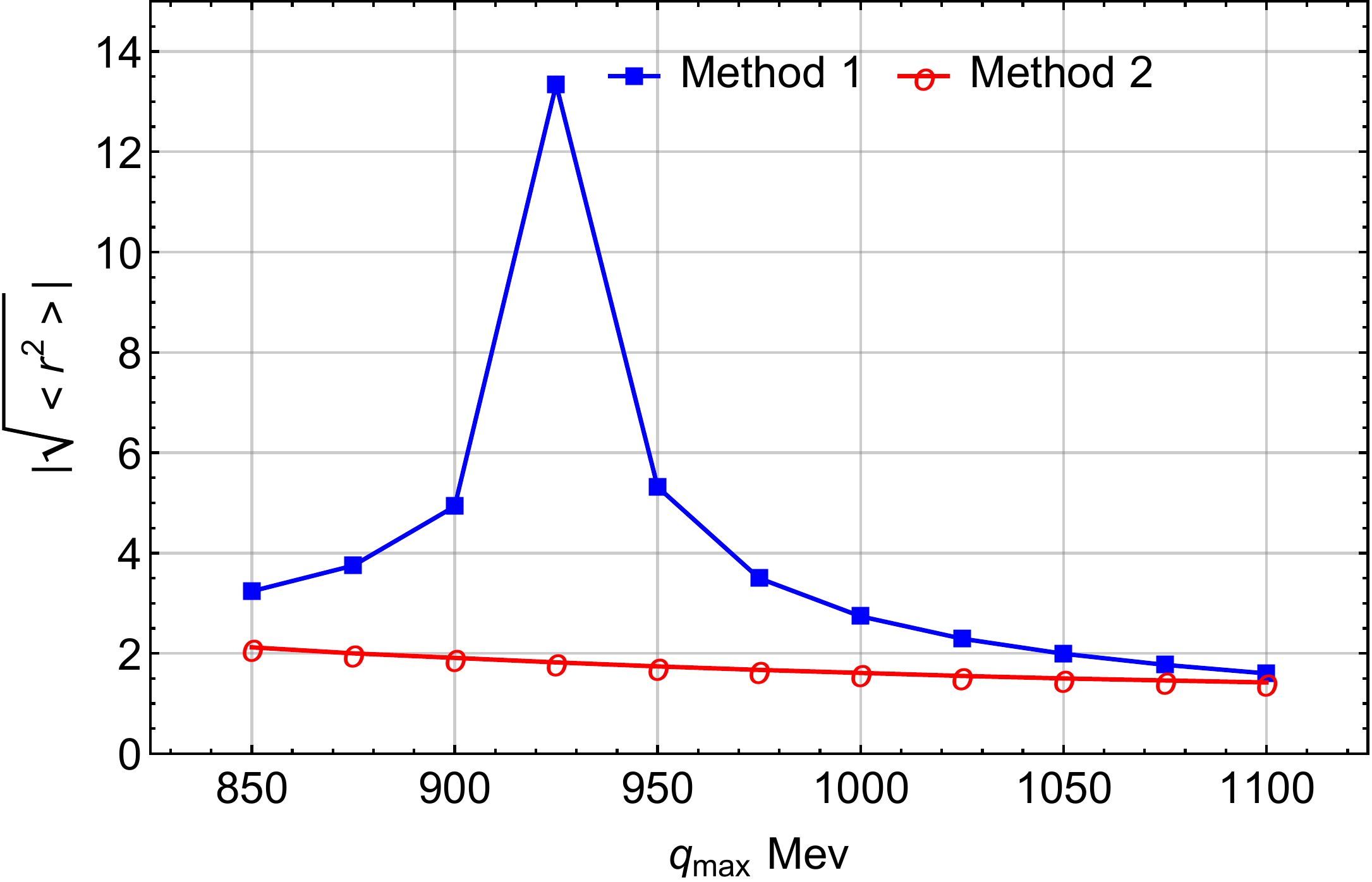} 
  \caption{\footnotesize The radii of $f_{0}(980)$.}
  \end{subfigure}
  \begin{subfigure}{.45\textwidth}
  \centering
  \includegraphics[width=1\linewidth]{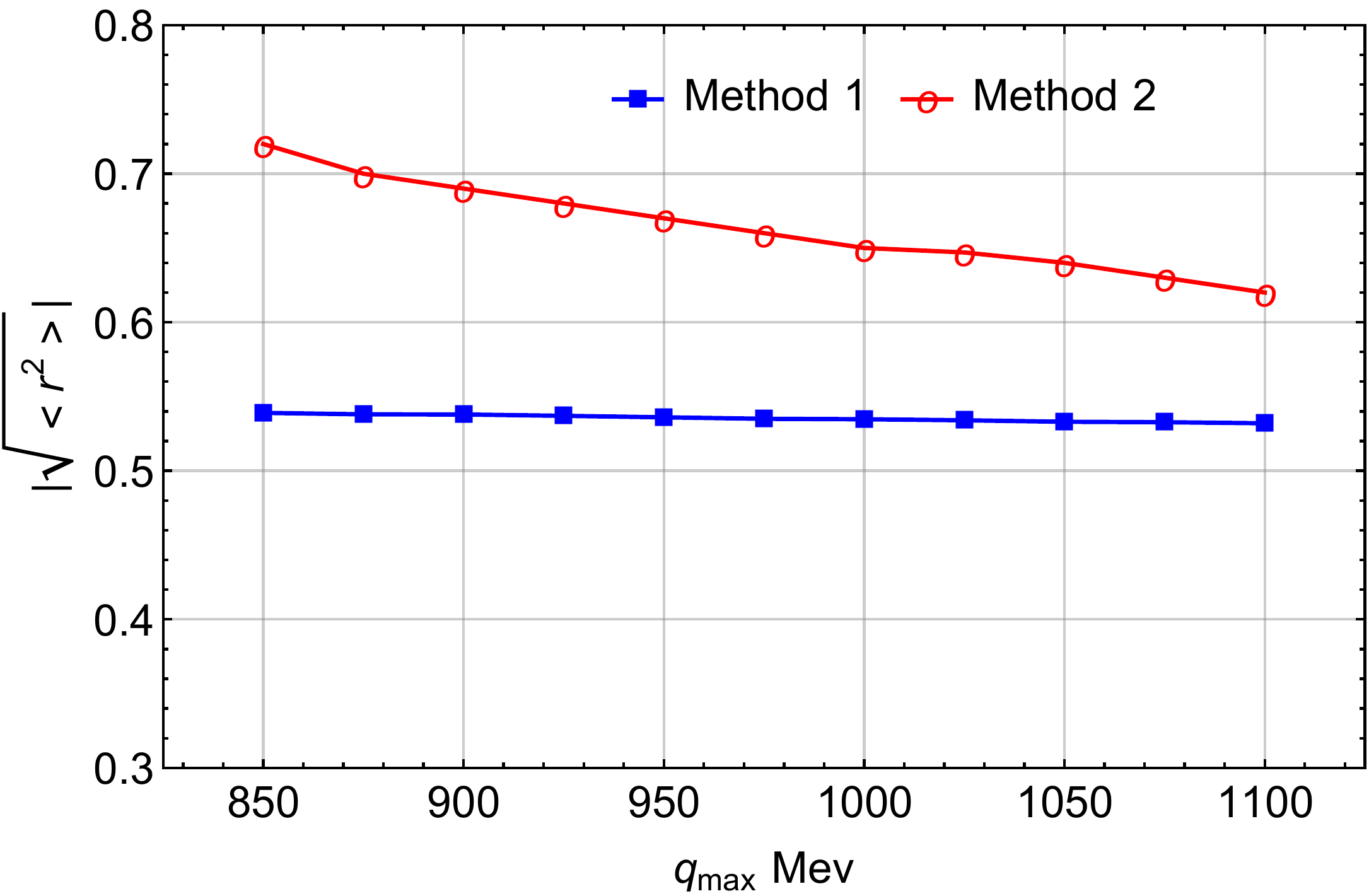} 
  \caption{\footnotesize The radii of $\sigma$.}
  \end{subfigure}
  \begin{subfigure}{.45\textwidth}
  \centering
   \includegraphics[width=1\linewidth]{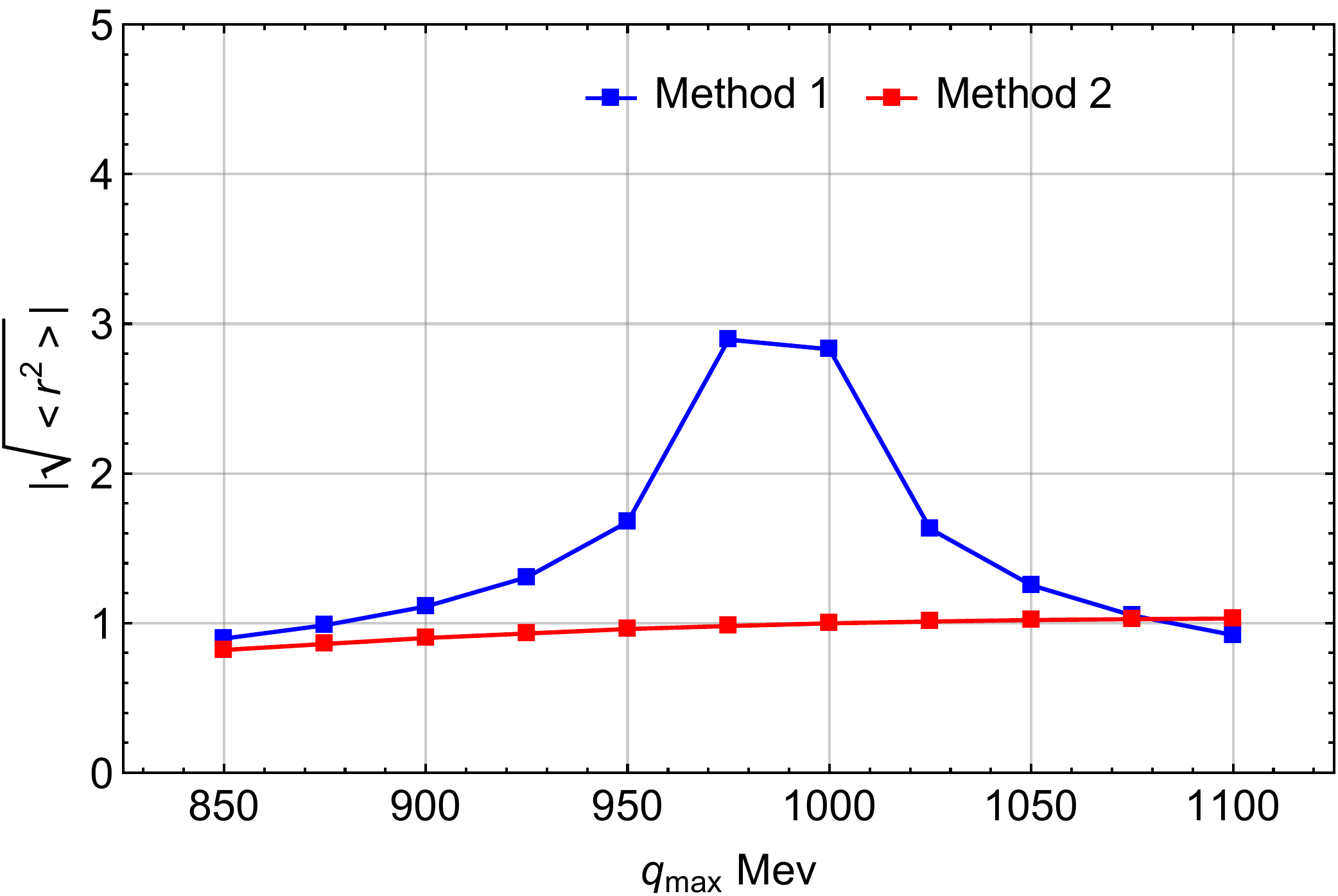} 
   \caption{\footnotesize The radii of $a_{0}(980)$.}
   \end{subfigure}
  \caption{Results of the radii by varying the cutoffs, where method 1 means the radii calculated with Eq.  \eqref{eq32}, and method 2 the one with Eq. \eqref{eq33}.}
    \label{rms1}
\end{figure}

To study the response of these states to the external sources, one need to know the form factor of these states. Thus, we evaluate the wave functions for them, and then, we can calculate the observables of the radii once we have their form factor. The wave functions of these state for all distances are shown in Fig. \ref{fig:fig9}, where the real parts and imaginary parts of the wave functions for the $f_{0}$,  $\sigma$ and $a_{0}$ states are given since the poles corresponding to these states are complex. From Fig. \ref{fig:fig9}, one can see that, up to about 4 fm, the wave functions for them become zero. Once we have the wave functions, we can investigate the radii of these states with Eq. (\ref{eq33}) which relate the wave functions at the origin, see Table. \ref{tab:tab5} with two cutoffs as above. As discuss before, we also can calculate the radii from the tail of the wave functions using Eq. (\ref{eq32}), as shown in Table \ref{tab:tab6}. From the results of Tables \ref{tab:tab5} and \ref{tab:tab6}, we can clearly see that the radii of the $f_{0}$ and $a_{0}$ states in two approaches of Eqs. (\ref{eq33}) and (\ref{eq32}) are larger than the typical hadronic scale $0.8 $ fm \cite{Sekihara}, whereas the one of the $\sigma$ state keeps in the the typical hadronic scale $\lesssim 0.8 $ fm. But, in Table \ref{tab:tab6}, we find that the one for $f_0(980)$ with cutoff $q_{max}=931$ MeV is much larger, $|\langle r^2 \rangle|=16.36$ fm, which is due to the corresponding pole closing to the $K\bar{K}$ threshold where the binding energy becomes zero, see Eq. (\ref{eq32}). Indeed, when we vary the cutoff and then change the positions of the corresponding poles closed to the threshold, the results of the first method with Eq. (\ref{eq32}) become unstable, as shown in Fig. \ref{rms1}. In Fig. \ref{rms1}, we can see that the results with the seconde method are much stable and the ones with the first method have singularities when the cutoff move the pole near to the threshold where the binding energy becomes zero. As discussed in Ref. \cite{Sekihara:2010uz}, the mean-squared radius is well defined with Eq. (\ref{eq33}) both for the bound states and the resonance states. Thus, at the end, we obtain $|\langle r^2 \rangle|_{f_0(980)} = 1.80 \pm 0.35$ fm, $|\langle r^2 \rangle|_{\sigma} = 0.68 \pm 0.05$ fm and $|\langle r^2 \rangle|_{a_0(980)} = 0.94 \pm 0.09$ fm, where we take the central value of the cutoff $q_{max}=931$ MeV within 15\% uncertainties.

\begin{table}
\center
\caption{The radii of states calculated with Eq. \eqref{eq33}.}
\begin{tabular}{|c|c|c|c|c|}
\hline
   Resonances        &   $ q_{max}=931$ MeV &  $|\langle r^2 \rangle|$       &  $ q_{max}=1080$ MeV   &   $|\langle r^2 \rangle|$       \\ \hline
  $f_{0}$    &        $1.42 + 1.10 i$  fm    & 1.80 fm      &   $1.31 + 0.62 i$ fm      &  1.45 fm   \\ \hline
  $\sigma$   &   $0.68 + 0.005 i$ fm      &  0.68 fm &       $ 0.63 + 0.04 i$  fm   &  0.63 fm  \\ \hline
  $a_{0}$   &       $0.83 + 0.44 i$ fm    & 0.94 fm     &   $0.96 + 0.35 i$ fm      &  1.03 fm       \\ \hline
\end{tabular}
\label{tab:tab5}
\end{table}

\begin{table}
\center
\caption{The radii of states evaluated with Eq. \eqref{eq32}.}
\begin{tabular}{|c|c|c|c|c|}
\hline
   Resonances        &   $ q_{max}=931$ MeV & $|\langle r^2 \rangle|$         &  $q_{max}=1080$ MeV   &  $|\langle r^2 \rangle|$   \\ \hline
  $f_{0}$     &       $16.32 + 1.20 i$  fm   & 16.36 fm  &  $1.73 + 0.13 i$  fm      &  1.73 fm  \\ \hline
  $\sigma$    &  $0.43 + 0.31 i$  fm      &  0.54 fm &        $0.44 + 0.29 i$  fm   & 0.53 fm  \\ \hline
  $a_{0}$    &       $0.56 - 1.25 i$ fm       & 1.37 fm    &   $0.96 + 0.36 i$ fm      &  1.02 fm      \\ \hline
\end{tabular}
\label{tab:tab6}
\end{table}

\subsection{Single channel approach}

\begin{figure}
\centering
  \includegraphics[width=.45\linewidth]{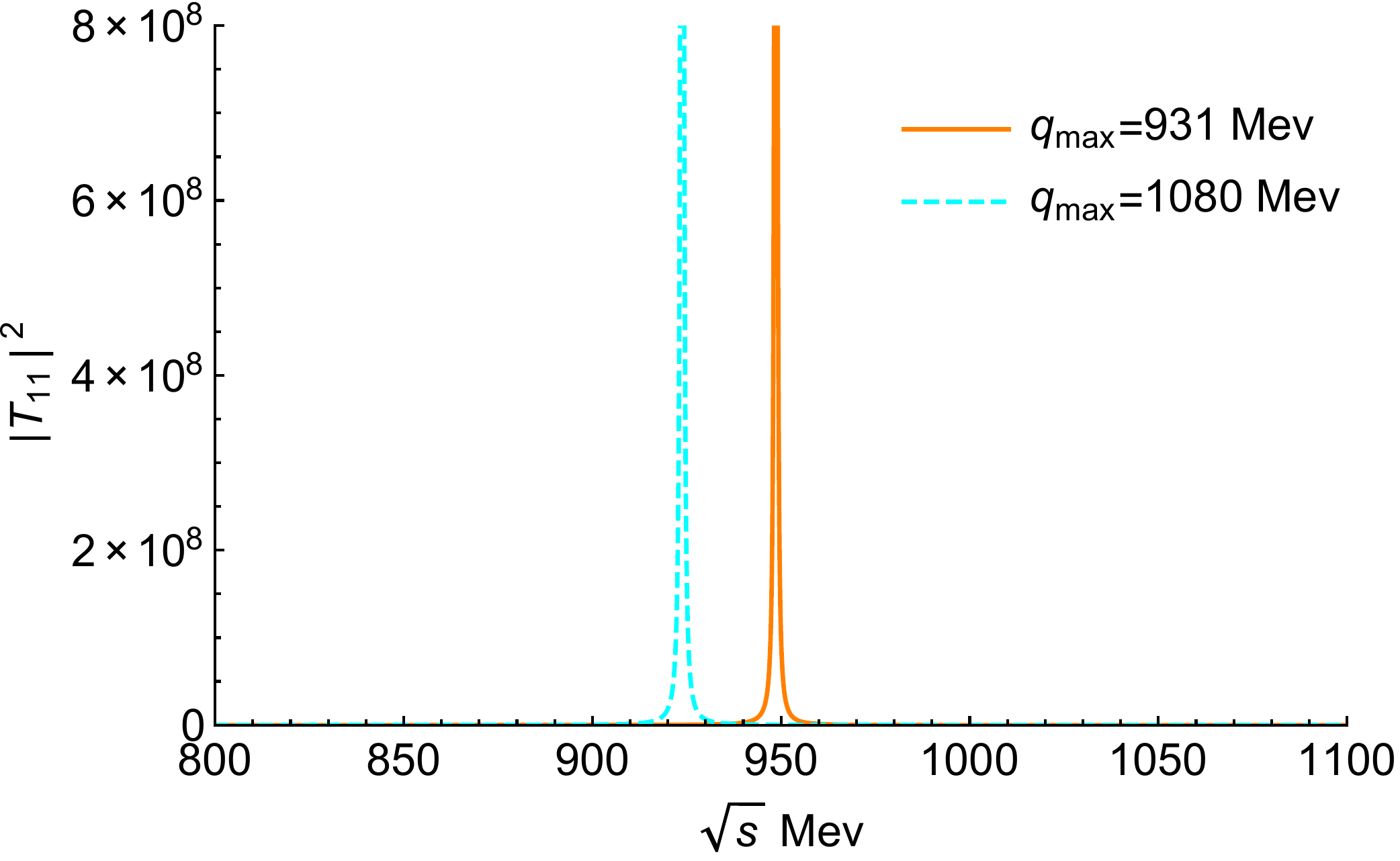} 
  \includegraphics[width=.45\linewidth]{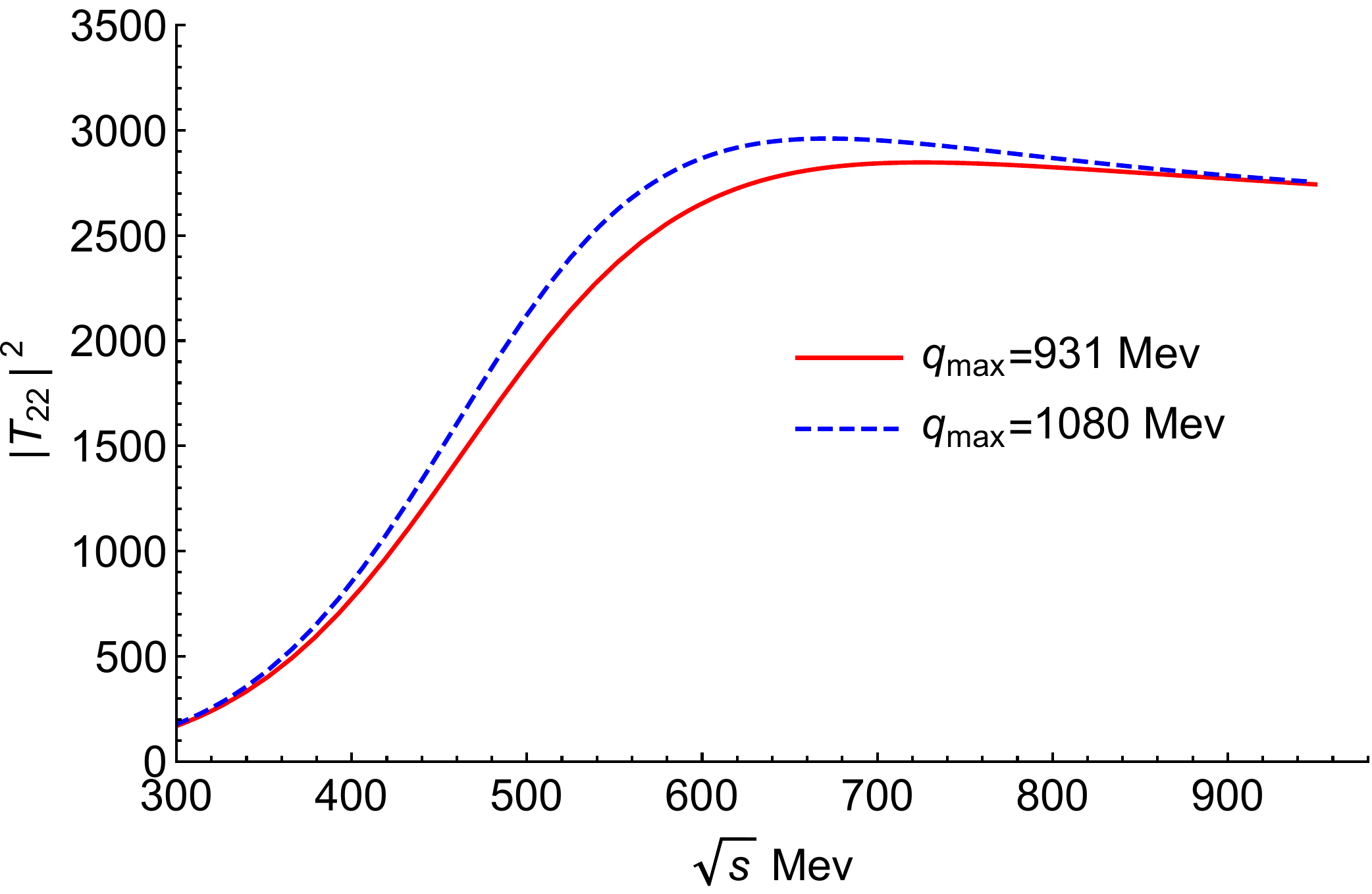} 
  \caption{\footnotesize Modulus squared of the $K\bar{K}$ (left) and $\pi\pi$ (right) scattering amplitudes.}
\label{fig:fig12}
\end{figure}

\begin{figure}
\begin{subfigure}{.45\textwidth}
  \centering
  \includegraphics[width=1\linewidth]{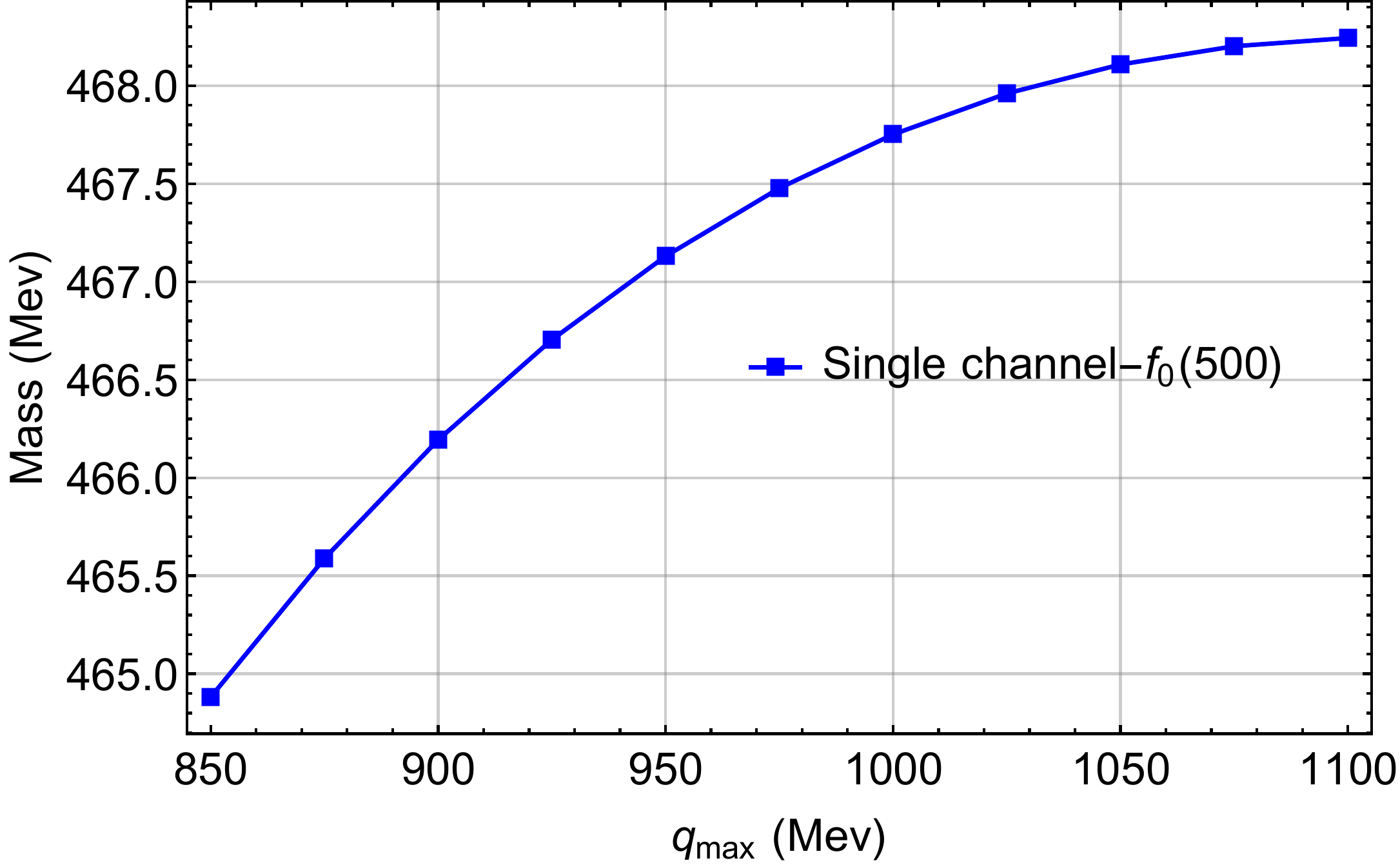} 
  \caption{\footnotesize Mass of the corresponding pole for $\sigma$ state.}
\end{subfigure} 
\begin{subfigure}{.45\textwidth}
  \centering
  \includegraphics[width=1\linewidth]{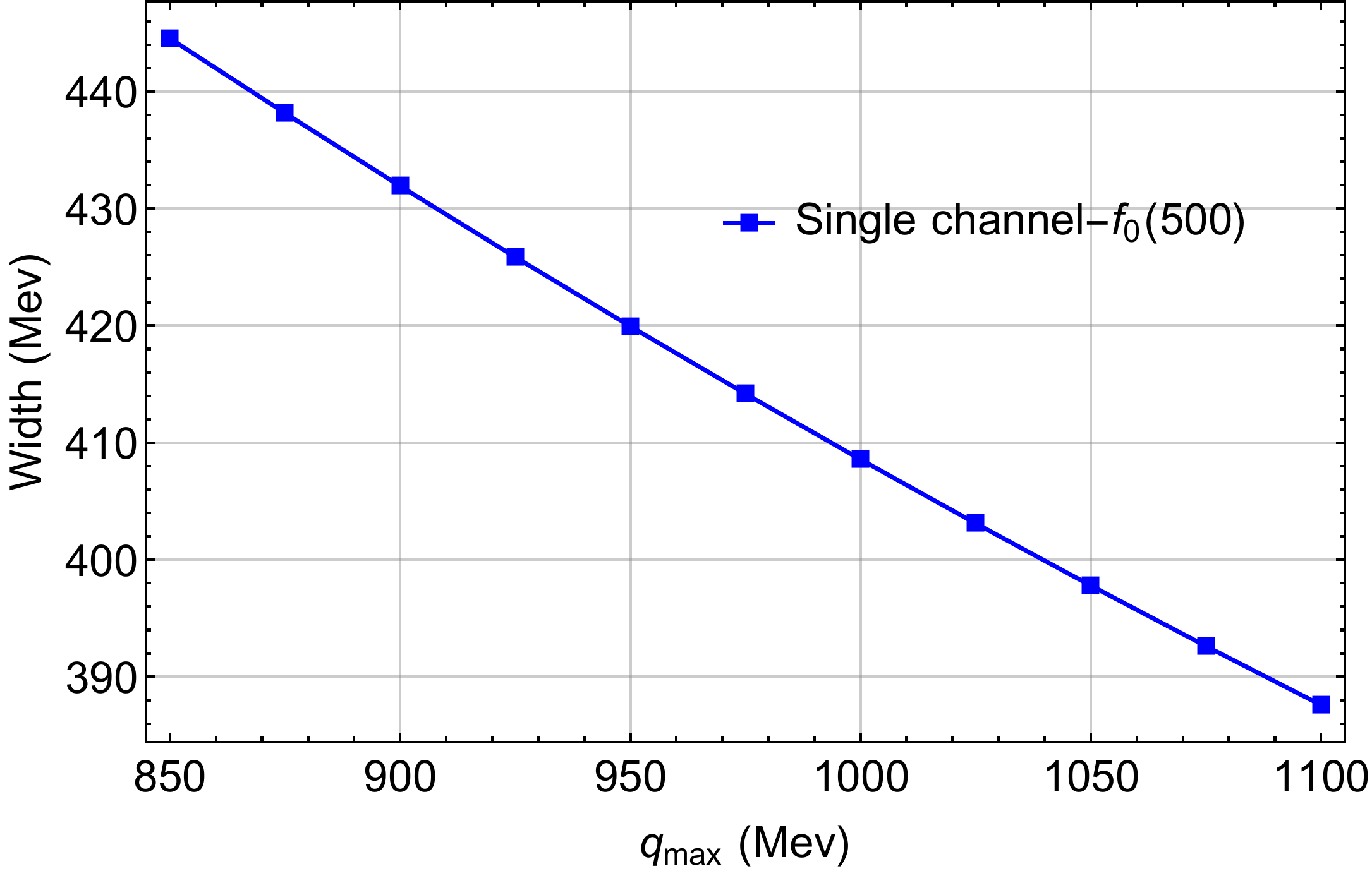} 
  \caption{\footnotesize Width of the corresponding pole for $\sigma$ state.}
\end{subfigure} 
\begin{subfigure}{.5\textwidth}
  \centering
  \includegraphics[width=1\linewidth]{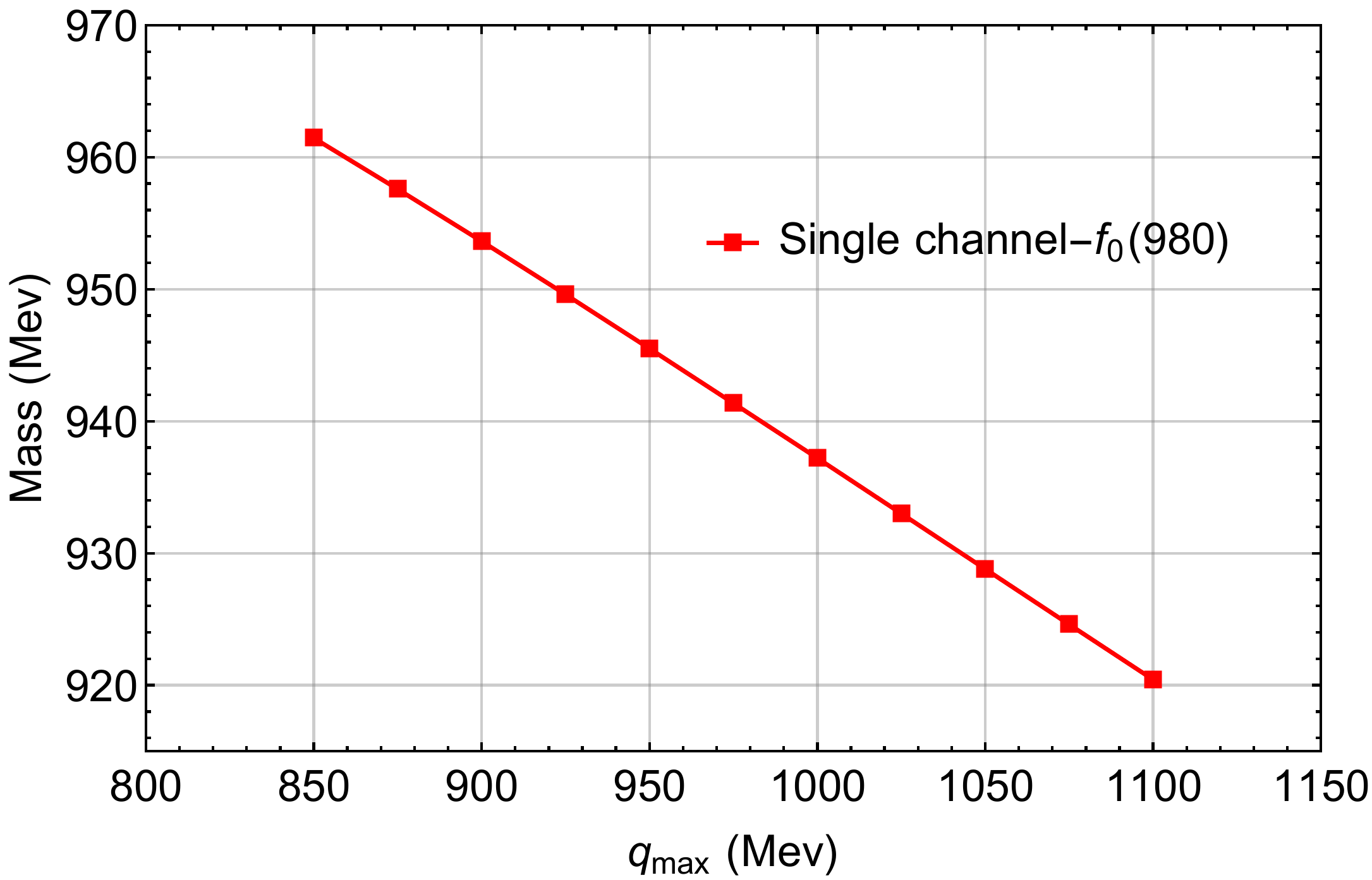} 
  \caption{\footnotesize Mass of the corresponding pole for $f_{0}(980)$ state.}
  \end{subfigure}  
  \caption{\footnotesize Results of the masses and widths for the states of $\sigma$ and $f_{0}(980)$ as a function of the cutoffs.}
\label{fig:fig10}
\end{figure}

\begin{figure}
\centering
  \includegraphics[width=.45\linewidth]{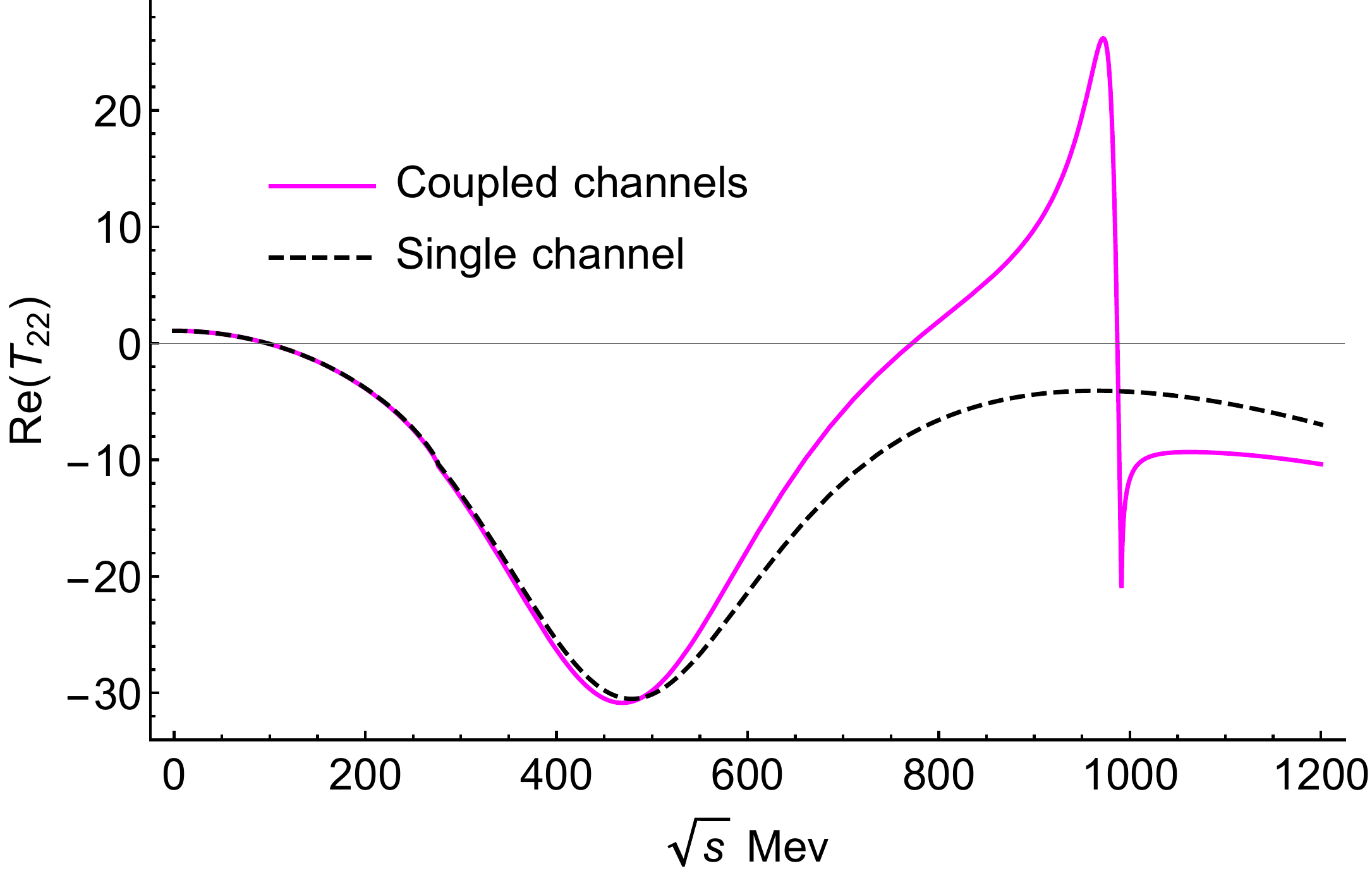} 
  \includegraphics[width=.45\linewidth]{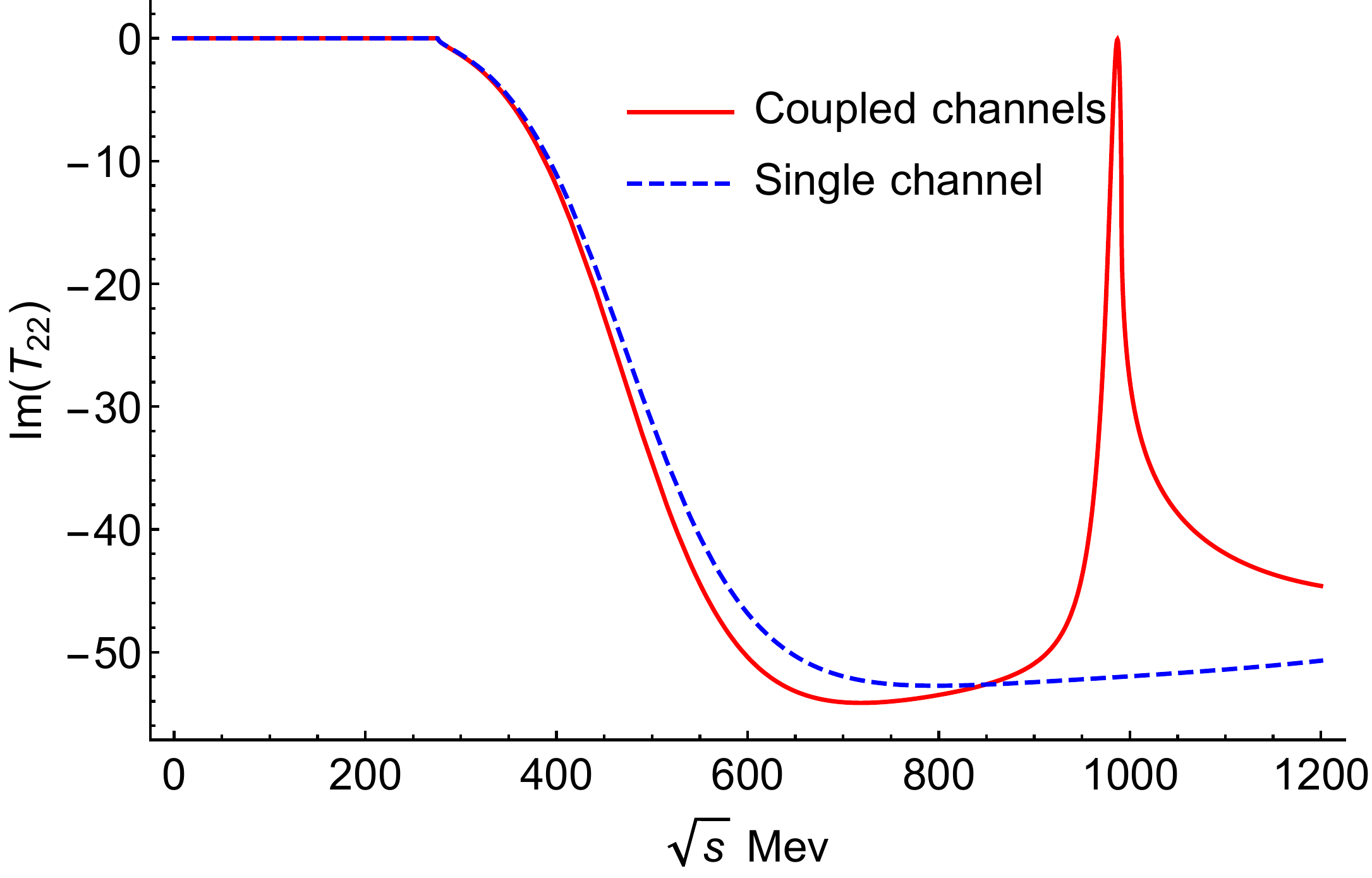} 
  \caption{Real (left) and imaginary (right) parts of the $\pi\pi$ scattering amplitude in coupled and single channels.}
\label{fig:fig11}
\end{figure}

In the previous section, we have investigated the properties of the $\sigma$, $f_{0}$, and $a_{0}$ states in the coupled channel formalism. For the sake of the completeness and the comprehensive understanding of these dynamically generated states, we continue to examine their properties in single channel interactions where one can make a further checking their dominant components. At first, we show the results of the modulus squared of the scattering amplitudes, $|T|^2$, in Fig. \ref{fig:fig12}, where one can see the sharp peak with nearly zero width in the $K\bar{K}$ channel on the left and the wide bump structure in the $\pi\pi$ channel on the right. Next, we search for the corresponding poles in second Riemann sheets. For the $\pi\pi$ channel interaction, as shown in Fig. \ref{fig:fig10} where we vary the cutoffs, we always find the pole in the second Riemann sheet above the threshold of which the mass changes weakly and the width varies not so much as the case of the coupled channel interactions. For the case of the $K\bar{K}$ channel, now the pole keeps below the threshold, and thus, has no width as a pure bound state  since there is no decay channel, see Fig. \ref{fig:fig10} (c), which are more bound compared with the results of coupled channel cases in Fig. \ref{fig:fig8}. Therefore, we can conclude that the $\sigma$ state is a resonance mainly formed by the $\pi\pi$ interaction and the one of the $f_{0}$ state is a bound state of the $K\bar{K}$ component as found in the coupled channel interactions above. To reveal more details, see Fig. \ref{fig:fig11} for the real and imaginary parts of the $\pi\pi$ scattering amplitudes in the coupled and the single channels, one can see that in the region of  the $\sigma$ state appeared, 400--700 MeV, the amplitudes are not affected so much by the coupled channel of $K\bar{K}$, which is a bit far away from the threshold of $K\bar{K}$. Indeed, the structure of the $f_0(980)$ state can be clearly seen closed to the threshold of $K\bar{K}$, as shown in Fig. \ref{fig:fig11}. However, in the isospin $I=1$ sector, the potential of the $K\bar{K}$ channel is too weak to create a pole in the second Riemann sheet when it decouples to the $\pi\eta$ channel, of which the potential is independent with the energy. This means that the coupled channel effects play much important role in the dynamical production of the $a_{0}$ state.

As in case of  the coupled channel interactions, we make a further studies of the compositeness, the wave functions and the radii. The results of the couplings are given in Table \ref{tab:tab7}, even though the strengths of the couplings have lost the relative meanings in the case of the single channel interaction. But, from the results of the compositeness, see Table \ref{tab:tab8}, with the couplings obtained, the compositeness for the $f_0(980)$ state is a bit smaller than the ones of the coupled channel cases, which is consistent with the results of the coupled channel cases in Table \ref{tab:tab3}.

\begin{table}
\center
\caption{Couplings of $\sigma$ and $f_{0}(980)$ states in a single channel.}
\begin{tabular}{|c|c|c|c|c|}
\hline
  $ q_{max} = 931$ MeV & $g_{K\bar{K}}g_{K\bar{K}}(\text{GeV}^{2}) $ & $|g_{K\bar{K}}|(\text{GeV}) $  & $g_{\pi\pi}g_{\pi\pi} (\text{GeV}^{2}) $ & $|g_{\pi\pi}|(\text{GeV}) $ \\ \hline
  $\sigma$ : $466.81 +212.21 i$  &   0              &   0                  &     $-4.41 + 7.77 i$         &   2.98 \\ \hline
  $f_{0}(980)$ : $948.62$      &   26.4              &   5.13                 &     0            &   0  \\ \hline
  $ q_{max} = 1080$ MeV                                                              \\ \hline
  $\sigma$ : $468.213 + 195.8 i$  &   0             &   0                  &     $-3.20 + 8.05 i$           &   2.942                       \\ \hline
  $f_{0}(980)$ : $923.77$      &   29.8              &   5.45                 &     0            &  0                    \\ \hline
\end{tabular}
\label{tab:tab7}
\end{table}

\begin{table}
\center
\caption{Compositeness of $\sigma$ and $f_{0}(980)$ states in single channel.}
\begin{tabular}{|c|c|c|c|c|}
\hline
  $ q_{max}=931$ MeV   &   $( 1- Z)_{K\bar{K}}$  &   $|( 1- Z)_{K\bar{K}}|$  &     $ (1- Z)_{\pi\pi} $  &   $ |(1- Z)_{\pi\pi}| $    \\ \hline
  $\sigma$ : $467.13 +209.968 i$  &   0         &   0                &     $-0.11 - 0.37 i$      &      0.39                       \\ \hline  
  $f_{0}(980)$ : $948.62$      &   0.62         &   0.62                 &     0     &      0   \\ \hline    
    $ q_{max}=1080 $ MeV                                                                            \\ \hline 
  $\sigma$ : $468.213 + 195.8 i$  &  0         &   0                &     $-0.13 - 0.36 i$      &      0.386                       \\ \hline
  $f_{0}(980)$ : $923.77$      &   0.52          &   0.52                 &     0     &      0   \\ \hline
\end{tabular}
\label{tab:tab8}
\end{table}

\begin{figure}
\begin{subfigure}{.45\textwidth}
  \centering
  \includegraphics[width=1\linewidth]{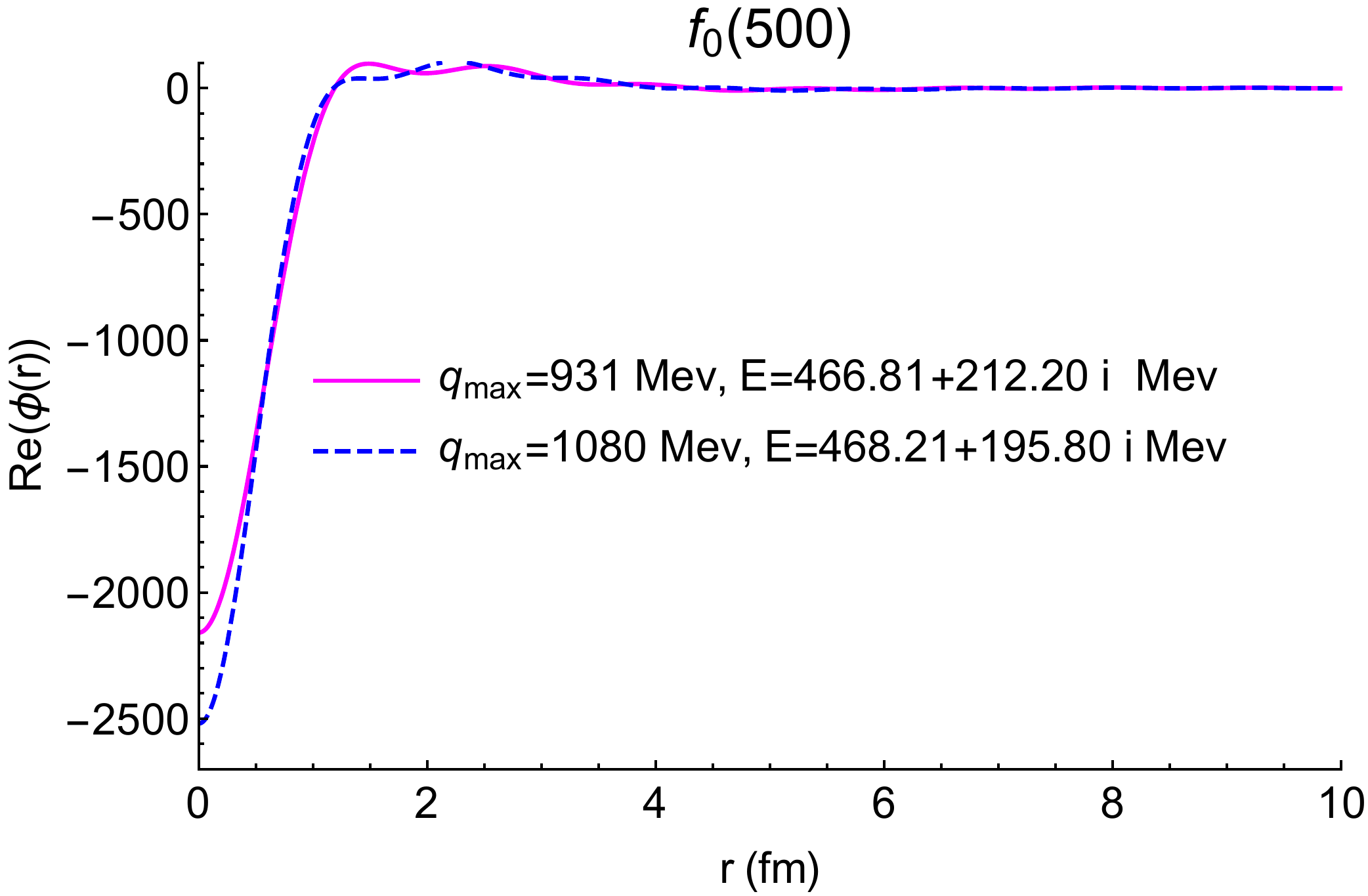} 
\end{subfigure} 
\begin{subfigure}{.45\textwidth}
  \centering
  \includegraphics[width=1\linewidth]{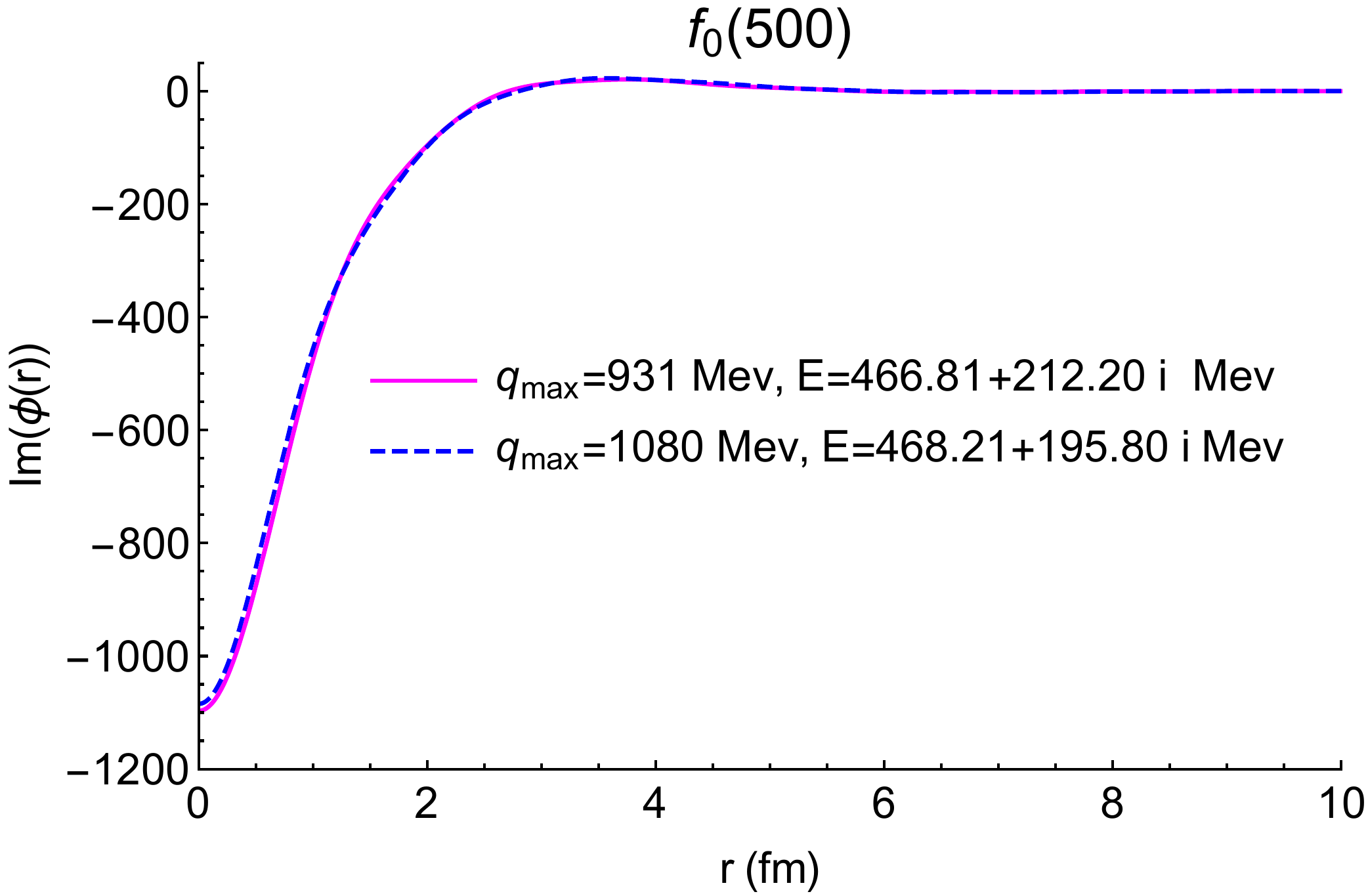} 
  \end{subfigure}
\caption{Wave function of $\sigma$ state in the $\pi\pi$ channel.}
\label{fig:fig13a}
\end{figure}

\begin{figure}
  \centering
  \includegraphics[width=0.6\linewidth]{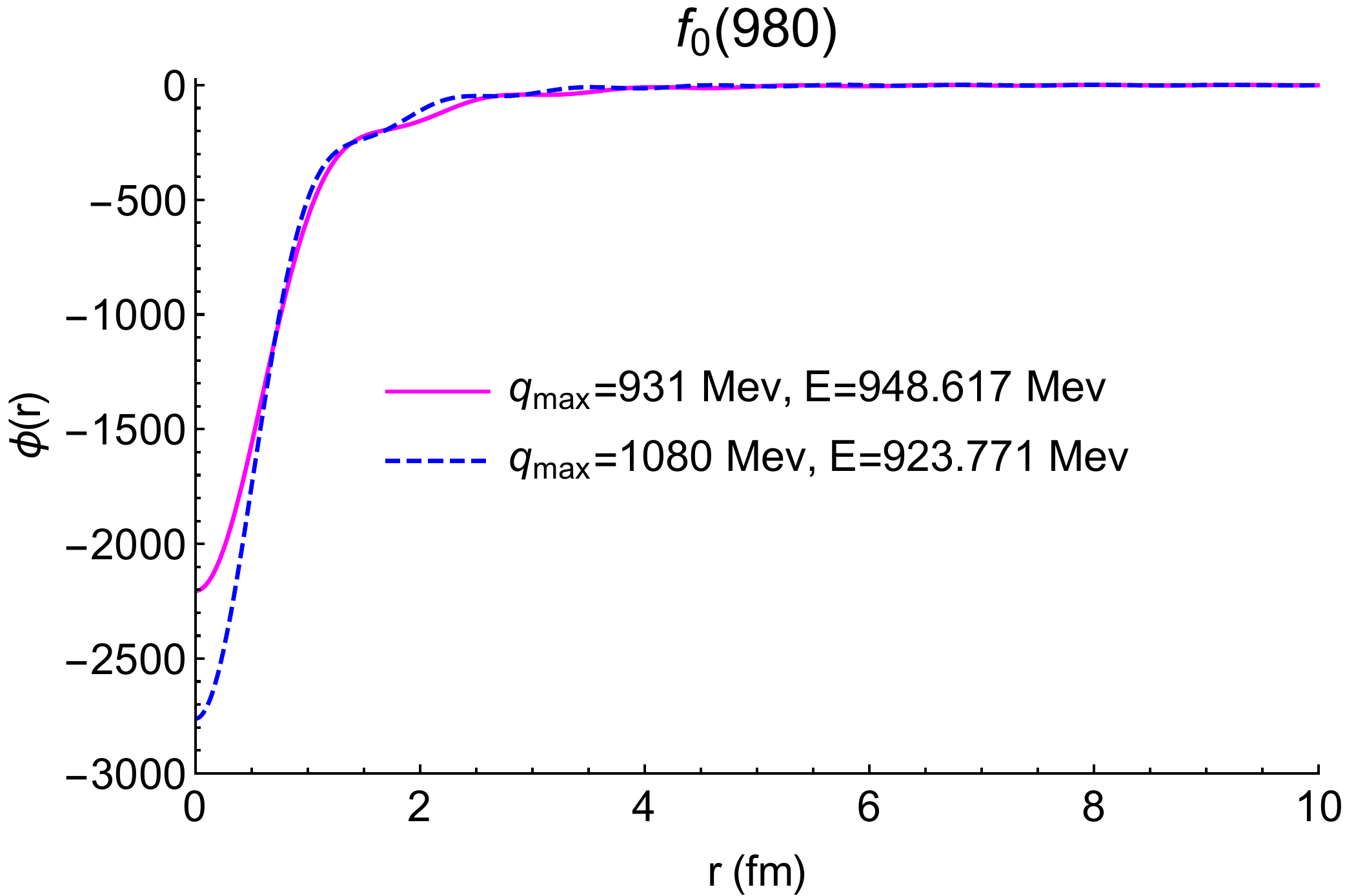}  
  \caption{Wave function of $f_{0}(980)$ state in the $K\bar{K}$ channel.}
\label{fig:fig13}
\end{figure}

\begin{figure}
  \centering
   \includegraphics[width=0.45\linewidth]{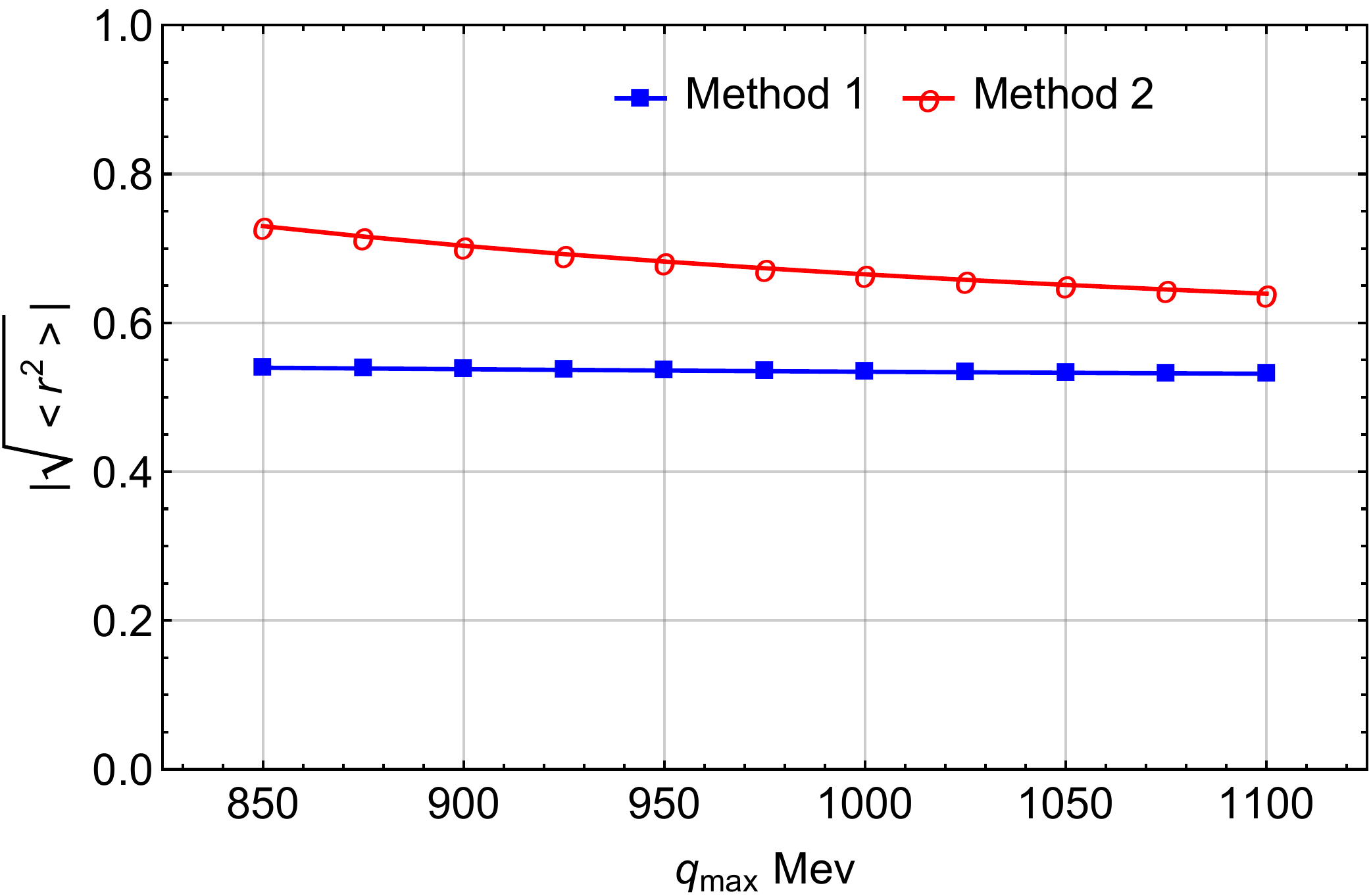} 
   \includegraphics[width=0.45\linewidth]{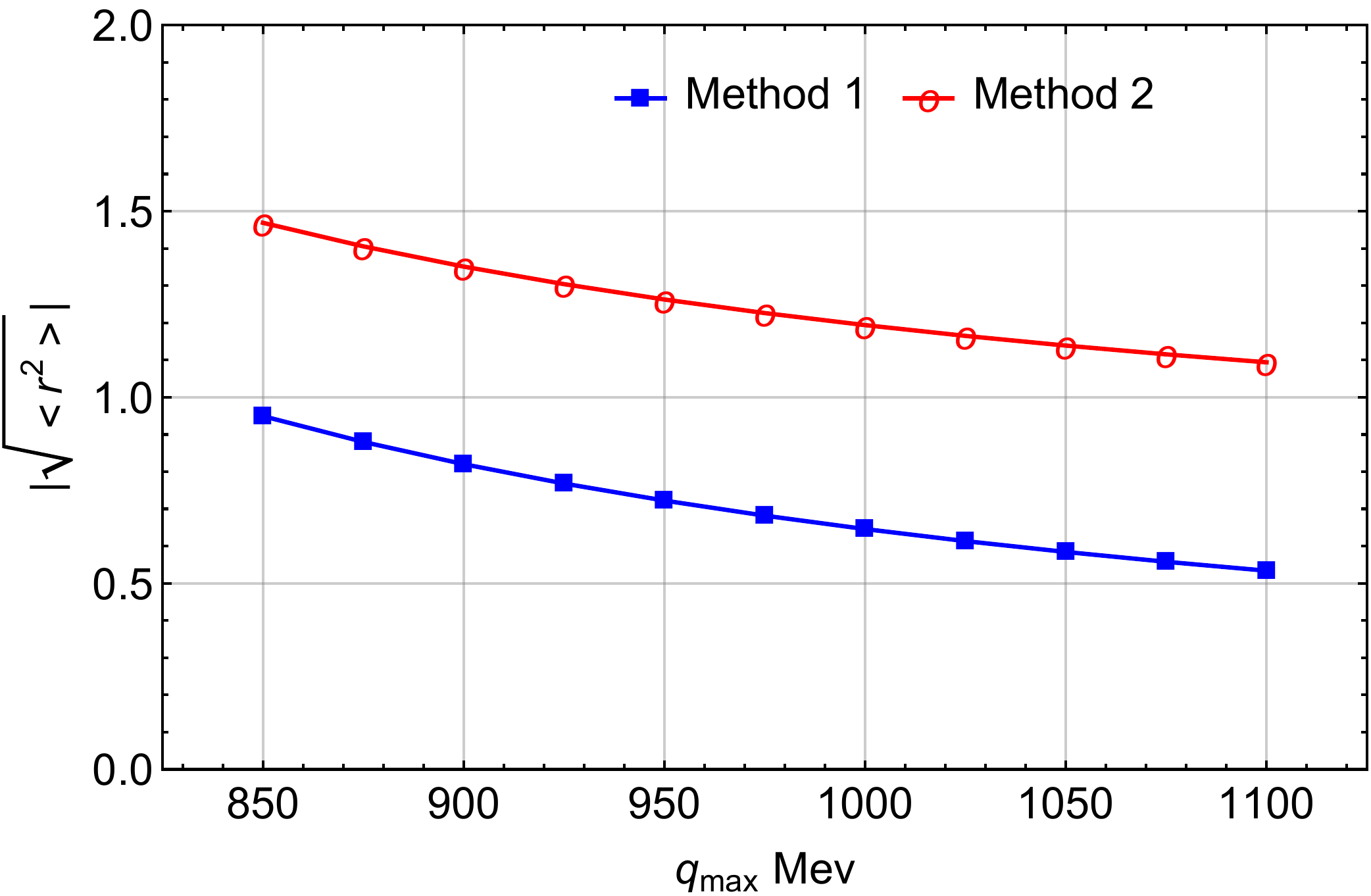} 
  \caption{Radii of the states $\sigma$ (left) and $f_0(980)$ (right) as a function of cutoffs in single channel case, where the methods 1 and 2 are referred the same as in Fig. \ref{rms1}.}
    \label{rms2}
\end{figure}

The wave functions of the $\sigma$ and $f_{0}(980)$ states are shown in Figs. \ref{fig:fig13a} and \ref{fig:fig13}, respectively. And their radii  calculated from the form factor and the tail of the wave functions are given in Tables. \ref{tab:rad1} and \ref{tab:rad2}, respectively, of which the trajectories with different cutoffs are shown in Fig. \ref{rms2}. The results of Tables. \ref{tab:rad1} and \ref{tab:rad2} are consistent with the ones obtained in the coupled channel cases, see Tables. \ref{tab:tab5} and \ref{tab:tab6}. Since now the  $f_{0}(980)$ state is pure $K\bar{K}$ bound state, the radii with the tail of the wave functions in Eq.\eqref{eq32} are well defined and always smaller than the ones evaluated from the form factor with Eq. \eqref{eq33}, compared the right part of Fig. \ref{rms2} with the sub-figure of Fig. \ref{rms1} (a).

\begin{table}
\center
\caption{Radii of the states calculated with Eq. \eqref{eq33} in the single channel case.}
\label{tab:rad1}
\begin{tabular}{|c|c|c|c|c|}
\hline
   Resonances               &  $ q_{max}=931$ MeV  &   $|\langle r^2 \rangle|$     &   $ q_{max}=1080$ MeV  &  $|\langle r^2 \rangle|$  \\ \hline
    $\sigma$   &   0.69 + 0.007 $i$ fm      &  0.69 fm &        0.64 + 0.03 $i$  fm   &  0.64 fm  \\ \hline
    $f_{0}(980)$   &   1.29 fm    &  1.29 fm     &  1.11 fm   &  1.11 fm  \\ \hline
\end{tabular}
\end{table}

\begin{table}
\center
\caption{Radii of the states evaluated with Eq. \eqref{eq32} in the single channel case.}
\label{tab:rad2}
\begin{tabular}{|c|c|c|c|c|}
\hline
   Resonances      &  $ q_{max}=931$ MeV    &  $|\langle r^2 \rangle|$   &   $ q_{max}=1080$ MeV &  $|\langle r^2 \rangle|$    \\ \hline
    $\sigma$   &   0.43 + 0.32 $i$ fm      &  0.54 fm &        0.43 + 0.30 $i$  fm   &  0.53 fm  \\ \hline
    $f_{0}(980)$         &  0.75 fm      &     0.75 fm       &   0.55 fm    &   0.55 fm   \\ \hline
\end{tabular}
\end{table}

\section{Conclusions}

In the present work, we investigate the properties of the $\sigma$, $f_{0}(980)$, and $a_{0}(980)$ states with the chiral unitary approach, where we use the formalisms of the coupled channels and the single channel. Within the isospin limit, two poles are found in the second Riemann sheet in isospin $I=0$ sector corresponding to the $\sigma$ and $f_{0}(980)$ resonances, and a pole in $I=1$ sector is found, which corresponds to the $a_{0}(980)$ state. In the case of the single channel calculations, we find the corresponding poles of the $\sigma$ and $f_{0}(980)$ states in the $\pi\pi$ and $K\bar{K}$ channels with $I=0$, respectively. However, in $I=1$ sector the potential of the $K\bar{K}$ channel is too weak to create a pole in the second Riemann sheet. When we vary the only one free parameter of the cutoff, these states are stably dynamically generated both in the coupled channel and the single channel formalism. Besides, we also predict the phase shifts in $I=1$ sector with the coupled channel formalism.

Furthermore, we studied the couplings, the compositeness, the wave functions,  and the mean-squared distance of these dynamically generated states in both the coupled channels and the single channel formalisms. From the results of the couplings and the compositeness, we conclude that the $f_{0}(980)$ state is essentially made by the $K\bar{K}$ component, which is about 80\%, and has very small parts of $\pi\pi$. However, the $\sigma$ state has the main contributions from the $\pi\pi$ channel, of which the component amounts to about 40\%, and has quite small quantity of the $K\bar{K}$ component. Thus, the $\sigma$ resonance has a large parts of something else except for the molecular components. For the case of the $a_{0}(980)$ state, the $\pi\eta$ channel has important contributions to its generations in the coupled channel interactions. Even though it is dominated by the $K\bar{K}$ component with 55\%, it also has large contributions of about 16\% from the $\pi\eta$ component. With the wave functions obtained, we calculate the radii of these states and get $|\langle r^2 \rangle|_{f_0(980)} = 1.80 \pm 0.35$ fm, $|\langle r^2 \rangle|_{\sigma} = 0.68 \pm 0.05$ fm and $|\langle r^2 \rangle|_{a_0(980)} = 0.94 \pm 0.09$ fm, which can be tested in the future experiments. Finally, from our results of the couplings, the compositeness, the wave functions  and the radii, we can conclude that the $f_{0}(980)$ state is mainly a $K\bar{K}$ bound state, the $\sigma$ state a resonance of $\pi\pi$ and the $a_{0}(980)$ state a loose $K\bar{K}$ bound state.

\section*{ACKNOWLEDGMENTS}

We thank Prof. B. S. Zou for the careful reading the script and the useful comments, and also acknowledge Profs.  N. N.  Achasov, J. M. Fr\`ere and K. Azizi for the useful comments and the valuable informations.

 \addcontentsline{toc}{section}{References}
\end{document}